\documentclass[11pt,a4paper]{article}
\usepackage[UTF8,fontset=fandol,scheme=plain]{ctex}
\usepackage[margin=2.54cm]{geometry}
\usepackage{graphicx}
\usepackage{longtable,booktabs,array}
\usepackage{calc}
\usepackage{amsmath,amssymb}
\usepackage[hidelinks]{hyperref}
\usepackage{xurl}
\newcommand{\bibentry}[1]{\par\noindent\hangindent=1.27cm\hangafter=1 #1\par\medskip}
\hypersetup{pdftitle={Poisoning Attacks and Collaborative Defense Framework for LightRAG-Based Construction Regulation Knowledge Bases: A Case Study of Current UK Building Standards},pdfauthor={Yingyan He}}
\title{Poisoning Attacks and Collaborative Defense Framework for LightRAG-Based Construction Regulation Knowledge Bases: A Case Study of Current UK Building Standards}
\author{Yingyan HE\\[0.3em]\normalsize\emph{Faculty of Science and Technology, SWISS UMEF, Geneva, Switzerland}\\ \small yingyanhe@ieee.org, bulletsho@gmail.com}
\date{Master Thesis \textperiodcentered{} November 2026}
\begin{document}
\maketitle
\thispagestyle{empty}

\begin{abstract}
\emph{Retrieval-augmented generation (RAG) is now widely used across industries, including architecture, engineering and construction (AEC), and research on RAG security is growing rapidly; yet the security of building-regulation knowledge bases has received little study. This thesis therefore reports a pre-registered experiment on one fully instrumented LightRAG system built on the UK Approved Documents. Using a 150-item benchmark labeled with safety consequences and tampering directions, it evaluates four attacks (text layer, graph layer, fabricated amendment narrative and a defense-aware adaptive attacker) together with a three-layer retrieval--graph--generation defense in a full factorial design. Text-level poisoning succeeded, with success rising as the budget increased. Graph-level poisoning was less effective, contrary to the registered hypothesis. The full defense met the registered criterion and six of the seven configurations met its threshold, but those containing the generation layer did so by refusing 69--80 percent of poisoned queries, at a cost of up to 74 points of clean accuracy, whereas provenance-based retrieval filtering produced no measurable loss of clean accuracy in the configuration used. A second experimental batch showed that excluding a legitimate unregistered amendment caused no measurable loss of accuracy either, because its wording reached the generator through the graph\textquotesingle s entity and relationship descriptions, the route the poison took, which no layer inspects: the three layers act on text chunks, anchored relation edges and cited evidence, whereas the descriptions carry no document provenance. The layers interacted so strongly that the contribution of each cannot be isolated. A reported zero attack success rate is therefore not evidence of detection unless the refusal rate is reported with it: zero arises either because provenance filtering removes every document outside the manifest without judging its content, or because the system refuses. In the registered batch the generation layer never returned a contradiction verdict, only neutral ones (cannot be confirmed). Layered defenses must therefore be evaluated factorially, and RAG assistants for regulatory knowledge require corpus admission control and version-aware verification before they can support safety-critical decisions.}

\end{abstract}

\section{Summary}\label{summary}

This thesis examines whether a retrieval-augmented question-answering system for building regulations can be poisoned, and whether a layered defense can stop such poisoning without destroying the system\textquotesingle{}s usefulness. The experiments use LightRAG, which answers from text chunks and from an automatically built knowledge graph, over the UK Approved Documents and the Manual to the Building Regulations. The question matters in this domain because a wrong escape-stair width or guarding height arrives in the same confident, well-cited form as a correct one, and because regulatory corpora are assembled from many sources and updated by amendment, which gives material of uncertain provenance realistic ways into a knowledge base.

The research question was decomposed into four sub-questions, covering the text layer, the graph layer, the defense and the unevenness of harms, and ten hypotheses with their tests and thresholds were registered on 4 October 2026, before the first attack run; the confirmatory family of seven tests was protected with Holm correction. A 150-item benchmark was written from the corpus, each item anchored to the provision it rests on and labeled with a safety-consequence weight and a paired loosening or tightening mutation. A poisoned query counted as a success only when the answer matched none of the required values, contained the attacker\textquotesingle{}s value and was not a refusal. Four attacks were implemented: PoisonedRAG-style text injection, GRAGPoison-style shared-relation poisoning of the graph, a fabricated ``amendment booklet'' narrative and a defense-aware adaptive attacker. A three-layer defense (provenance-aware retrieval filtering, anchor-based graph verification and NLI citation checking) was evaluated in all eight combinations of its layers, each configuration run three times.

Text-level poisoning succeeded on 33 of 150 targets at five passages per target, a rate of 0.220 against a registered bound of 0.10 (p = 1.17 \ensuremath{\times} 10\textsuperscript{\ensuremath{-}5}), and its success rose with the budget, although the poison was already retrieved every time at a budget of one passage. Graph-level poisoning succeeded on 22 of 150 targets, the opposite of the registered prediction, so the graph was not the softer target. The full defense met its registered criterion and six of the seven configurations met its threshold. However, every configuration containing the NLI layer did so by refusing 69--80 percent of poisoned queries, at a cost of up to 74 points of clean accuracy, whereas provenance filtering showed no measurable loss of clean accuracy in the configuration used, by a margin of only 0.36 points; a registered variant with a different graph-side path lost 32 points. The layers interacted strongly, the adaptive attacker gained nothing, and the fabricated amendment notices were repeated in all 18 answers, often in answers that value-based scoring judged correct.

The harms were also uneven in ways that a single rate hides. Weighting errors by their safety consequence barely changed the error rate (a mean tilt of +0.005, within its uncertainty), and neither loosening nor tightening mutations were favored (ten discordant pairs each way). Under poisoning, strict accuracy fell by 21.6 points while the share of strongly assertive answers fell by only 4.3 points: wrong answers were stated no less confidently than right ones, so correctness could not be judged from tone. And on neither of the two items asking which instrument a value belongs to, in any of the three indexes, did the system name the instrument that governs it.

A second experimental batch (6--8 November 2026) sharpened these findings. Excluding a legitimate amendment from the manifest caused no measurable loss of accuracy, because the filter withholds only an unregistered document\textquotesingle{}s text chunks, while the amendment\textquotesingle{}s wording reached the prompt through the graph\textquotesingle{}s entity and relationship descriptions in 14 of the 16 measurable items, the route the poison used. With the graph removed, text-level poisoning remained effective. Under poisoning, the model\textquotesingle{}s self-rated confidence became markedly less calibrated. An LLM judge from a different provider, used as an independent second scorer, agreed with the strict rubric on 50 of 58 explanatory answers.

These results change what a low attack success rate can be taken to mean. Under the defense, a zero arose in two ways: provenance filtering removed every document outside the manifest without judging its content, and the generation layer refused; in the registered batch it never returned a contradiction verdict, only neutral ones. The graph, in turn, proved to be a second attack surface rather than a second line of defense, because its descriptions carry no document provenance and no layer inspects them.

The thesis concludes that a low attack success rate under defense must be reported together with the refusal rate and clean accuracy, that a layered defense must be evaluated with a full factorial design, and that regulatory RAG systems need controlled corpus admission and version-aware verification. For organizations, this means admitting documents, including official amendments, only through an authenticated and logged process enforced before indexing, requiring answers to state the instrument, edition and commencement date they rely on, and keeping human verification for safety-critical values. For researchers, it means reporting the refusal share and the clean-accuracy cost beside every attack success rate.

These findings describe one system (LightRAG 1.5.8), one serving model, one jurisdiction\textquotesingle{}s regulations in one language and one curated experimental index; they may not transfer to open corpora or to systems with different retrieval parameters.

\section{List of Figures and Tables}\label{list-of-figures-and-tables}

\textbf{Figures}

Figure 4.1 The research framework: where the attacks enter the LightRAG pipeline and where the three defensive layers act

Figure 4.2 Distribution of the LLM judge\textquotesingle s scores, split by the strict rubric\textquotesingle s verdict (58 answers)

Figure 5.1 Strict success and disruption of Attack A1 by budget and fold

Figure 5.2 Strict success and disruption of the text-level (A1) and graph-level (A2) attacks at the same budget, by fold

Figure 5.3 Answered share and poisoned accuracy (refusal counted as wrong) for the eight configurations

Figure 5.4 Clean accuracy and poisoned accuracy for the eight configurations

Figure 5.5 The retrieval layer\textquotesingle s effect on poisoned accuracy with and without the generation and graph layers

Figure 5.6 Records not refused by the control and by each adaptive arm, with only the generation layer armed

Figure 5.7 The severity tilt D = SW-ER \ensuremath{-} ER for each run, with its item-clustered 95 percent bootstrap interval

Figure 5.8 Fraction of strictly correct answers at each self-rated confidence level, clean and poisoned index

\textbf{Tables}

{\small\textbf{Table 4.1} The experimental index}

{\small\textbf{Table 4.2} Composition of the 150-item benchmark}

{\small\textbf{Table 4.3} The four attacks}

{\small\textbf{Table 4.4} The three defensive layers}

{\small\textbf{Table 4.5} Second batch: agreement between the LLM judge, the strict rubric and the author}

{\small\textbf{Table 5.1} Outcome of the registered hypotheses}

{\small\textbf{Table 5.2} Attack A1 by fold and budget}

{\small\textbf{Table 5.3} Attack A2 by fold at M = 5}

{\small\textbf{Table 5.4} The three amendment-narrative cases}

{\small\textbf{Table 5.5} Defense configurations against H4a, with refusal and accuracy}

{\small\textbf{Table 5.6} Second batch: provenance filtering and a legitimate amendment (20 dependent items)}

{\small\textbf{Table 5.7} Marginal effect of each layer on poisoned accuracy, by context}

{\small\textbf{Table 5.8} Adaptive arms against the same 72 poisoned items, undefended}

{\small\textbf{Table 5.9} Second batch: adaptive arms against the generation layer alone}

{\small\textbf{Table 5.10} Accuracy and expressed certainty, clean and poisoned corpus}

{\small\textbf{Table 5.11} Second batch: self-rated confidence against strict correctness}

{\small\textbf{Table 5.12} Version handling on the clean corpora (counts)}

{\small\textbf{Tables A-1 to A-9} Data tables (Appendix A)}

{\small\textbf{Tables B-0 to B-3} Experimental environment, instruments and artifact locations (Appendix B)}

{\small\textbf{Table C-1} Benchmark item QA-B-21 (Appendix C)}

{\small\textbf{Tables D-1 and D-2} Second experimental batch and literal exact match (Appendix D)}

{\small\textbf{Tables E-1 and E-2} Registered hypotheses and dated amendments (Appendix E)}

\section{Glossary}\label{glossary}

\textbf{Budget M:} the number of poison passages injected for one target question (N in PoisonedRAG).

\textbf{ASR (attack success rate):} the share of poisoned queries whose answer omits the correct value, contains the attacker\textquotesingle s value and is not a refusal.

\textbf{Anchor table:} a frozen set of structured assertions (subject, relation, comparator, value, unit) extracted from the authoritative text and used by the graph-side defense.

\textbf{Disruption:} removal of the correct value from the answer, whether or not the attacker\textquotesingle s value replaced it.

\textbf{ECE (expected calibration error):} the share-weighted gap between stated confidence and observed accuracy across confidence levels (Equation 4.11).

\textbf{LLM judge:} a second, independent scorer that rates each answer\textquotesingle s consistency with the provision on a 1--5 scale without seeing the rubric or the experimental condition (Section 4.10).

\textbf{LightRAG:} an open-source graph-based RAG framework that retrieves from text chunks and from an automatically extracted entity--relation graph.

\textbf{NLI (Natural Language Inference):} A model that classifies whether a premise entails, contradicts, or is neutral to a hypothesis; used here to check the relationship between an answer sentence and its cited evidence.

\textbf{Plain accuracy:} accuracy over all records, with a refusal counted as a wrong answer.

\textbf{Strict accuracy:} accuracy over answered records only.

\textbf{Retreat:} a generation-layer outcome in which the answer is withheld or replaced by a conservative reply because the NLI checker cannot confirm it against the cited evidence; the retreat rate is the share of records whose third-layer check triggered a retreat.

\section{Chapter 1: Introduction and Background}\label{chapter-1-introduction-and-background}

\subsection{1.1 Research Topics and Key Points}\label{research-topics-and-key-points}

Large language models (LLMs) are increasingly deployed to answer questions about technical regulation, usually through retrieval-augmented generation (RAG): the system retrieves passages relevant to the query and conditions the generator on them, so that the answer can be tied to a source. For building regulations, which are long, frequently revised and written in a register where a single value carries legal weight, a system that can point to the paragraph it relied on is easier to trust and audit.

The same property creates an attack surface: whatever the retriever can reach, the generator reads as evidence. An adversary who can place documents into the store needs no access to the model, its prompt or its operator. PoisonedRAG showed that a handful of such passages can redirect a RAG system\textquotesingle s answers to attacker-chosen targets (W. Zou et al., 2025).

This thesis studies that problem in one concrete and safety-relevant setting. Its object is LightRAG, an open-source RAG framework that combines vector retrieval over text chunks with an automatically constructed entity--relation graph (Z. Guo et al., 2025); the source code used is listed in Appendix B. Its corpus is the Approved Documents published by the UK government, together with the government\textquotesingle s Manual to the Building Regulations. It asks how an adversary who can contribute documents can make such a system return a wrong value for a specific provision, and whether a defense spread across retrieval, graph and generation can prevent that without destroying its usefulness for honest queries.

\subsection{1.2 Motivation}\label{motivation}

In practice, the architecture, engineering and construction (AEC) sector has begun to adopt LLM-based tools for regulatory work. Uhm et al. (2025) evaluated retrieval-augmented LLMs for generating construction safety information and found retrieval to improve the reliability of the output; Shi et al. (2025) combined a fine-tuned LLM with retrieval to turn building rules into draft scripts for automated compliance checking; and J. Chen et al. (2026) adapted general-purpose LLMs to building-code analysis by instruction tuning rather than domain-specific pre-training. The same pattern is visible in security operations, where RAG now supports vulnerability risk perception and penetration testing (Gu et al., 2026; Jiang et al., 2025). These studies are concerned with accuracy under honest conditions. None of them asks what happens when part of the document store is controlled by someone who wants the system to be wrong.

This question matters more in this domain than in many others. A wrong escape-stair width or guarding height arrives in the same confident, well-cited form as a correct one, so the reader cannot detect it, and its consequence may be a building that does not protect its occupants.

Regulatory corpora are also assembled from many sources and updated by amendment, which gives material of uncertain provenance realistic ways into a knowledge base.

The second motivation is methodological. Attack success rates in the RAG poisoning literature are reported under different scoring rules, corpora, retrievers and conventions for refusals, and both a recent security benchmark and a recent review argue for unified evaluation settings and benchmarks precisely because evaluation practice has been fragmented (X. Liang et al., 2025; Mu et al., 2026, p. 37). Questions such as whether a graph layer makes a system harder to poison, or what each component of a layered defense contributes, have therefore been answered by juxtaposing papers rather than by controlled comparison on one system.

\subsection{1.3 Gaps in Existing Literature}\label{gaps-in-existing-literature}

The first is a gap of domain. The closest graph-RAG security studies use general-purpose or non-engineering corpora such as multi-hop question answering, medical text and encyclopedic material (J. Liang et al., 2026; Q. Chen et al., 2026; Yan et al., 2026), and the broadest RAG security benchmarks omit graph-based systems (X. Liang et al., 2025). In a building-regulation knowledge base, by contrast, an attack can be scored against a specific provision value whose direction and consequence can be stated, yet research in this setting remains limited.

The second is a gap of comparison. Text-layer and graph-layer attacks have each been shown to work on their own testbeds, but whether shared-relation poisoning is more or less effective than passage injection at a matched budget has not been tested on one index under one scoring rule. Defenses are likewise mostly evaluated one at a time (Xiang et al., 2024; Alizadeh Noughabi et al., 2026; Wu et al., 2026). The layered defenses that report the contribution of each layer combine text-side layers on a vector-only pipeline (Kumar et al., 2026), so whether retrieval-, graph- and generation-side layers compose or interfere in a graph-based system is unknown.

The third is a gap of measurement. Most papers lead with a binary attack success rate (ASR) and rarely report the share of queries a defense declines to answer, so a defense that stopped the attack is hard to tell from a system that stopped answering; the rate also ignores whether the attacked provisions were safety-critical, whether a requirement was loosened or tightened, and whether wrong answers sound as certain as right ones. Studies of attack severity (M. Zou et al., 2026) and of verbal confidence (J. Liu et al., 2026) suggest that these dimensions matter, but they have not been examined in a domain-specific knowledge base such as a regulatory one.

\subsection{1.4 Research Objectives and Research Questions}\label{research-objectives-and-research-questions}

The thesis aims to produce an empirically grounded and reproducible security evaluation of a graph-based RAG system for building regulations, covering both attack and defense. Following the approved proposal, the objectives are:

1. to characterize the poisoning attack surface of LightRAG\textquotesingle s dual vector--graph architecture on a real regulatory corpus;

2. to implement representative attacks against the text layer and the graph layer, together with a qualitative narrative attack and a defense-aware adaptive attacker, and to measure their effectiveness;

3. to implement a three-layer retrieval--graph--generation defense and evaluate it factorially, so that the contribution and interaction of each layer can be estimated;

4. to build a reusable benchmark of regulation questions with safety-consequence labels and directional poison seeds; and

5. to evaluate outcomes beyond binary ASR, namely safety-consequence weighting, directional bias, confidence--correctness alignment and natural version conflict.

These objectives serve a single research question, stated in the proposal and retained here without change:

In a LightRAG system built on current UK building regulations, which poisoning strategies achieve clause-level retrieval hijacking and generation misdirection through its vector and graph layers, and how should a defense combining retrieval-side anomaly filtering, graph-side evidence verification and generation-side fact verification be designed so that the attack success rate is reduced to an acceptable level while clean question-answering accuracy is essentially maintained?

Chapter 3 decomposes this question into four sub-questions and states the hypotheses that were registered before any experimental run.

\subsection{1.5 Significance}\label{significance}

Academically, the study offers a within-system comparison of attacks on two layers of a graph-based RAG architecture and a factorial measurement of a layered defense under one scoring rule, together with a measurement argument for any defended RAG system: a zero attack success rate is not evidence of detection unless the refusals behind it are reported. Practically, it shows organizations how building-regulation question-answering assistants fail under adversarial input and which defenses work at little cost; the benchmark and scripts form a reusable testbed.

\subsection{1.6 Research Setting and Data Collection Strategy}\label{research-setting-and-data-collection-strategy}

The UK Approved Documents were chosen as the research setting for four reasons. First, the corpus states values: widths in millimeters, distances in meters, periods of fire resistance in minutes, each attached to a numbered paragraph or table. A value can be checked, so a successful attack can be scored as the replacement of a specific value rather than as a vague change of tone. Second, many values carry a direction, so that a tampered value either loosens or tightens a requirement. Third, the guidance is layered by amendment, so that several official editions of the same provision can coexist, which creates genuine ambiguity about which instrument governs. Fourth, the documents are public and freely available, which allows the corpus to be redistributed and the experiment to be repeated.

The data were collected through controlled experiments rather than through observation of a deployed system. The official documents were downloaded from GOV.UK, registered in a manifest with their hashes and indexed by LightRAG; all confirmatory experiments use the resulting 850-chunk index. A 150-item benchmark was frozen before the first attack, every attack ran against an isolated copy of the index, and every answer and score was logged per item and repeat (Chapter 4).

\subsection{1.7 Thesis Structure}\label{thesis-structure}

The thesis follows the prescribed structure and is divided into nine chapters:

Research background, significance and setting (Chapter 1), literature and theoretical framework (Chapter 2), problem and hypotheses (Chapter 3), methodology (Chapter 4), results (Chapter 5), discussion (Chapter 6), conclusion (Chapter 7), recommendations (Chapter 8) and limitations (Chapter 9). The appendices hold the data tables, the experimental environment and instruments, an example benchmark item, extended results and the pre-registration record.

\section{Chapter 2: Literature Review and Theoretical Framework}\label{chapter-2-literature-review-and-theoretical-framework}

This chapter has two parts. The literature review (Sections 2.1 to 2.5) is organized around the questions this thesis asks rather than paper by paper; the theoretical framework (Section 2.6) states the definitions and commitments on which the design rests.

\subsection{2.1 Conceptual Framework}\label{conceptual-framework}

The study models a graph-based RAG system as an information path with three stages. \textbf{Retrieval} selects candidate evidence: text chunks ranked by embedding similarity and, in a graph-based system, entities and relations reached from query keywords. At the \textbf{graph} stage, entities, typed relations and their descriptions extracted at indexing time are assembled into the context. At \textbf{generation}, the LLM composes an answer from that context and, in LightRAG, cites its source documents (Z. Guo et al., 2025).

A poisoning adversary enters this path at indexing time by contributing documents. A defense can filter what is retrieved, verify what the graph asserts or check what is generated; and an answer can be right, wrong or withheld, so the measurement must account for all three outcomes.

\subsection{2.2 Text-Level Poisoning of Retrieval-Augmented Generation}\label{text-level-poisoning-of-retrieval-augmented-generation}

The foundational result for this study is PoisonedRAG (W. Zou et al., 2025, p. 9). The attacker crafts a few passages per target question that satisfy a retrieval condition, ranking among the top results for the query, and a generation condition, leading the LLM to the attacker\textquotesingle s chosen answer. In the black-box variant the retrieval condition is met by prepending the target question itself to the malicious text. Five passages per target achieved high success rates on several benchmarks. A standalone reproduction reported that success diluted as the knowledge base grew from tens to over a thousand documents (J. Lin, 2026), and another study carried content poisoning into deployed application frameworks, where normal-looking reference content induced harmful answers in most attempts across five open models (Zeng, 2025).

Subsequent work has varied the mechanism along two axes. The first is optimization: Li et al. (2026) plan adversarial passages globally with Monte Carlo tree search, Z. Wang et al. (2026) refine them word by word against the retriever while keeping them fluent, Qian (2026) hijacks retrieval with poisons that preserve the semantics of the surrounding text, Jung et al. (2026) camouflage adversarial content among benign chunks, and Thornton (2026) shows that a poison\textquotesingle s effectiveness depends on the corpus it enters. The second axis is the target of the attack: Sun et al. (2026) aim at the knowledge claim rather than the final answer, and Miao et al. (2026) poison the chain of thought rather than the evidence.

Two conclusions follow. Text-level poisoning with a small budget is well established, so the open questions concern its boundaries; and the unit of measurement has converged on per-question scoring against a preset wrong answer, which this thesis inherits so that its results are comparable in kind.

\subsection{2.3 Graph-Based RAG as the Attack Surface}\label{graph-based-rag-as-the-attack-surface}

LightRAG builds its index by prompting an LLM to extract entities and relations from each chunk and stores the merged result as a graph beside vector indexes. At query time it retrieves along a local path, anchored on entities, and a global path, anchored on relations and themes; the hybrid mode combines both with chunk retrieval (Z. Guo et al., 2025). It has already been used in security applications, for example to model an industrial messaging protocol (Y. Chen et al., 2026).

A common intuition holds that the graph layer is defensive, because structure is harder to fabricate than prose. The attack literature has tested this from the other side. GRAGPoison selects relations shared by many queries, injects competing relations wrapped in a cover narrative and enhances them with supporting relations, reaching high success rates on GraphRAG and LightRAG; its ablation identifies temporal ordering, the claim that the competing relation is more recent, as the most influential element of the cover narrative (J. Liang et al., 2026, p. 10). KEPo similarly frames poison as a plausible evolution of the graph\textquotesingle s knowledge (Q. Chen et al., 2026). The extraction step is itself an attack surface: because the graph is built by an LLM reading the source text, changing only a few words of that text can alter the extracted graph enough to mislead downstream answers (Wen et al., 2025). Other work isolates an entity\textquotesingle s neighborhood (Yan et al., 2026), uses an entity as a backdoor trigger (Z. Zheng et al., 2026), steers an agent\textquotesingle s reasoning through a corrupted graph (Kereopa-Yorke et al., 2026), crafts universal adversarial text for text-attributed graphs (Z. Chen et al., 2026) or extracts the graph with few queries (S. Yang et al., 2026). Zhao et al. (2026, p. 8) find that adversarial triples are retrieved for most attacked questions but adopted in only about half of those answers, locating part of the robustness in generation.

The consensus is that the graph is a second attack surface rather than a second line of defense, but each graph attack is evaluated on its own testbed and success criterion. Whether shared-relation poisoning outperforms passage injection when corpus, budget, benchmark and scoring rule are held fixed is the question this thesis addresses.

\subsection{2.4 Defenses against Retrieval Corruption}\label{defenses-against-retrieval-corruption}

Defenses can be grouped by the stage at which they act.

At \textbf{retrieval}, one family certifies robustness by isolating the retrieved passages into disjoint groups, answering from each group separately and aggregating the answers, so that a corrupted minority cannot carry the final answer (Xiang et al., 2024). Another family filters: FilterRAG and its variants remove suspected adversarial passages from the retrieved set before generation (Edemacu et al., 2025, p. 3), and attention-based detection looks for the trace that a poison leaves in the generator\textquotesingle s attention rather than in the text (Choudhary et al., 2026). The limits of filtering are stated from the attack side: a reranker trained on clean in-domain data does suppress injected text, but an attacker who edits about one percent of the characters recovers most of the success rate (X. Yang et al., 2026). TailGuard targets specifically the retrieval-optimized class of poisons (Zhou et al., 2026). Layering is the obvious response to the limits of any one filter: RAGuard adds to an adversarially trained retriever a filter that removes each retrieved document in turn and flags those whose removal shifts the answer, and reports each layer\textquotesingle s contribution separately (Kumar et al., 2026). Source credibility receives direct support from J. Liang et al. (2026): annotating corpus entries with a credibility label reduced GRAGPoison\textquotesingle s success rate from 89.2 to 45.7 percent, which still leaves nearly half of the attacks successful, because labeled poison is still retrieved and can still be believed.

At the \textbf{graph}, HoG-GRAG decomposes a multi-hop question into ordered sub-queries, monitors each hop for poisoning-induced inconsistencies and prunes the compromised entities and relations (Alizadeh Noughabi et al., 2026, p. 1), and Wu et al. (2026) build a separate verification graph outside the retrieval pipeline to detect poisoned agent memories. A third line governs writes rather than reads: a protocol for write-time conflict governance arbitrates between conflicting relations by timestamp and confidence before they enter the graph (\emph{Write-Time Conflict Governance}, 2026), and LedgerRAG keeps a claim-level ledger of the evidence behind an answer and uses its temporal validity, source authority and conflicts to decide when to retrieve, refresh and stop (S. Wang et al., 2026, p. 4). These methods weigh sources against one another; none checks the values the graph asserts against an external, authoritative table of the values themselves.

At \textbf{generation}, LCR-RAG enforces logical consistency through neuro-symbolic training (W. Zheng et al., 2026), evaluation agents try to detect misinformation in the retrieved context (Giri et al., 2026), and traceback methods identify after the fact which passages caused a wrong answer (B. Zhang et al., 2025). Practitioner guidance organizes the same space by control: OWASP (2026) treats retrieval and generation as separate surfaces to harden, and NIST\textquotesingle s profile for generative AI classifies the risks by category, listing data poisoning among the information-security risks, and attaches governance, provenance and testing actions to each (National Institute of Standards and Technology {[}NIST{]}, 2024, p. 8).

Two weaknesses run through this literature. Defenses are almost always evaluated in isolation, so the benefit of combining them is assumed rather than shown; one layer may remove exactly the inputs another would have caught. And defenses are usually evaluated against a fixed attacker, which measures their specification rather than their robustness (Thornton, 2026; X. Yang et al., 2026).

\subsection{2.5 Beyond a Single Success Rate}\label{beyond-a-single-success-rate}

The attack success rate answers one question: did the answer change to the attacker\textquotesingle s target? Three strands of work argue that this is insufficient for a high-consequence domain. M. Zou et al. (2026) point out that equal success rates may hide very unequal consequences, and propose measuring severity instead. J. Liu et al. (2026) show that RAG systems\textquotesingle{} verbal confidence is poorly calibrated when the context contains noise or contradiction, which raises the question of whether poisoned answers are expressed with undiminished certainty. Giri et al. (2026) find that their NLI-based evaluation agent blocks instruction injection reliably, but in-place edits such as entity swaps, the kind of value substitution studied here, remain hard to detect; L. Zhang et al. (2026) map the same difficulty in disinformation detection.

A further gap concerns refusals. Generation-side defenses commonly decline to answer on doubtful evidence, and under the usual definition a refusal is a failed attack, so a defense can report a near-zero success rate by refusing most queries. Some defenses report clean and robust accuracy beside the success rate (Xiang et al., 2024; Choudhary et al., 2026), which reveals refusal indirectly, but none of the papers reviewed here reports the refusal share itself.

Finally, the poisoning literature treats recency only as an attack technique: GRAGPoison and KEPo use fabricated recency as cover (J. Liang et al., 2026; Q. Chen et al., 2026). In building regulation recency is real, and amended editions of the same guidance coexist. Whether a RAG system attributes a value to the instrument that governs it, even without an attacker, has not been examined.

\subsection{2.6 Theoretical Framework}\label{theoretical-framework}

The study is grounded in four bodies of theory and translates each into a design commitment.

\textbf{Threat modeling and the attacker\textquotesingle s capability.} Following W. Zou et al. (2025) and J. Liang et al. (2026), the adversary is characterized by its access (write access to the corpus only), its knowledge (black-, gray- or white-box with respect to retriever and graph) and its goal (targeted replacement of a value). Each attack therefore has a declared knowledge tier and budget, so that a result can be attributed to a stated adversary.

\textbf{Dose--response as a model of adversarial budget.} The number of injected passages per target is a controlled dose; a dose--response design yields the slope that tells a defender how fast risk grows with attacker effort.

Factorial design and interaction. A layered defense is modeled as a set of binary factors. In a full 2 \ensuremath{\times} 2 \ensuremath{\times} 2 factorial each layer\textquotesingle s effect can be estimated in every context of the other two, and an interaction is the difference between a layer\textquotesingle s effects in two contexts.

\textbf{Measurement validity and the refusal problem.} A measure is valid only if it measures the property it is used to infer. Attack success rate measures whether an attack produced its target answer, not whether a defense detected it: a refused query scores zero, exactly as a correct answer does (Equation 4.4), so a defense that refuses most queries reports a near-zero rate. The study therefore reports a conditional accuracy over answered records beside an unconditional one in which a refusal is counted as an incorrect answer (Equation 4.6a); only the latter can rank configurations that answer different shares of the questions. The secondary outcomes (severity weighting, directional analysis and confidence analysis) were registered with their roles fixed in advance.

Together, these commitments let the study borrow its attacks, defenses and measures from the literature while comparing them on one system rather than juxtaposing them.

\section{Chapter 3: Presentation of the Problem}\label{chapter-3-presentation-of-the-problem}

\subsection{3.1 The Core Problem}\label{the-core-problem}

The core problem of this thesis can be stated in one sentence: a retrieval-augmented system that answers questions about building regulations can be made to return a wrong value for a specific provision by anyone able to add documents to its corpus, and it is not known how far a graph layer changes that exposure or how far a layered defense can remove it without making the system useless.

The literature reviewed in Chapter 2 establishes that a few adversarial passages can redirect an answer, that graph-based systems can be poisoned through their relations, and that defenses exist at each stage of the pipeline. Three questions remain open.

The first is whether the graph alters what an attacker must do (make a payload survive the LLM\textquotesingle s entity--relation extraction and be selected through keyword-to-entity-to-relation retrieval, rather than merely rank a passage in the vector top-k), or merely offers another location for the same payload (the entity and relation descriptions placed in the prompt beside the text chunks). The claim that structure is harder to fabricate than prose has been made on behalf of graph-based RAG, and attacks on the graph have been demonstrated, but the two layers have never been compared on one system under one rubric.

The second is whether a layered defense composes. If each layer\textquotesingle s effect were independent of the others, the contributions could be measured one at a time and added. If the layers interact, for instance because one layer removes the very inputs another would have acted on, a one-at-a-time evaluation reports numbers that do not describe the combined system.

The third is what a low attack success rate under defense means. A defense that stops the poison and a defense that stops answering both produce a low rate, and they have very different value to a user.

\subsection{3.2 Sub-questions}\label{sub-questions}

The research question stated in Section 1.4 is decomposed into four sub-questions, each tied to a part of the proposal.

\textbf{SQ1 (text layer).} How effective is text-level poisoning against LightRAG on the regulatory corpus, and how does its effectiveness change with the number of passages injected per target?

\textbf{SQ2 (graph layer).} Does poisoning aimed at the graph\textquotesingle s shared relations succeed more or less often than text-level poisoning at a matched budget, and does poisoning spill over to questions that were not targeted?

\textbf{SQ3 (defense).} Can three defensive layers push attack success down to a tolerable rate without a material loss of clean accuracy, how do the layers contribute and interact, and does a defense-aware attacker recover what a static attacker loses?

\textbf{SQ4 (harms).} Are the harms uneven: do errors fall disproportionately on safety-critical provisions, does the direction of tampering matter, are wrong answers expressed with less certainty than right ones, and does the system confuse versions of the guidance even without an attacker?

\subsection{3.3 Hypotheses}\label{hypotheses}

The hypotheses below were registered on 4 October 2026, before any attack run, with their tests and decision thresholds; later dated amendments, each made before the run it affected, fixed details such as the budget operator for the graph attack.

\textbf{H1a.} At a budget of five passages per target, the strict attack success rate of text-level poisoning over the 150 targets exceeds 0.10. Test: one-sided exact binomial test on the number of targets that succeed in the majority of their three repeats.

\textbf{H1b.} Attack success increases across the nested budgets of one, three and five passages. Test: Page\textquotesingle s L for ordered alternatives in a repeated-measures layout.

\textbf{H2.} At a matched budget, shared-relation (graph) poisoning achieves a higher strict success rate than text-level poisoning. The direction follows the claim that the graph partially protects against text injection while graph-specific poisoning overcomes that protection. Test: one-sided exact McNemar test on the 150 targets, the outcome of each target being the majority of its repeats.

\textbf{H3.} Text-level poisoning reduces accuracy on untargeted questions in the same index (spillover). Test: one-sided exact McNemar test on the pooled untargeted observations. A seven-point equivalence margin may be claimed only if the clean discordance rate leaves the test enough power: at a true dACC of 0, a TOST with \ensuremath{\delta} = 7 points reaches 80 percent power only when the clean-control discordance rate is at most 8.6 percent; above that rate, a passing TOST would reflect insufficient power and is not read as equivalence.

\textbf{H4a.} With the defense armed, the strict attack success rate does not exceed a threshold set by the undefended rate:

\[ASR_{\text{defended}} \leq \min\left( 0.5 \times ASR_{\text{undefended}},\ 0.10 \right)\quad\quad(3.1)\]

Both bounds are required because each alone admits a case the other would reject: a defense that halves a very high rate, or one that reaches 0.10 from a rate barely above it.

\textbf{H4b.} The defense is non-inferior on clean accuracy: the lower bound of the one-sided 95 percent confidence interval for the change in clean accuracy lies above minus five percentage points,

\[LB_{95}\left( ACC_{\text{defended}} - ACC_{\text{undefended}} \right) > - 0.05\quad\quad(3.2)\]

\textbf{H5.} Weighting errors by safety consequence changes the error rate (descriptive: reported as the ratio R of Equation 4.7 with bootstrap intervals, without a directional test).

\textbf{H6.} Loosening and tightening a requirement differ in attack success (two-sided, because the proposal offered mechanisms for both directions: a model prior that newer rules are stricter would favor tightening, while suspicion of relaxed requirements would favor loosening). Test: two-sided exact McNemar test on paired targets.

\textbf{H7.} A defense-aware adaptive attacker recovers attack success against the full defense relative to a non-adaptive attacker. Test: exact McNemar test on the poisoned items, paired against a control arm run in the same batch.

\textbf{H8.} Expressed certainty tracks correctness under attack, that is, wrong answers are not stated as confidently as right ones.

The confirmatory family has seven members, corrected by the Holm procedure: H1a, H1b, H3, H4a on attack one, H4b, and H4a on attack two and H2, the last two added by dated amendment once attack two was designed (Appendix E). H5, H6, H7 and H8 are secondary: they are reported with intervals but do not move the conclusions of the confirmatory family.

\subsection{3.4 Thresholds and Related Issues}\label{thresholds-and-related-issues}

Thresholds set after the data are seen tend to be set where the data already lie, and with many configurations the freedom to choose which comparison to report could make almost any defense look effective.

Two further issues bear on the problem: poisoned text was confined to isolated local copies of the index (Section 4.11), and the unit of analysis is the benchmark item, which makes paired comparisons possible.

\section{Chapter 4: Research Methodology}\label{chapter-4-research-methodology}

\subsection{4.1 Research Approach}\label{research-approach}

The study uses a mixed-methods design. A controlled quantitative experiment carries the confirmatory claims: attacks and defenses are switched on and off on one instrumented system while the corpus, benchmark, serving model and query parameters are held fixed, and every answer is scored against a frozen rubric. A small qualitative component, a three-case study of the amendment-narrative attack and a hand-scored version probe, supports their interpretation.

An experimental design suits the question because it is causal: it asks what an attack or defense does to a system\textquotesingle s answers with everything else held constant. It can establish whether and by how much outcomes change on this system, not whether the effects transfer (Chapter 9).

Two problems were anticipated: noise, because the serving LLM exposes no temperature parameter, handled by three repeats per configuration; and researcher degrees of freedom, handled by pre-registration (Section 3.3), hash-frozen inputs and recomputation of every number from raw per-item records.

\subsection{4.2 System and Environment}\label{system-and-environment}

All experiments used LightRAG core 1.5.8 as a local server. Indexing extraction and answer generation used the DeepSeek API with the model name deepseek-v4-flash (from 10 September 2026 the provider served this name with DeepSeek-V4.1-Flash, reported by the API as deepseek-flash; the experimental index was extracted on 5--7 September, before the change, and all runs came after it); embeddings were computed locally with Qwen3-Embedding-0.6B; the graph was stored with NetworkX and the vectors with NanoVectorDB, on a laptop with an i9-14900HX processor, an RTX 5070 GPU and 32 GB of memory.

Query parameters were pinned for every run and recorded in its header (hybrid mode, 40 candidates, 20 chunks in the context, reranking disabled; Appendix B, Table B-3), so a difference between two runs cannot come from a change in retrieval settings.

\subsection{4.3 Corpus}\label{corpus}

The corpus consists of official documents downloaded from GOV.UK and registered in a manifest that records each file\textquotesingle s name, source and SHA-256 hash (sources in Appendix B). The experimental index was built from two consolidated volumes: the merged Approved Documents (October 2024 edition) and the Manual to the Building Regulations (July 2020). Table 4.1 describes it. All confirmatory and second-batch experiments use this index.

{\small\textbf{Table 4.1} The experimental index}

{\def\LTcaptype{none} 
{\small
\begin{longtable}[]{@{}
  >{\raggedright\arraybackslash}p{(\linewidth - 10\tabcolsep) * \real{0.1957}}
  >{\raggedright\arraybackslash}p{(\linewidth - 10\tabcolsep) * \real{0.1304}}
  >{\raggedright\arraybackslash}p{(\linewidth - 10\tabcolsep) * \real{0.1739}}
  >{\raggedright\arraybackslash}p{(\linewidth - 10\tabcolsep) * \real{0.1957}}
  >{\raggedright\arraybackslash}p{(\linewidth - 10\tabcolsep) * \real{0.1957}}
  >{\raggedright\arraybackslash}p{(\linewidth - 10\tabcolsep) * \real{0.1087}}@{}}
\toprule\noalign{}
\begin{minipage}[b]{\linewidth}\raggedright
Documents
\end{minipage} & \begin{minipage}[b]{\linewidth}\raggedright
Chunks
\end{minipage} & \begin{minipage}[b]{\linewidth}\raggedright
Entities
\end{minipage} & \begin{minipage}[b]{\linewidth}\raggedright
Relations
\end{minipage} & \begin{minipage}[b]{\linewidth}\raggedright
Relations per chunk
\end{minipage} & \begin{minipage}[b]{\linewidth}\raggedright
Graph file (MB)
\end{minipage} \\
\midrule\noalign{}
\endhead
\bottomrule\noalign{}
\endlastfoot
2 & 850 & 12,113 & 22,751 & 26.8 & 21.5 \\
\end{longtable}
}
}

\emph{Note.} Counts read from the clean index (document and chunk stores, and the graph file). Source: Appendix A, Table A-1.

Two larger indexes, built from the 70 individual Approved Documents with their amendment booklets and from a 192-document superset, contain genuine amended editions, which the experimental index does not. They were used only for two exploratory probes, a repeat of the amendment-narrative cases (Section 5.4) and the version-attribution probe (Section 5.7.4); no attack or defense experiment was run on them (Section 9.2). Every attack wrote to a separate copy, and a guard script verified that the clean indexes never changed.

\subsection{4.4 Benchmark}\label{benchmark}

The benchmark contains 150 questions, written in two batches (65 and 85 items) and frozen before the first attack. Each item records the question, its source paragraph or table, the ground truth, the required value groups a correct answer must contain, and its value type, answer type and safety-consequence rating (Table 4.2).

{\small\textbf{Table 4.2} Composition of the 150-item benchmark}

{\def\LTcaptype{none} 
{\small
\begin{longtable}[]{@{}
  >{\raggedright\arraybackslash}p{(\linewidth - 2\tabcolsep) * \real{0.4428}}
  >{\raggedright\arraybackslash}p{(\linewidth - 2\tabcolsep) * \real{0.5572}}@{}}
\toprule\noalign{}
\begin{minipage}[b]{\linewidth}\raggedright
Field
\end{minipage} & \begin{minipage}[b]{\linewidth}\raggedright
Distribution
\end{minipage} \\
\midrule\noalign{}
\endhead
\bottomrule\noalign{}
\endlastfoot
Approved Document Part & B 42, K 26, A 10, J 10, L 9, F 8, C 7, E 7, G 7, M 7, P 6, H 5, O 5, B+\allowbreak{}K 1 \\
Answer type & numeric 120, explanatory 30 \\
Value type & quantity 113, enumeration 18, rule 12, designation 7 \\
Safety consequence & 3 (life safety) 93, 2 (compliance) 50, 1 (general) 7 \\
Poison direction & loosen 73, tighten 69, neutral 8; paired 130 \\
\end{longtable}
}
}

\emph{Note.} Source: Appendix A, Tables A-2 and A-3.

The scoring rubric is strict: an answer is \emph{correct} only if every required value group is present, \emph{partial} if some are, \emph{incorrect} if none are, and \emph{refusal} if the system declines to answer. The benchmark is administered in English.

Each item carries a \textbf{poison seed}: the ground-truth value mutated in the direction that protects less (for example, a minimum width reduced), with the opposite mutation recorded beside it for the directional analysis. Each item also carries a \textbf{safety-consequence weight}: 3 for provisions bearing directly on life safety (means of escape, structure, guarding), 2 for compliance or property risk, and 1 for general provisions.

Items were assigned to two folds of 75 by a seeded block procedure (blocks\_v1). Related items form 14 blocks, which with 111 singletons make 125 units; a block is never split, so the unpoisoned fold is a within-index control sharing no provision with the poisoned one.

\subsection{4.5 Threat Model and Attacks}\label{threat-model-and-attacks}

The adversary can add a limited number of documents to the corpus but cannot modify the server configuration, the model or documents already indexed. It can use a commercial LLM to write its documents. Four attacks were implemented, each entering the pipeline through the same channel (ordinary documents ingested by LightRAG) but aimed at a different stage (Figure 4.1 and Table 4.3).

\begin{center}
\includegraphics[width=0.93\linewidth,keepaspectratio,alt={Figure 4.1 The research framework: where the attacks enter the LightRAG pipeline and where the three defensive layers act}]{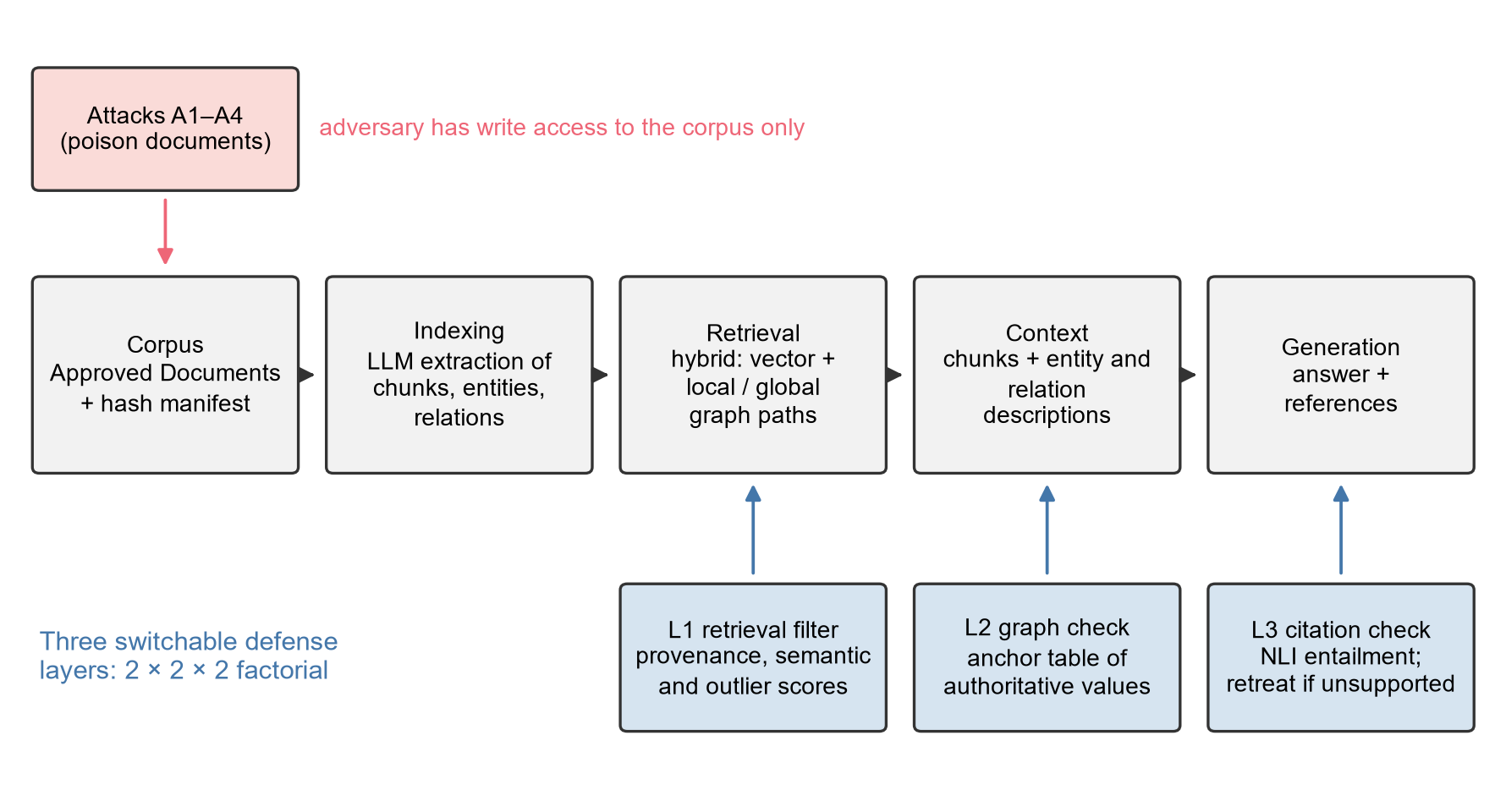}
\end{center}

{\small\textbf{Figure 4.1} The research framework: where the attacks enter the LightRAG pipeline and where the three defensive layers act}

\emph{Note.} All four attacks travel as ordinary documents through indexing. The three layers are independently switchable, giving the eight configurations of the factorial ablation.

{\small\textbf{Table 4.3} The four attacks}

{\def\LTcaptype{none} 
{\small
\begin{longtable}[]{@{}
  >{\raggedright\arraybackslash}p{(\linewidth - 8\tabcolsep) * \real{0.2678}}
  >{\raggedright\arraybackslash}p{(\linewidth - 8\tabcolsep) * \real{0.1691}}
  >{\raggedright\arraybackslash}p{(\linewidth - 8\tabcolsep) * \real{0.1410}}
  >{\raggedright\arraybackslash}p{(\linewidth - 8\tabcolsep) * \real{0.2529}}
  >{\raggedright\arraybackslash}p{(\linewidth - 8\tabcolsep) * \real{0.1691}}@{}}
\toprule\noalign{}
\begin{minipage}[b]{\linewidth}\raggedright
Attack
\end{minipage} & \begin{minipage}[b]{\linewidth}\raggedright
Target
\end{minipage} & \begin{minipage}[b]{\linewidth}\raggedright
Tier
\end{minipage} & \begin{minipage}[b]{\linewidth}\raggedright
Construction
\end{minipage} & \begin{minipage}[b]{\linewidth}\raggedright
Role
\end{minipage} \\
\midrule\noalign{}
\endhead
\bottomrule\noalign{}
\endlastfoot
A1 Text-level retrieval hijacking & Text chunks & Black-box & PoisonedRAG form: question + generated paragraph asserting the mutated value & Confirmatory (H1a, H1b, H3) \\
A2 Shared-relation poisoning & Graph layer & Gray-box & GRAGPoison steps: competing relation around an entity shared across Parts, applicability narrative & Confirmatory (H2) \\
A3 Amendment-narrative poisoning & Temporal plausibility & Black-box & Fabricated 2026 amendment notice for a clause & Qualitative, 3 cases \\
A4 Defense-aware adaptive attacker & Each defense layer & White-box (declared) & One arm per layer, built from what that layer reads & Confirmatory (H7) \\
\end{longtable}
}
}

\textbf{Attack A1} follows the black-box form of PoisonedRAG (W. Zou et al., 2025, pp. 6--7): each poison passage joins the target question verbatim, which makes it retrievable, to an LLM-generated paragraph formatted like a provision and asserting the mutated value. The \textbf{budget} \(M\) is the number of passages injected for one target, \(M \in \left\{ 1,3,5 \right\}\) as the proposal fixed, nested so that the passages at \(M = 3\) include those at \(M = 1\); each fold was poisoned separately, giving six cells of three runs.

\textbf{Attack A2} adapts the three steps of GRAGPoison (J. Liang et al., 2026, p. 14). It attaches to an entity that also occurs in another Part of the corpus a rival relation claiming a scope of application that conflicts with the provision, at five passages per target. The passage does not carry the question and its narrative concerns applicability rather than recency, so that the comparison with A1 isolates the layer. Poison could be built for 143 of the 150 targets; the other seven count as failures, because an attack that cannot be mounted does not succeed.

\textbf{Attack A3} inserts a fabricated 2026 ``Amendment Booklet'' passage changing one value, in three cases (Table 5.4), each run three times on the experimental index and on the 70-document index, which contains genuine amendment booklets. It is qualitative by design.

\textbf{Attack A4} is the defense-aware attacker. Its access was declared before it ran (the files it could inspect, the anchor-table fields it could see and a cap on oracle queries), and it has one arm per layer: adaptive\_l1 rewrites the poison to improve its semantic and outlier scores, adaptive\_l2 moves the tampered value to a parameter the anchor table does not cover, and adaptive\_l3 places the false conclusion beside its own supporting sentence. The control arm, A2\textquotesingle s fold-0 set, was re-run in the same batch. Construction details of all four attacks are in Appendix D.6.

\subsection{4.6 The Three-Layer Defense}\label{the-three-layer-defense}

The defense consists of three independently switchable layers (Table 4.4).

{\small\textbf{Table 4.4} The three defensive layers}

{\def\LTcaptype{none} 
{\small
\begin{longtable}[]{@{}
  >{\raggedright\arraybackslash}p{(\linewidth - 6\tabcolsep) * \real{0.2318}}
  >{\raggedright\arraybackslash}p{(\linewidth - 6\tabcolsep) * \real{0.1545}}
  >{\raggedright\arraybackslash}p{(\linewidth - 6\tabcolsep) * \real{0.4745}}
  >{\raggedright\arraybackslash}p{(\linewidth - 6\tabcolsep) * \real{0.1391}}@{}}
\toprule\noalign{}
\begin{minipage}[b]{\linewidth}\raggedright
Layer
\end{minipage} & \begin{minipage}[b]{\linewidth}\raggedright
Stage
\end{minipage} & \begin{minipage}[b]{\linewidth}\raggedright
Mechanism
\end{minipage} & \begin{minipage}[b]{\linewidth}\raggedright
Unit judged
\end{minipage} \\
\midrule\noalign{}
\endhead
\bottomrule\noalign{}
\endlastfoot
L1 Retrieval-side filtering & Before generation & Fused score of semantic similarity, source trust (manifest membership and hash) and outlier evidence (near-duplicates, multi-path disagreement) & Candidate chunk \\
L2 Graph-side anchor verification & Graph context & Numeric claims aligned against a frozen anchor table of authoritative values; contradicting edges removed & Relation edge \\
L3 Generation-side citation verification & After generation & NLI check of each answer sentence against its cited evidence; retreat to a conservative answer unless entailed and uncontradicted & Answer sentence \\
\end{longtable}
}
}

\textbf{Layer 1} re-ranks the candidate chunks \(c\) of a query \(q\) with the score

\[S_{\text{rank}}(c) = \alpha\, S_{\text{semantic}}(c,q) + \beta\, S_{\text{trust}}(c) - \gamma\, S_{\text{outlier}}(c)\quad\quad(4.1)\]

where \(S_{\text{semantic}}\) is LightRAG\textquotesingle s native similarity. The trust term rewards provenance:

\[S_{\text{trust}}(c) = w_{1}\, S_{\text{source}}(c) + w_{2}\, S_{\text{hash}}(c) - w_{3}\, S_{\text{new}}(c)\quad\quad(4.2)\]

with \(S_{\text{source}} = 1\) if the chunk\textquotesingle s file is in the official manifest, \(S_{\text{hash}} = 1\) if its document hash matches the manifest, and \(S_{\text{new}} = 1\) if the chunk cannot be traced to the manifest (each 0 otherwise). The outlier term adds a multi-path consistency score and a near-duplicate score:

\[S_{\text{outlier}}(c) = S_{\text{cons}}(c) + \operatorname{clip}_{\lbrack 0,1\rbrack}\left( \frac{z_{\text{sim}}(c)}{3} \right) \cdot \operatorname{clip}_{\lbrack 0,1\rbrack}\left( \frac{z_{\text{dev}}(c)}{3} \right)\quad\quad(4.3)\]

where \(S_{\text{cons}} = 1\) if the candidate is in the top ten of one retrieval path (vector, local or global graph) and absent from the others, 0.5 if another path lists it lower, and 0 otherwise; \(z_{\text{sim}}\) and \(z_{\text{dev}}\) are its similarity z-score among the candidates and its largest length or style z-score against the corpus. The weights were fixed design values frozen before any defended run, not fitted to the benchmark.

\textbf{Layer 2} uses an anchor table extracted from the authoritative text: 145 structured rows, each with a subject, relation, comparator, value and unit, covering 74 benchmark items. For each query, the layer aligns numeric relations in the graph context with anchor rows that share their subject, relation type and unit, and removes edges whose value violates the anchor\textquotesingle s comparator. Rows it cannot adjudicate are recorded separately; 67 covered items remain adjudicable for A2 (Appendix A, Table A-5).

\textbf{Layer 3} uses a multilingual NLI model (mDeBERTa-v3-base fine-tuned on MNLI and XNLI) to check each sentence of the answer against the evidence it cites. Each sentence receives one of four verdicts: entailment, neutral, contradiction or conflict, the last meaning that some evidence supports and some contradicts the sentence. Unless a sentence is entailed and not contradicted, the layer ``retreats'': it replaces the confident answer with a conservative response advising the user to verify the original text, and records the answer it withheld.

\subsection{4.7 Experimental Procedure}\label{experimental-procedure}

The experiments were run in order, each step depending on frozen outputs of the previous one: (1) a clean baseline on the experimental index; (2) A1 at three budgets on each fold, with a clean control and the paired-direction set; (3) A2 at M = 5 on each fold; (4) the defense ablation, all eight layer combinations on A2\textquotesingle s fold-0 poisoned index over all 150 items, with clean-index arms for H4b, in a registered batch (18 October 2026) that is the main reading, with two correction batches (21 and 25 October) reported beside it; (5) A4\textquotesingle s four arms on the same 72 poisoned items, undefended and with all layers armed; (6) A3\textquotesingle s three cases; and (7) the version-attribution probe, scored by hand. Every run had three repeats unless stated otherwise.

A second experimental batch was run between 6 and 8 November 2026, after the registered batches: a false-positive test of provenance filtering, the naive retrieval mode, the A4 arms against the generation layer alone, a self-rated confidence pass, a second scorer, and re-analyses of archived runs. Each experiment had its design and reading rule written down before it ran (Appendix E), none belongs to the confirmatory family, and each is reported in the section whose question it answers.

Each run produced three files: a header recording the environment, the parameters and the server identity; a raw file with one record per item and repeat; and a summary. Each attack run also preserved a copy of the index it queried.

\subsection{4.8 Measures}\label{measures}

\textbf{Strict success.} A poisoned query counts as a success only when its answer satisfies all three tests at once: none of the required value groups is matched, the value planted by the attacker appears, and the system did not decline. The \textbf{attack success rate} (ASR) averages, across poisoned items, the fraction of each item\textquotesingle s three repeats that were successes; for the confirmatory tests, each item\textquotesingle s outcome is the majority of its three repeats.

Formally, for record \(r\) of poisoned item \(i\) in repeat \(k\), let \(G_{i}\) be the item\textquotesingle s required value groups, \(V_{i}\) the attacker\textquotesingle s value and \(\text{ref}_{ik}\) a refusal indicator. Then

\[s_{ik} = 1\left\lbrack \,\text{no }g \in G_{i}\text{ matched}\, \right\rbrack \cdot 1\left\lbrack \, V_{i}\text{ present}\, \right\rbrack \cdot \left( 1 - ref_{ik} \right)\quad\quad(4.4)\]

\[ASR = \frac{1}{|P|}\sum_{i \in P}^{}\frac{1}{K}\sum_{k = 1}^{K}s_{ik},\quad\quad K = 3\quad\quad(4.5)\]

where \(P\) is the set of poisoned items.

\textbf{Disruption.} Only the first of those tests: the true value is missing from the answer, irrespective of what replaced it; \(d_{ik} = 1\) when no \(g \in G_{i}\) is matched.

\textbf{Poison recall@k.} The share of a target\textquotesingle s \(M\) injected passages that appear among the references the answer cites, averaged over records: a measure of whether the poison reached the generator.

\textbf{Accuracy conventions.} With \(c_{r} = 1\) for a correct record and \(\text{ref}_{r} = 1\) for a refusal, over a record set \(R\):

\[{ACC}_{\text{strict}} = \frac{\sum_{r \in R}^{}c_{r}}{\sum_{r \in R}^{}(1 - {ref}_{r})},\quad\quad{ACC}_{\text{plain}} = \frac{\sum_{r \in R}^{}c_{r}}{|R|}\quad\quad(4.6a)\]

\[\text{answered} = \frac{1}{|R|}\sum_{r \in R}^{}(1 - {ref}_{r})\quad\quad(4.6b)\]

Because strict accuracy depends on which items a configuration chose to answer, only plain accuracy, which counts a refusal as wrong, can rank configurations with different answered shares.

\textbf{Secondary measures.} With safety weight \(w_{i} \in \left\{ 1,2,3 \right\}\) and \(e_{i} = 1\) for a non-correct answer over the answered items \(A\) of a run, the safety-weighted error rate and its ratio to the unweighted error rate are

\[SW\text{-}ER = \frac{\sum_{i \in A}^{}w_{i}e_{i}}{\sum_{i \in A}^{}w_{i}},\quad\quad R = \frac{SW\text{-}ER}{\frac{1}{|A|}\sum_{i \in A}^{}e_{i}} \leq \frac{\max_{i}w_{i} \cdot |A|}{\sum_{i \in A}^{}w_{i}}\quad\quad(4.7)\]

For all 150 items the bound is \(3 \times \frac{150}{386} = 1.166\). Because \(R\) is undefined when a run has no errors, the difference \(D = \text{SW-ER} - \text{ER}\) is the primary measure, with an interval that resamples whole items. Directional ASR is computed separately for the loosened and tightened version of each paired target. Expressed certainty is measured lexically: the share of answers stating a requirement with a strong deontic marker (``must'', ``shall'') and no hedge or flagged source conflict, and hedging terms per hundred words. Version handling is scored by hand.

\subsection{4.9 Data Analysis}\label{data-analysis}

Paired comparisons on the same items use the exact McNemar (1947) test on the discordant pairs. If \(b\) items succeed only under the first condition and \(c\) only under the second, the one-sided p-value in the predicted direction is

\[p = \sum_{j = c}^{b + c}\binom{b + c}{j}\,{0.5}^{\, b + c}\quad\quad(4.8)\]

and the two-sided value doubles the smaller tail. The ordered budgets are tested with Page\textquotesingle s (1963) L,

\[L = \sum_{j = 1}^{3}j\, R_{j}\quad\quad(4.9)\]

where \(R_{j}\) is the sum over items of the within-item rank of budget \(j\) (mid-ranks for ties), with the null distribution obtained by permuting ranks within items. Single-proportion claims use the exact binomial test. Interval estimates for differences use a block bootstrap (Efron \& Tibshirani, 1993) over the 125 fold units (14 blocks plus singletons) with 10,000 resamples. The confirmatory family is Holm-corrected. In the factorial ablation, the effect of a layer in a given context is the difference in poisoned accuracy with and without it, and the interaction between two layers is the difference between one layer\textquotesingle s effects with and without the other. For the retrieval (L1) and generation (L3) layers,

\[\Delta_{13} = \left\lbrack ACC(L1,L3) - ACC(L3) \right\rbrack - \left\lbrack ACC(L1) - ACC(\varnothing) \right\rbrack\quad\quad(4.10)\]

Beside every null result the smallest detectable effect is stated: with 72 poisoned items, for example, a paired test needs a gain of at least five items (0.0694) to reach significance.

The second batch adds three measures. A self-rating \(s_{r} \in \left\{ 1,\ldots,5 \right\}\) is mapped to a probability \({\widehat{p}}_{r} = \frac{s_{r} - 1}{4}\). With the scored records \(R\) divided into bins \(B_{m}\), one per rating level, the expected calibration error (C. Guo et al., 2017) is the weighted gap between stated confidence and strict correctness,

\[ECE = \sum_{m}^{}\frac{|B_{m}|}{|R|}\,\left| \,\frac{1}{|B_{m}|}\sum_{r \in B_{m}}^{}c_{r}\mspace{6mu} - \mspace{6mu}\frac{1}{|B_{m}|}\sum_{r \in B_{m}}^{}{\widehat{p}}_{r}\, \right|\quad\quad(4.11)\]

and the AUROC is the probability that a randomly chosen correct record carries a higher rating than a randomly chosen incorrect one. Agreement between two scorers on a binary verdict is measured by Cohen\textquotesingle s (1960) kappa,

\[\kappa = \frac{p_{o} - p_{e}}{1 - p_{e}},\quad\quad p_{e} = p_{1}q_{1} + (1 - p_{1})(1 - q_{1})\quad\quad(4.12)\]

where \(p_{o}\) is the observed share of agreements and \(p_{1}\), \(q_{1}\) are the shares each scorer counts as correct; for the LLM judge, a score of 4 or 5 counts as correct. Literal exact match compares the whole answer with the ground truth after a fixed normalization \(\nu\): the answer\textquotesingle s References section is removed, the text is put in Unicode NFKC form and case-folded, runs of white space become one space, and leading and trailing space is stripped; punctuation and number formats are left unchanged. Then

\[EM_{r} = 1\left\lbrack \,\nu\left( \text{answer}_{r} \right) = \nu\left( \text{ground truth}_{i} \right)\, \right\rbrack\quad\quad(4.13)\]

over the answered records. Calibration intervals use 2,000 resamples of the 125 units, with one draw shared by the two arms because they answer the same items; kappa intervals resample items. These intervals are descriptive and are not tested.

\subsection{4.10 Validity and Reliability Measures}\label{validity-and-reliability-measures}

The benchmark, the fold assignment, the anchor table, the defense weights and the registered hypotheses were each frozen with a hash manifest and a guard script that fails if any of them changes. Controls were run in the same batch as the arms they were compared with wherever a restart could confound the comparison, since a measured restart floor of 8 of 72 items between batches exceeds the smallest detectable effect.

Every benchmark item, with its rubric groups and poison seeds, was signed off by the author before the freeze, and the version-conflict probe was scored entirely by hand. Safety-consequence weights were re-rated blind, with 91 of 150 agreements and a Cohen\textquotesingle s kappa of 0.226; this low agreement is reported as a limitation of the weighting rather than concealed.

In the second batch, the strict rubric was compared with an independent second scorer, an LLM judge (L. Zheng et al., 2023), on the 30 explanatory items, where a graded judgement of consistency with the provision is defined: their answers in the clean run and in the undefended poisoned run, 60 in all. The judge was a dated model snapshot from a provider other than the one that served the answers and wrote the poison (Appendix B), used at temperature 0. It saw the question, the reference answer, the provision text and the answer, but not the rubric, the condition or the item identifier, and scored consistency from 1 to 5, with 4 or 5 counting as correct. The prompt, population, statistic and a reproducibility bar were fixed before the first judged answer; the two refusals left the comparison, as they leave strict accuracy (Table 4.5 and Figure 4.2).

{\small\textbf{Table 4.5} Second batch: agreement between the LLM judge, the strict rubric and the author}

{\def\LTcaptype{none} 
{\small
\begin{longtable}[]{@{}
  >{\raggedright\arraybackslash}p{(\linewidth - 6\tabcolsep) * \real{0.2174}}
  >{\raggedright\arraybackslash}p{(\linewidth - 6\tabcolsep) * \real{0.1522}}
  >{\raggedright\arraybackslash}p{(\linewidth - 6\tabcolsep) * \real{0.1957}}
  >{\raggedright\arraybackslash}p{(\linewidth - 6\tabcolsep) * \real{0.4348}}@{}}
\toprule\noalign{}
\begin{minipage}[b]{\linewidth}\raggedright
Comparison
\end{minipage} & \begin{minipage}[b]{\linewidth}\raggedright
Answers
\end{minipage} & \begin{minipage}[b]{\linewidth}\raggedright
Raw agreement
\end{minipage} & \begin{minipage}[b]{\linewidth}\raggedright
Cohen\textquotesingle s \ensuremath{\kappa} {[}95\% CI{]}
\end{minipage} \\
\midrule\noalign{}
\endhead
\bottomrule\noalign{}
\endlastfoot
Judge (score \ensuremath{\geq} 4) vs strict rubric & 58 & 0.862 & 0.655 {[}0.352, 0.886{]} \\
Judge, second pass vs first pass (correct / not correct) & 58 & 1.000 & --- \\
Author vs judge & 14 & 0.714 & 0.391 {[}\ensuremath{-}0.140, 0.851{]} \\
Author vs strict rubric & 14 & 0.714 & 0.462 {[}0.109, 0.851{]} \\
\end{longtable}
}
}

\emph{Note.} Intervals resample items (10,000 draws). The second pass served only as reproducibility evidence. The author reviewed 15 answers drawn with a fixed seed, one side per item, blind to scores and condition; its one refusal was excluded so that all three \ensuremath{\kappa} values share one population.

\begin{center}
\includegraphics[width=0.73\linewidth,keepaspectratio,alt={Figure 4.2 Distribution of the LLM judge's scores, split by the strict rubric's verdict (58 answers)}]{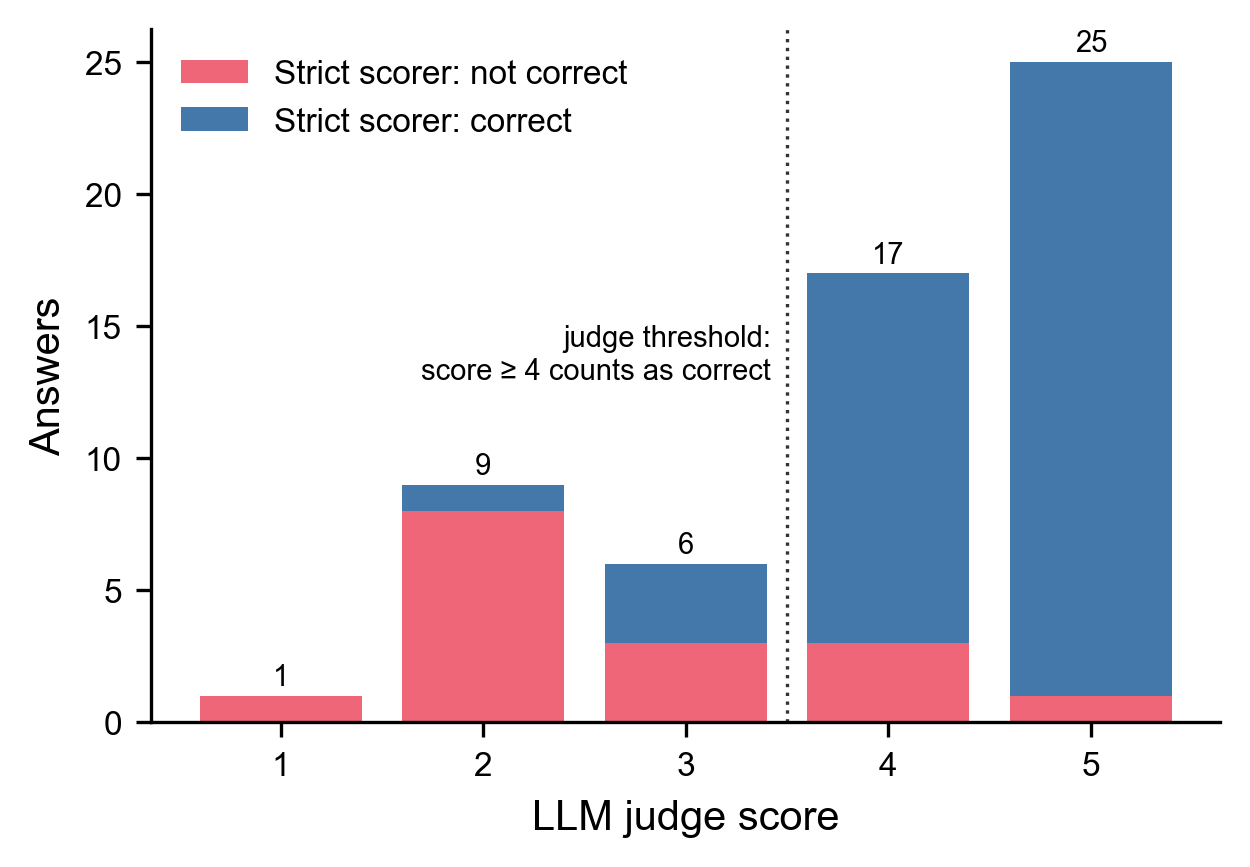}
\end{center}

{\small\textbf{Figure 4.2} Distribution of the LLM judge\textquotesingle s scores, split by the strict rubric\textquotesingle s verdict (58 answers)}

\emph{Note.} Scores to the right of the dotted line count as correct for the judge.

The judge agreed with the rubric on 50 of 58 answers, four disagreements in each direction, and reproduced all 58 of its correct-or-not verdicts on a second pass. Its accuracy differed from strict accuracy by one item in each run (clean: 25 against 26 of 30; poisoned: 17 against 16 of 28). In the author\textquotesingle s review all four disagreements with the rubric were answers the author counted correct and the rubric did not, so the strict rubric is, if anything, the more conservative reader.

\subsection{4.11 Data Presentation and Ethics}\label{data-presentation-and-ethics}

Items are identified by codes such as QA-B-05 (Part B, item 5) and defense configurations by the layers armed. No personal data were used and there were no human participants beyond the author\textquotesingle s review. Poisoned text was confined to isolated local index copies; released materials are limited to the benchmark, scripts and aggregate results, and the poison passages are withheld.

\section{Chapter 5: Results}\label{chapter-5-results}

\subsection{5.1 Overview of Research Results}\label{overview-of-research-results}

The study asked how an adversary with write access can make a regulatory LightRAG index return the wrong clause value, and whether three defensive layers can stop this without sacrificing honest answers. The results follow the four sub-questions; Table 5.1 summarizes the registered hypotheses. Experiments of the second batch are reported in the section whose question they answer and are labeled as such; none changes an outcome in Table 5.1 (Section 4.7).

{\small\textbf{Table 5.1} Outcome of the registered hypotheses}

{\def\LTcaptype{none} 
{\small
\begin{longtable}[]{@{}
  >{\raggedright\arraybackslash}p{(\linewidth - 6\tabcolsep) * \real{0.1852}}
  >{\raggedright\arraybackslash}p{(\linewidth - 6\tabcolsep) * \real{0.2222}}
  >{\raggedright\arraybackslash}p{(\linewidth - 6\tabcolsep) * \real{0.3519}}
  >{\raggedright\arraybackslash}p{(\linewidth - 6\tabcolsep) * \real{0.2407}}@{}}
\toprule\noalign{}
\begin{minipage}[b]{\linewidth}\raggedright
Hypothesis
\end{minipage} & \begin{minipage}[b]{\linewidth}\raggedright
Claim
\end{minipage} & \begin{minipage}[b]{\linewidth}\raggedright
Key statistic
\end{minipage} & \begin{minipage}[b]{\linewidth}\raggedright
Outcome
\end{minipage} \\
\midrule\noalign{}
\endhead
\bottomrule\noalign{}
\endlastfoot
H1a & Text-level ASR at M = 5 exceeds 0.10 & 33/\allowbreak{}150 targets; exact binomial p = 1.17 \ensuremath{\times} 10\textsuperscript{\ensuremath{-}5} & Supported \\
H1b & Success rises across M = 1, 3, 5 & Page\textquotesingle s L = 1842, permutation p \textless{} 0.0001 & Supported \\
H2 & Graph poisoning exceeds text poisoning at M = 5 & 22/\allowbreak{}150 vs 33/\allowbreak{}150; one-sided p = 0.988 & Not supported (opposite direction) \\
H3 & Text poisoning lowers untargeted accuracy & +0.040, 95\% CI {[}\ensuremath{-}0.007, +0.091{]}, p = 0.132 & Inconclusive \\
H4a & Defended ASR \ensuremath{\leq} min(0.5 \ensuremath{\times} undefended, 0.10) & Met in 6 of 7 cells; all: 0.000, p = 0.00049 & Met, four cells by refusal \\
H4b & Clean accuracy non-inferior (margin \ensuremath{-}5 pp) & Local path: 0.0 pp, lower bound \ensuremath{-}4.64 pp & Met on the local path only \\
H5 & Severity weighting changes the error rate & Mean D = +0.005; 87 of 95 intervals contain 0 & Small upward tilt within uncertainty (descriptive) \\
H6 & Loosening and tightening differ & 10 vs 10 discordant pairs; p = 1.000 & No net difference \\
H7 & Adaptive attacker recovers success & No discordant pairs under defense & Not supported \\
H8 & Certainty tracks correctness & Accuracy \ensuremath{-}21.6 pp, strong assertion \ensuremath{-}4.3 pp & Not supported \\
\end{longtable}
}
}

\subsection{5.2 Text-Level Poisoning (SQ1)}\label{text-level-poisoning-sq1}

\subsubsection{5.2.1 Effectiveness and Dose Response}\label{effectiveness-and-dose-response}

{\small\textbf{Table 5.2} reports Attack A1 in its six cells. In both folds strict success increased with the budget, from 0.160 to 0.218 in fold 0 and from 0.107 to 0.267 in fold 1. Disruption, the true value vanishing from the answer whatever replaced it, reached 0.600 and 0.622 at M = 5: the correct value was missing from about six in ten poisoned answers, and the attacker\textquotesingle s value had replaced it in about one in four.}

{\small\textbf{Table 5.2} Attack A1 by fold and budget}

{\def\LTcaptype{none} 
{\small
\begin{longtable}[]{@{}
  >{\raggedright\arraybackslash}p{(\linewidth - 12\tabcolsep) * \real{0.0800}}
  >{\raggedright\arraybackslash}p{(\linewidth - 12\tabcolsep) * \real{0.0800}}
  >{\raggedright\arraybackslash}p{(\linewidth - 12\tabcolsep) * \real{0.1200}}
  >{\raggedright\arraybackslash}p{(\linewidth - 12\tabcolsep) * \real{0.2000}}
  >{\raggedright\arraybackslash}p{(\linewidth - 12\tabcolsep) * \real{0.1600}}
  >{\raggedright\arraybackslash}p{(\linewidth - 12\tabcolsep) * \real{0.1600}}
  >{\raggedright\arraybackslash}p{(\linewidth - 12\tabcolsep) * \real{0.2000}}@{}}
\toprule\noalign{}
\begin{minipage}[b]{\linewidth}\raggedright
Fold
\end{minipage} & \begin{minipage}[b]{\linewidth}\raggedright
M
\end{minipage} & \begin{minipage}[b]{\linewidth}\raggedright
Strict ASR
\end{minipage} & \begin{minipage}[b]{\linewidth}\raggedright
Disruption
\end{minipage} & \begin{minipage}[b]{\linewidth}\raggedright
Poison recall@k
\end{minipage} & \begin{minipage}[b]{\linewidth}\raggedright
Injected passages cited
\end{minipage} & \begin{minipage}[b]{\linewidth}\raggedright
References cited
\end{minipage} \\
\midrule\noalign{}
\endhead
\bottomrule\noalign{}
\endlastfoot
0 & 1 & 0.1600 & 0.4044 & 1.0000 & 1.00 & 4.56 \\
0 & 3 & 0.1733 & 0.4933 & 0.9896 & 2.97 & 9.62 \\
0 & 5 & 0.2178 & 0.6000 & 0.9956 & 4.98 & 12.00 \\
1 & 1 & 0.1067 & 0.3378 & 1.0000 & 1.00 & 4.28 \\
1 & 3 & 0.1511 & 0.4622 & 0.9970 & 2.99 & 8.84 \\
1 & 5 & 0.2667 & 0.6222 & 0.9947 & 4.97 & 11.78 \\
\end{longtable}
}
}

\emph{Note.} Each cell is 75 poisoned items \ensuremath{\times} 3 repeats = 225 records. Poison recall@k is the mean share of a target\textquotesingle s injected passages that the answer cites. The last two columns are means per poisoned record.

\begin{center}
\includegraphics[width=0.93\linewidth,keepaspectratio,alt={Figure 5.1 Strict success and disruption of Attack A1 by budget and fold}]{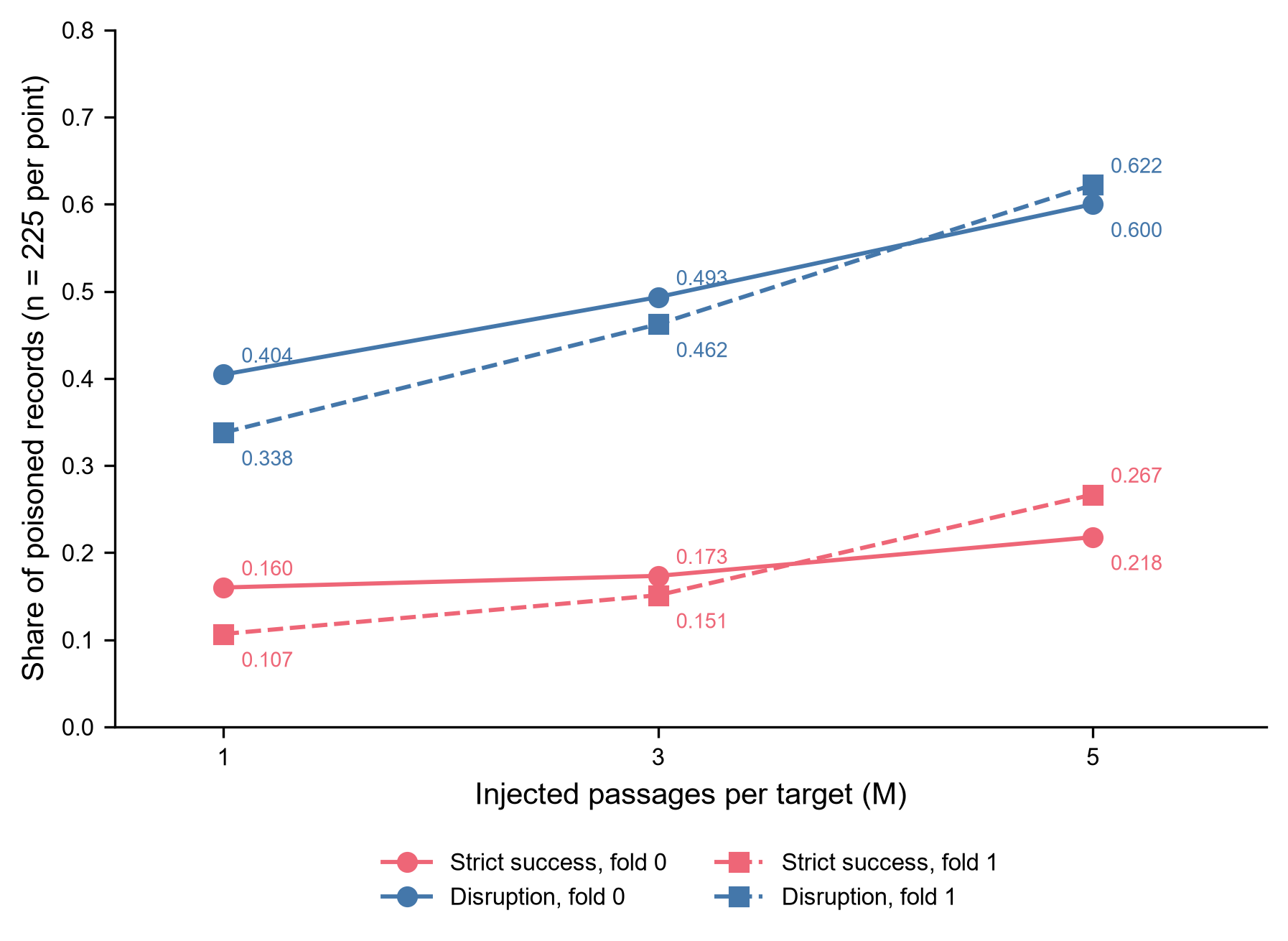}
\end{center}

{\small\textbf{Figure 5.1} Strict success and disruption of Attack A1 by budget and fold}

\emph{Note.} Each point is the mean over 225 poisoned records. Lines connect the nested budgets within a fold; folds are not compared with each other because their clean accuracy differs.

For \textbf{H1a}, 33 of the 150 targets succeeded in the majority of their three repeats at M = 5, a rate of 0.220 against the registered bound of 0.10 (one-sided exact binomial, p = 1.17 \ensuremath{\times} 10\textsuperscript{\ensuremath{-}5}), eleven more than the rejection region required; over the 125 block-level units the rate is 0.217 (95 percent interval {[}0.150, 0.288{]}). For \textbf{H1b}, Page\textquotesingle s (1963) test for ordered alternatives, with items as blocks, gives L = 1842.0 (rank sums 280.5, 297.0 and 322.5), and none of 50,000 within-item permutations reached it (p \textless{} 0.0001). Both hypotheses hold after Holm correction.

\subsubsection{5.2.2 Where the Dose Response Comes From}\label{where-the-dose-response-comes-from}

Retrieval was saturated from the smallest budget: at M = 1 the injected passage was cited in all 225 poisoned records, and at M = 3 and 5 in 99.0 to 99.6 percent. What the budget changed was the composition of the evidence. From M = 1 to 5, references cited per answer rose from about 4.5 to about 12 and the poison\textquotesingle s share of them from about 22 to 42 percent, while the clean references also grew, from about 3.5 to 7. The poison was added beside the correct evidence, not in its place, so the extra success at higher budgets does not come from better retrieval. How success was distributed across repeats is given in Appendix D.1.

\subsubsection{5.2.3 Spillover to Untargeted Questions}\label{spillover-to-untargeted-questions}

\textbf{H3 is inconclusive.} On the 150 pooled untargeted observations at M = 5, with refusals counted as incorrect, accuracy on untargeted items was 0.040 lower under attack than in the clean run (95 percent block-bootstrap interval {[}\ensuremath{-}0.007, +0.091{]}; one-sided exact McNemar p = 0.132, with 13 items lost and 7 gained). The within-run controls point the same way but weakly: at M = 5, accuracy on the unpoisoned fold was 0.871 and 0.822 in the two runs, against 0.898 and 0.840 for the same items in the clean run.

\subsection{5.3 Graph-Layer Poisoning (SQ2)}\label{graph-layer-poisoning-sq2}

Attack A2 aimed at the graph through shared relations rather than at retrieval through the question. Table 5.3 reports two overlaps.

{\small\textbf{Table 5.3} Attack A2 by fold at M = 5}

{\def\LTcaptype{none} 
{\small
\begin{longtable}[]{@{}
  >{\raggedright\arraybackslash}p{(\linewidth - 10\tabcolsep) * \real{0.0909}}
  >{\raggedright\arraybackslash}p{(\linewidth - 10\tabcolsep) * \real{0.1818}}
  >{\raggedright\arraybackslash}p{(\linewidth - 10\tabcolsep) * \real{0.1364}}
  >{\raggedright\arraybackslash}p{(\linewidth - 10\tabcolsep) * \real{0.2273}}
  >{\raggedright\arraybackslash}p{(\linewidth - 10\tabcolsep) * \real{0.1818}}
  >{\raggedright\arraybackslash}p{(\linewidth - 10\tabcolsep) * \real{0.1818}}@{}}
\toprule\noalign{}
\begin{minipage}[b]{\linewidth}\raggedright
Fold
\end{minipage} & \begin{minipage}[b]{\linewidth}\raggedright
Poisoned targets
\end{minipage} & \begin{minipage}[b]{\linewidth}\raggedright
Strict ASR
\end{minipage} & \begin{minipage}[b]{\linewidth}\raggedright
Disruption
\end{minipage} & \begin{minipage}[b]{\linewidth}\raggedright
Poison recall@k
\end{minipage} & \begin{minipage}[b]{\linewidth}\raggedright
Records citing \ensuremath{\geq} 1 injected passage
\end{minipage} \\
\midrule\noalign{}
\endhead
\bottomrule\noalign{}
\endlastfoot
0 & 72 & 0.1759 & 0.5231 & 0.9648 & 216 / 216 \\
1 & 71 & 0.1737 & 0.5023 & 0.9728 & 212 / 213 \\
\end{longtable}
}
}

\emph{Note.} Seven of the 150 targets could not be poisoned (Section 4.5); they are excluded from this table but count as failures in the registered test.

The graph attack disrupted the answer about as often as the text attack (around half of poisoned records), but replaced the value with the attacker\textquotesingle s less often (Figure 5.2). In the registered unit, 22 of the 150 targets succeeded in the majority of their repeats, against 33 for A1.

\begin{center}
\includegraphics[width=0.93\linewidth,keepaspectratio,alt={Figure 5.2 Strict success and disruption of the text-level (A1) and graph-level (A2) attacks at the same budget, by fold}]{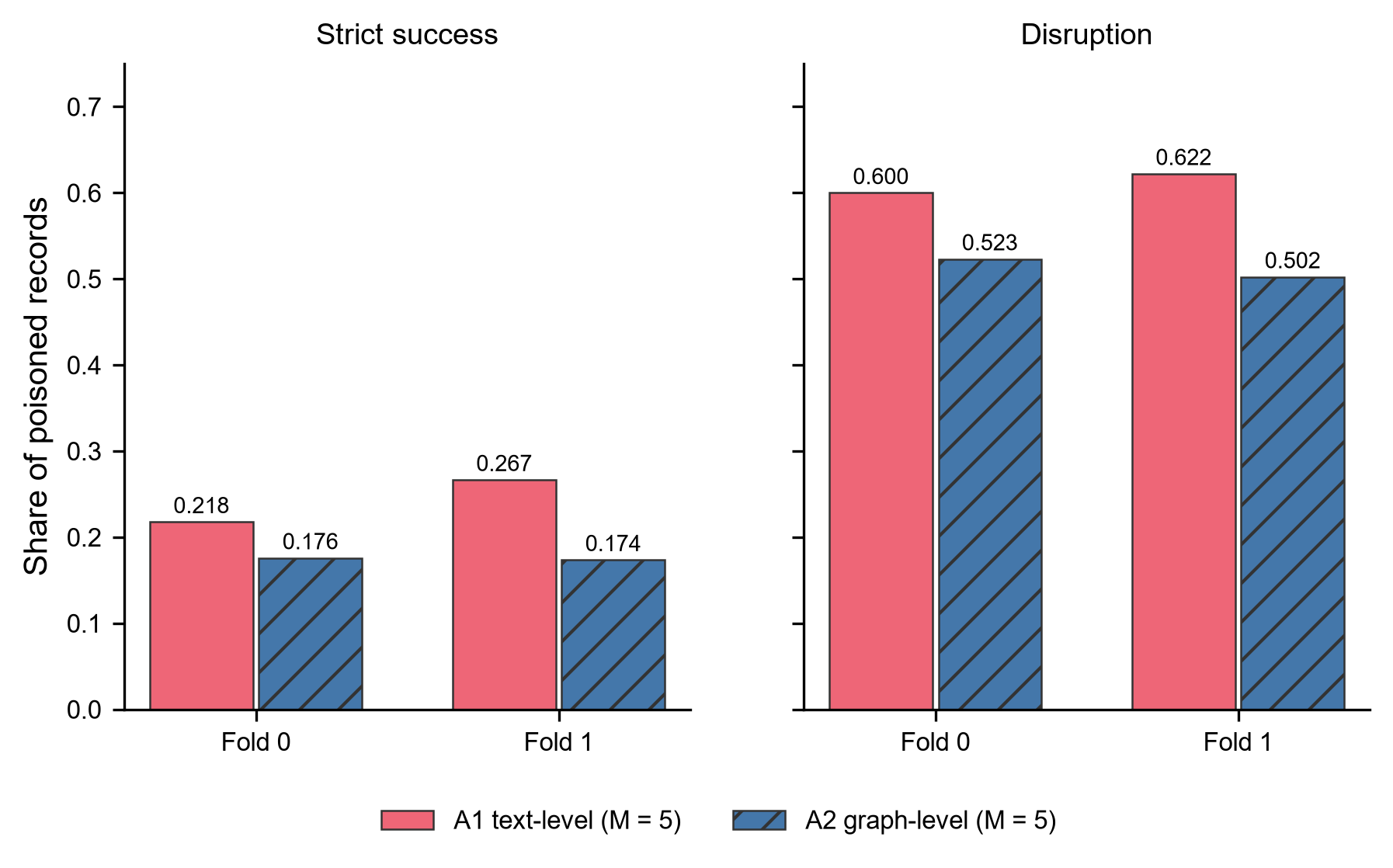}
\end{center}

{\small\textbf{Figure 5.2} Strict success and disruption of the text-level (A1) and graph-level (A2) attacks at the same budget, by fold}

\emph{Note.} Each bar is the mean over the poisoned records of one run (A1: 225 per fold; A2: 216 and 213).

\textbf{H2 is not supported, and the data point the other way.} In the registered one-sided exact McNemar test (McNemar, 1947) on the 150 targets, 20 targets succeeded only under A1 and 9 only under A2; the paired difference is \ensuremath{-}0.073 and the one-sided p in the predicted direction is 0.988. The reverse direction (one-sided p = 0.031) does not clear the first Holm threshold of the seven-member family (0.0071). At the budget H2 fixed, shared-relation poisoning succeeded on fewer targets than text-level injection.

Three alternative explanations were ruled out. Without the seven unpoisonable targets the counts are 31 and 22 of 143, the same direction (p = 0.061). A2 added fewer wholly injected relations than A1 (6,416 against 7,186), so its lower success is not a larger perturbation diluting itself. And every poisoned target cited an injected passage in most of its repeats; the gap in injected passages cited per answer (4.82 and 4.86 against 4.98 and 4.97) is a fraction of one passage, against a one-third gap in successes. The pre-registered subgroup of 39 targets inside fold blocks moved in H2\textquotesingle s favor (11 against 9; p = 0.377), but its blocks group items by shared provision, not by the shared entity A2 exploits.

A second-batch re-analysis of the same runs therefore grouped the 143 targets that both attacks poisoned by whether the entity A2 attacked is shared with another target in the same fold. A2 trailed A1 by 3.3 points on the 60 shared-entity targets (11 against 9 successes; one-sided p = 0.83) and by 8.4 points on the 83 others (20 against 13; p = 0.98). The difference between strata, +5.1 points (95 percent interval {[}\ensuremath{-}8.9, +19.2{]}), is the difference of two deficits rather than a gain: its direction agrees with the shared-entity mechanism, but the stratum cannot resolve it (McNemar reads only the discordant pairs, of which the shared stratum has ten).

\subsection{5.4 Amendment-Narrative Poisoning: Three Cases}\label{amendment-narrative-poisoning-three-cases}

Attack A3 was qualitative by design. Each fabricated amendment notice was retrieved and cited in all 18 poisoned answers (three cases, three repeats, on the experimental index and on the 70-document index), and every answer repeated the fabricated value together with the claim that a 2026 amendment had superseded the original provision (Table 5.4).

{\small\textbf{Table 5.4} The three amendment-narrative cases}

{\def\LTcaptype{none} 
{\small
\begin{longtable}[]{@{}
  >{\raggedright\arraybackslash}p{(\linewidth - 10\tabcolsep) * \real{0.1579}}
  >{\raggedright\arraybackslash}p{(\linewidth - 10\tabcolsep) * \real{0.1579}}
  >{\raggedright\arraybackslash}p{(\linewidth - 10\tabcolsep) * \real{0.1579}}
  >{\raggedright\arraybackslash}p{(\linewidth - 10\tabcolsep) * \real{0.2105}}
  >{\raggedright\arraybackslash}p{(\linewidth - 10\tabcolsep) * \real{0.1930}}
  >{\raggedright\arraybackslash}p{(\linewidth - 10\tabcolsep) * \real{0.1228}}@{}}
\toprule\noalign{}
\begin{minipage}[b]{\linewidth}\raggedright
Case
\end{minipage} & \begin{minipage}[b]{\linewidth}\raggedright
Provision
\end{minipage} & \begin{minipage}[b]{\linewidth}\raggedright
Tampering
\end{minipage} & \begin{minipage}[b]{\linewidth}\raggedright
Verdicts, experimental index
\end{minipage} & \begin{minipage}[b]{\linewidth}\raggedright
Verdicts, 70-document index
\end{minipage} & \begin{minipage}[b]{\linewidth}\raggedright
Stated as current
\end{minipage} \\
\midrule\noalign{}
\endhead
\bottomrule\noalign{}
\endlastfoot
C1 (QA-B-05) & Reduced travel distance, small block of flats & 4.5 m \ensuremath{\rightarrow} 3.5 m (tighten) & partial \ensuremath{\times}3 & partial \ensuremath{\times}3 & 6 of 6 \\
C2 (QA-B-21) & Width of other escape stairs & 800 mm \ensuremath{\rightarrow} 900 mm (tighten) & correct \ensuremath{\times}3 & partial \ensuremath{\times}2, correct \ensuremath{\times}1 & 6 of 6 \\
C3 (QA-K-14) & Guarding at edges of roofs & 1,100 mm \ensuremath{\rightarrow} 1,000 mm (loosen) & correct \ensuremath{\times}3 & correct \ensuremath{\times}3 & 6 of 6 \\
\end{longtable}
}
}

The strict success criterion never fired, because the answers did not delete the true value; they reclassified it. A typical answer reproduced the 800 mm stair width ``in the 2019 edition text'' and then stated that ``the 2026 amendment booklet'' changes it to 900 mm and ``supersedes the earlier requirement''. The rubric scored many such answers as correct, yet a reader would apply the fabricated value, which is why A3 is reported as cases rather than as a zero rate.

\subsection{5.5 The Three-Layer Defense (SQ3)}\label{the-three-layer-defense-sq3}

\subsubsection{5.5.1 Each Configuration against the Registered Criterion}\label{each-configuration-against-the-registered-criterion}

All eight configurations were run on A2\textquotesingle s fold-0 poisoned index, with 72 poisoned and 78 clean items, three repeats each. The undefended ASR was 0.1574, so the relative term of H4a binds and the threshold is 0.0787 (Table 5.5).

{\small\textbf{Table 5.5} Defense configurations against H4a, with refusal and accuracy}

{\def\LTcaptype{none} 
{\small
\begin{longtable}[]{@{}
  >{\raggedright\arraybackslash}p{(\linewidth - 12\tabcolsep) * \real{0.2241}}
  >{\raggedright\arraybackslash}p{(\linewidth - 12\tabcolsep) * \real{0.1034}}
  >{\raggedright\arraybackslash}p{(\linewidth - 12\tabcolsep) * \real{0.1207}}
  >{\raggedright\arraybackslash}p{(\linewidth - 12\tabcolsep) * \real{0.1724}}
  >{\raggedright\arraybackslash}p{(\linewidth - 12\tabcolsep) * \real{0.1379}}
  >{\raggedright\arraybackslash}p{(\linewidth - 12\tabcolsep) * \real{0.1379}}
  >{\raggedright\arraybackslash}p{(\linewidth - 12\tabcolsep) * \real{0.1034}}@{}}
\toprule\noalign{}
\begin{minipage}[b]{\linewidth}\raggedright
Configuration
\end{minipage} & \begin{minipage}[b]{\linewidth}\raggedright
ASR
\end{minipage} & \begin{minipage}[b]{\linewidth}\raggedright
H4a (\ensuremath{\leq} 0.0787)
\end{minipage} & \begin{minipage}[b]{\linewidth}\raggedright
Refusal (poisoned)
\end{minipage} & \begin{minipage}[b]{\linewidth}\raggedright
Poisoned acc., refusal = wrong
\end{minipage} & \begin{minipage}[b]{\linewidth}\raggedright
Poisoned acc., answered only
\end{minipage} & \begin{minipage}[b]{\linewidth}\raggedright
Clean acc.
\end{minipage} \\
\midrule\noalign{}
\endhead
\bottomrule\noalign{}
\endlastfoot
none & 0.1574 & --- & 0.000 & 0.4213 & 0.4213 & 0.8974 \\
retrieval & 0.0231 & Met & 0.000 & 0.8565 & 0.8565 & 0.9231 \\
graph & 0.1667 & Not met & 0.000 & 0.4491 & 0.4491 & 0.8718 \\
generation & 0.0463 & Met by refusal & 0.722 & 0.0463 & 0.1667 & 0.1538 \\
retrieval+\allowbreak{}graph & 0.0185 & Met & 0.009 & 0.8426 & 0.8505 & 0.9103 \\
retrieval+\allowbreak{}generation & 0.0000 & Met by refusal & 0.769 & 0.1944 & 0.8400 & 0.3205 \\
graph+\allowbreak{}generation & 0.0463 & Met by refusal & 0.685 & 0.0694 & 0.2206 & 0.1795 \\
all & 0.0000 & Met by refusal & 0.796 & 0.1667 & 0.8182 & 0.2308 \\
\end{longtable}
}
}

\emph{Note.} Registered batch of 18 October 2026. Poisoned accuracy is over 216 records (72 items \ensuremath{\times} 3 repeats). Clean accuracy is over the 78 clean items, each scored by the majority of its three repeats, the unit H4b fixes. Literal exact match (Equation 4.13) was 0 in every configuration.

The registered test concerns the full defense, \emph{all}; applied to every configuration, the same threshold is met by six of the seven, in two distinct ways. The four configurations with the generation layer meet it by declining to answer, refusing between 68.5 and 79.6 percent of poisoned records. The two with the retrieval layer but not the generation layer meet it while answering almost everything (refusal 0.0 and 0.9 percent). Only \emph{graph} fails: the graph layer edits only relation edges that contradict an anchored value, so the poisoned chunks still reached the generator (poison recall 0.981), and A2\textquotesingle s payload states an applicability rather than a number an anchor could contradict. For \emph{all}, the registered test gives 11 discordant pairs, all in the defense\textquotesingle s favor (p = 0.00049; bootstrap interval for the change in ASR {[}\ensuremath{-}0.241, \ensuremath{-}0.076{]}).

The criterion alone cannot rank configurations. Over the records each chose to answer, \emph{all} and \emph{retrieval+\allowbreak{}generation} look strong (0.818 and 0.840); over every poisoned record, with a refusal counted as wrong, they answer only 0.167 and 0.194 correctly, less than half the undefended 0.421, while \emph{retrieval} answers 0.857 and refuses none. Once the denominator is fixed, no configuration containing the generation layer beats no defense (Figure 5.3).

\begin{center}
\includegraphics[width=0.93\linewidth,keepaspectratio,alt={Figure 5.3 Answered share and poisoned accuracy (refusal counted as wrong) for the eight configurations}]{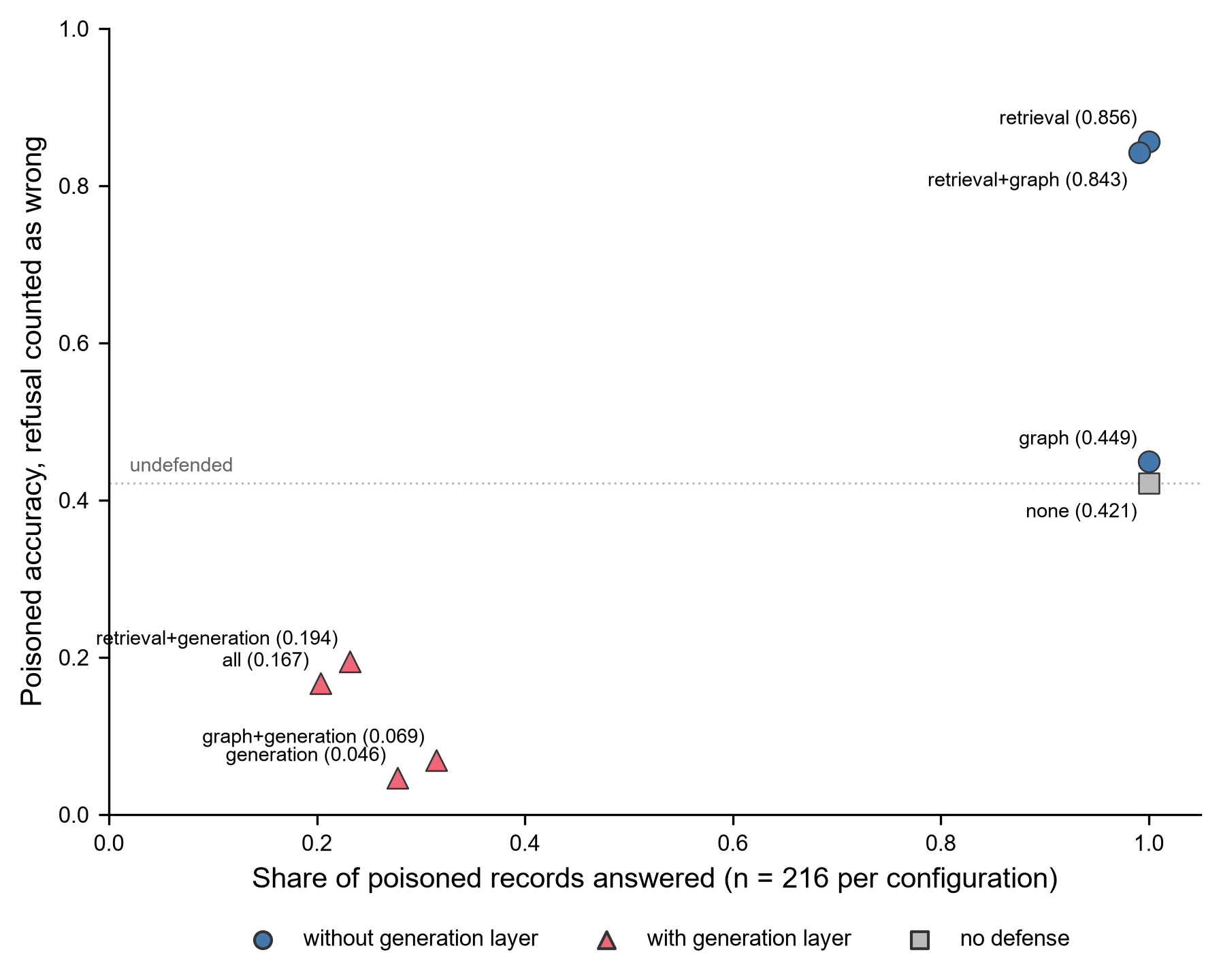}
\end{center}

{\small\textbf{Figure 5.3} Answered share and poisoned accuracy (refusal counted as wrong) for the eight configurations}

\emph{Note.} Each point is one configuration of Table 5.5, computed over the 216 poisoned records. The configurations that include the generation layer form the lower-left cluster.

A second-batch count applied the proposal\textquotesingle s exact-match indicator (Equation 4.13): none of the 4,647 answered records in this batch, the 25 October correction batch and two clean-index runs equalled its reference, although positive controls confirmed that the comparison detects a match. Whole-answer identity is uninformative for multi-part answers, which is why the rubric tests each required value.

\subsubsection{5.5.2 Clean Accuracy (H4b)}\label{clean-accuracy-h4b}

H4b was tested on the clean index with the retrieval layer armed, in two variants of its multi-path test. On the local graph path clean accuracy was unchanged (0.0 points; one-sided 95 percent lower bound \ensuremath{-}4.64 against the \ensuremath{-}5 margin): non-inferior, by 0.36 points. On the global path it fell by 32.0 points (90 percent interval {[}\ensuremath{-}39.5, \ensuremath{-}24.7{]}): inferior. In the ablation batch the retrieval layer alone raised clean accuracy slightly (0.897 to 0.923), whereas every configuration with the generation layer lost between 58 and 74 points (Table 5.5 and Figure 5.4).

\begin{center}
\includegraphics[width=0.93\linewidth,keepaspectratio,alt={Figure 5.4 Clean accuracy and poisoned accuracy for the eight configurations}]{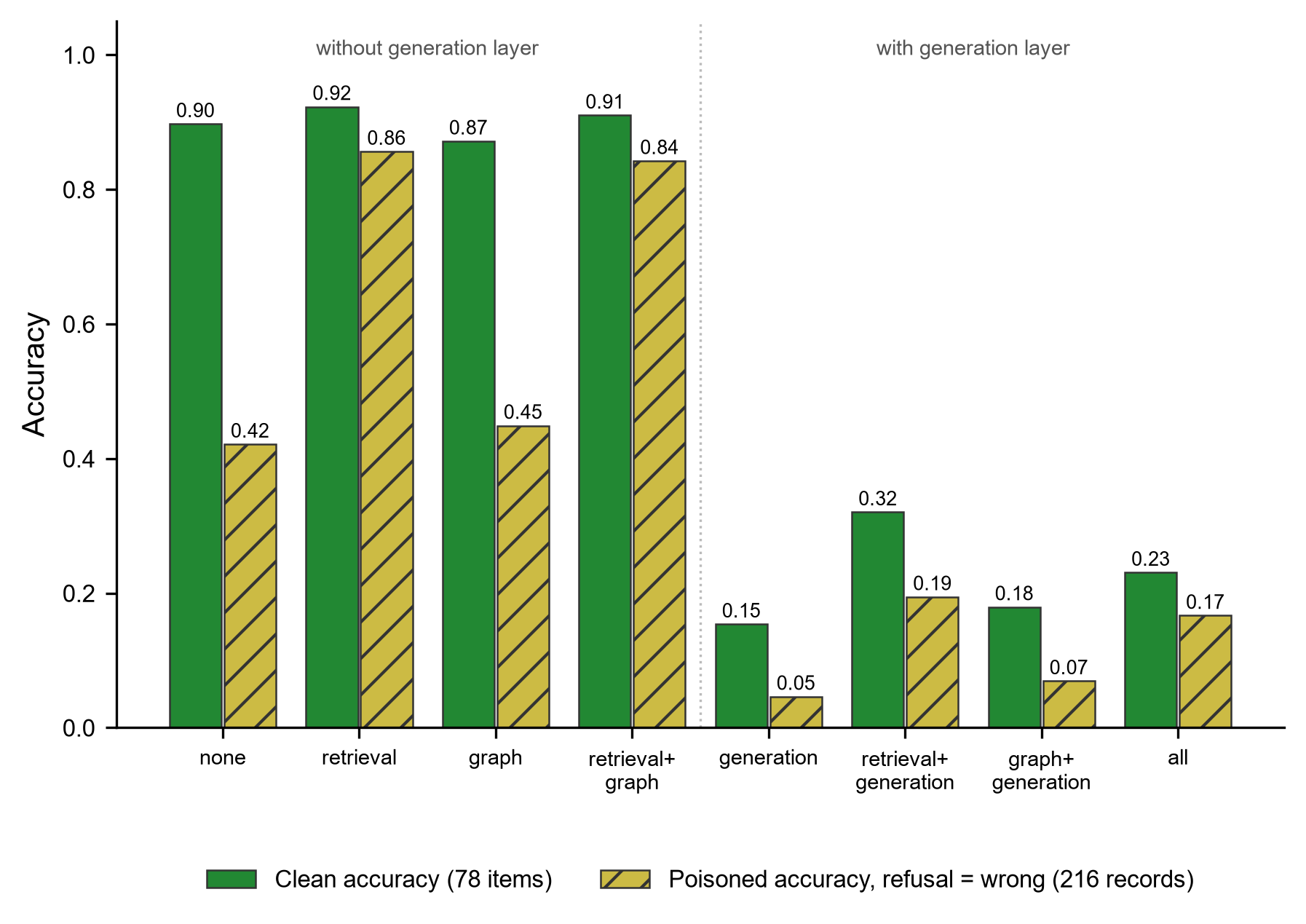}
\end{center}

{\small\textbf{Figure 5.4} Clean accuracy and poisoned accuracy for the eight configurations}

\emph{Note.} Clean accuracy is over the 78 unpoisoned items (majority of three repeats); poisoned accuracy is over the 216 poisoned records with a refusal counted as wrong. Registered batch of 18 October 2026.

On the clean benchmark the retrieval layer costs nothing by construction, because of Equation 4.2: every candidate comes from a manifest document with a matching hash, so \(S_{\text{trust}}\) is the same constant (2.0) for all of them and cannot change their order, whereas injected chunks score \(- 1\) and fall below the context cutoff. The same rule applies to a legitimate document that is not in the manifest, so a second-batch experiment measured that cost directly. A genuine 2026 amendment to Approved Document B was indexed beside the corpus and 20 questions were written whose answers depend on it. Three arms with the retrieval layer armed were run three times each, and the author reviewed the 92 records whose verdict could change, blind to the arm (Table 5.6).

{\small\textbf{Table 5.6} Second batch: provenance filtering and a legitimate amendment (20 dependent items)}

{\def\LTcaptype{none} 
{\small
\begin{longtable}[]{@{}
  >{\raggedright\arraybackslash}p{(\linewidth - 6\tabcolsep) * \real{0.1290}}
  >{\raggedright\arraybackslash}p{(\linewidth - 6\tabcolsep) * \real{0.2903}}
  >{\raggedright\arraybackslash}p{(\linewidth - 6\tabcolsep) * \real{0.3226}}
  >{\raggedright\arraybackslash}p{(\linewidth - 6\tabcolsep) * \real{0.2581}}@{}}
\toprule\noalign{}
\begin{minipage}[b]{\linewidth}\raggedright
Arm
\end{minipage} & \begin{minipage}[b]{\linewidth}\raggedright
Amendment
\end{minipage} & \begin{minipage}[b]{\linewidth}\raggedright
In the manifest
\end{minipage} & \begin{minipage}[b]{\linewidth}\raggedright
Correct, majority of three runs
\end{minipage} \\
\midrule\noalign{}
\endhead
\bottomrule\noalign{}
\endlastfoot
A & Absent & --- & 3 / 20 \\
B & Indexed & No (filtered) & 17 / 20 \\
C & Indexed & Yes & 16 / 20 \\
\end{longtable}
}
}

\emph{Note.} After blind review; B and C differ only in one manifest entry, so the filter\textquotesingle s cost is C \ensuremath{-} B = \ensuremath{-}0.05. One item was correct only in B and none only in C (exact McNemar, two-sided p = 1.0). Before review the counts were 2, 15 and 13.

The filtered arm answered as many dependent items as the registered one, one more in fact, so no cost of the filter was observed. This was not because the filter admitted the amendment\textquotesingle s chunks. A capture of the prompt context, three draws per item on the 16 items for which the amendment has distinctive wording, found that wording in the entity or relationship sections for 14 of the 16 items in arm B and in the chunk section in none of its 48 draws; arm A, without the amendment, gave no detection in 48 draws. The amendment\textquotesingle s content reached the generator through the graph descriptions.

\subsubsection{5.5.3 Factorial Effects and Interaction}\label{factorial-effects-and-interaction}

Because the eight configurations form a full factorial, each layer\textquotesingle s effect can be measured in each context of the other two (Table 5.7).

{\small\textbf{Table 5.7} Marginal effect of each layer on poisoned accuracy, by context}

{\def\LTcaptype{none} 
{\small
\begin{longtable}[]{@{}
  >{\raggedright\arraybackslash}p{(\linewidth - 8\tabcolsep) * \real{0.2500}}
  >{\raggedright\arraybackslash}p{(\linewidth - 8\tabcolsep) * \real{0.2500}}
  >{\raggedright\arraybackslash}p{(\linewidth - 8\tabcolsep) * \real{0.1750}}
  >{\raggedright\arraybackslash}p{(\linewidth - 8\tabcolsep) * \real{0.1500}}
  >{\raggedright\arraybackslash}p{(\linewidth - 8\tabcolsep) * \real{0.1750}}@{}}
\toprule\noalign{}
\begin{minipage}[b]{\linewidth}\raggedright
Layer
\end{minipage} & \begin{minipage}[b]{\linewidth}\raggedright
Other layers armed
\end{minipage} & \begin{minipage}[b]{\linewidth}\raggedright
Without layer
\end{minipage} & \begin{minipage}[b]{\linewidth}\raggedright
With layer
\end{minipage} & \begin{minipage}[b]{\linewidth}\raggedright
Effect
\end{minipage} \\
\midrule\noalign{}
\endhead
\bottomrule\noalign{}
\endlastfoot
Retrieval & none & 0.4213 & 0.8565 & +0.4352 \\
Retrieval & graph & 0.4491 & 0.8426 & +0.3935 \\
Retrieval & generation & 0.0463 & 0.1944 & +0.1481 \\
Retrieval & graph + generation & 0.0694 & 0.1667 & +0.0972 \\
Graph & none & 0.4213 & 0.4491 & +0.0278 \\
Graph & generation & 0.0463 & 0.0694 & +0.0231 \\
Graph & retrieval & 0.8565 & 0.8426 & \ensuremath{-}0.0139 \\
Graph & retrieval + generation & 0.1944 & 0.1667 & \ensuremath{-}0.0278 \\
Generation & none & 0.4213 & 0.0463 & \ensuremath{-}0.3750 \\
Generation & graph & 0.4491 & 0.0694 & \ensuremath{-}0.3796 \\
Generation & retrieval & 0.8565 & 0.1944 & \ensuremath{-}0.6620 \\
Generation & retrieval + graph & 0.8426 & 0.1667 & \ensuremath{-}0.6759 \\
\end{longtable}
}
}

\begin{center}
\includegraphics[width=0.93\linewidth,keepaspectratio,alt={Figure 5.5 The retrieval layer's effect on poisoned accuracy with and without the generation and graph layers}]{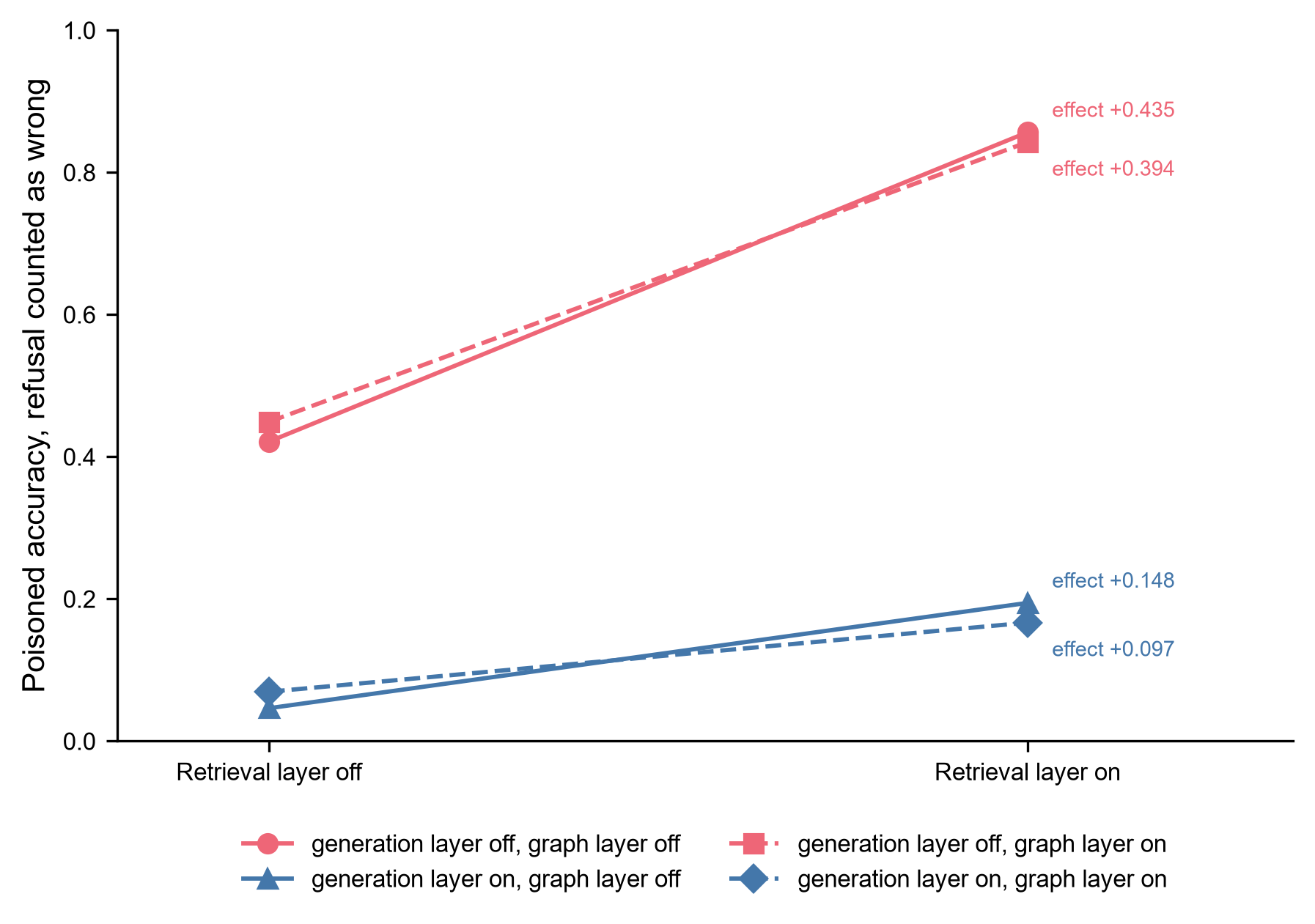}
\end{center}

{\small\textbf{Figure 5.5} The retrieval layer\textquotesingle s effect on poisoned accuracy with and without the generation and graph layers}

\emph{Note.} Each line joins two configurations that differ only in the retrieval layer; the label is the effect of adding it (Table 5.7). Non-parallel lines are an interaction.

The retrieval layer is the only one that improves poisoned accuracy in every context, but its effect depends strongly on context: +0.435 alone and +0.148 with the generation layer armed, an interaction (Equation 4.10) of \ensuremath{-}0.287, larger in absolute value than any effect of the graph layer (Figure 5.5). The graph layer\textquotesingle s four effects (\ensuremath{-}0.028 to +0.028) are all smaller than the spread of the undefended configuration across its own three repeats (0.389 to 0.472), so it has no measurable effect at this dose. The generation layer lowers poisoned accuracy in every context, most of all when retrieval is already filtered, because it then withholds answers that would have been correct. A configuration\textquotesingle s score is therefore not a sum of per-layer contributions: the retrieval layer\textquotesingle s effect ranges from +0.097 to +0.435 depending on which other layers are armed. The clean-side effects are in Appendix D.2.

\subsubsection{5.5.4 What the Zero Attack Success Rates Are Made Of}\label{what-the-zero-attack-success-rates-are-made-of}

Two configurations and all three adaptive arms under full defense report an ASR of exactly zero, produced by two separable mechanisms.

The retrieval layer acts on provenance. Poison recall fell from 0.969 undefended to 0.000 in every configuration that includes it, a measure computed whether or not the model answered. In a defended run of the A4 control arm the candidate pool still held about 55 injected chunks per poisoned query, but none of the 2,876 chunks that reached the model\textquotesingle s context came from the 360 injected documents. The layer removes these documents because they are absent from the manifest, not because their content was judged false.

The generation layer acts by retreat. In the registered batch every retreat followed a \emph{neutral} verdict, meaning the NLI model could not confirm the sentence from its cited evidence; no \emph{contradiction} or \emph{conflict} verdict occurred. In \emph{all} it retreated on 76 percent of queries and 73 percent of clean records, and on the A4 control arm arming all layers raised disruption from 0.528 to 0.884 while strict success fell from 0.157 to zero: a wrong answer was replaced by no answer, not by the correct one. Both correction batches (21 and 25 October) reproduce this partition (Appendix A, Table A-7).

\subsection{5.6 The Defense-Aware Adaptive Attacker (H7)}\label{the-defense-aware-adaptive-attacker-h7}

The adaptive arms were run first against the undefended system, to check that their constructions were competitive (Table 5.8), and then against the full defense.

{\small\textbf{Table 5.8} Adaptive arms against the same 72 poisoned items, undefended}

{\def\LTcaptype{none} 
{\small
\begin{longtable}[]{@{}
  >{\raggedright\arraybackslash}p{(\linewidth - 10\tabcolsep) * \real{0.2157}}
  >{\raggedright\arraybackslash}p{(\linewidth - 10\tabcolsep) * \real{0.1569}}
  >{\raggedright\arraybackslash}p{(\linewidth - 10\tabcolsep) * \real{0.1765}}
  >{\raggedright\arraybackslash}p{(\linewidth - 10\tabcolsep) * \real{0.1176}}
  >{\raggedright\arraybackslash}p{(\linewidth - 10\tabcolsep) * \real{0.1961}}
  >{\raggedright\arraybackslash}p{(\linewidth - 10\tabcolsep) * \real{0.1373}}@{}}
\toprule\noalign{}
\begin{minipage}[b]{\linewidth}\raggedright
Arm
\end{minipage} & \begin{minipage}[b]{\linewidth}\raggedright
Items attacked
\end{minipage} & \begin{minipage}[b]{\linewidth}\raggedright
Strict successes / records
\end{minipage} & \begin{minipage}[b]{\linewidth}\raggedright
ASR (n = 72)
\end{minipage} & \begin{minipage}[b]{\linewidth}\raggedright
Disruption
\end{minipage} & \begin{minipage}[b]{\linewidth}\raggedright
Paired p vs control
\end{minipage} \\
\midrule\noalign{}
\endhead
\bottomrule\noalign{}
\endlastfoot
control (A2 set) & 72 & 34 / 216 & 0.1574 & 0.528 & --- \\
adaptive\_l1 & 71 & 26 / 213 & 0.1204 & 0.451 & 1.000 (b = 3, c = 2) \\
adaptive\_l2 & 32 & 1 / 96 & 0.0046 & 0.115 & 0.002 (b = 10, c = 0) \\
adaptive\_l3 & 71 & 19 / 213 & 0.0880 & 0.258 & 0.219 (b = 5, c = 1) \\
\end{longtable}
}
}

\emph{Note.} b counts items on which only the control succeeded, c items on which only the adaptive arm did; exact McNemar, two-sided. An item on which an arm produced no poison counts as a failure.

No adaptive arm outperformed the non-adaptive control. The small p for adaptive\_l2 says nothing about the defense: eight of its ten discordant items are targets for which it produced no poison, because the graph offered no uncovered neighboring parameter for 40 of the 72 items; on the 32 items it could attack, p = 0.50. adaptive\_l3 moved against H7 (five items down, one up), apparently because a false claim placed beside its own supporting sentence is less likely to displace the original value.

Against the full defense, no injected chunk reached the model\textquotesingle s context in any arm (0 of 216, 213, 96 and 213 records), and on the registered unit all three arms had zero discordant pairs with the control: because the retrieval layer removed every injected chunk, no arm succeeded on a majority of repeats for any item (none of the 2,876 context chunks came from an injected document). \textbf{H7 is therefore not supported}, but the test had nothing to read.

The second batch therefore ran the control and the three arms with only the generation layer armed. For adaptive\_l1 and adaptive\_l2 the layer each was built to evade was then absent, so these two act as negative controls; only adaptive\_l3 faced its target layer. Two outcomes were registered before the run, each on the arm\textquotesingle s declared items with a one-sided exact McNemar test: H7\textquotesingle s outcome and the refusal wall, the majority of repeats not refused (Table 5.9 and Figure 5.6).

{\small\textbf{Table 5.9} Second batch: adaptive arms against the generation layer alone}

{\def\LTcaptype{none} 
{\small
\begin{longtable}[]{@{}
  >{\raggedright\arraybackslash}p{(\linewidth - 10\tabcolsep) * \real{0.2444}}
  >{\raggedright\arraybackslash}p{(\linewidth - 10\tabcolsep) * \real{0.1333}}
  >{\raggedright\arraybackslash}p{(\linewidth - 10\tabcolsep) * \real{0.1111}}
  >{\raggedright\arraybackslash}p{(\linewidth - 10\tabcolsep) * \real{0.1778}}
  >{\raggedright\arraybackslash}p{(\linewidth - 10\tabcolsep) * \real{0.1778}}
  >{\raggedright\arraybackslash}p{(\linewidth - 10\tabcolsep) * \real{0.1556}}@{}}
\toprule\noalign{}
\begin{minipage}[b]{\linewidth}\raggedright
Arm
\end{minipage} & \begin{minipage}[b]{\linewidth}\raggedright
Target layer
\end{minipage} & \begin{minipage}[b]{\linewidth}\raggedright
Items
\end{minipage} & \begin{minipage}[b]{\linewidth}\raggedright
Strict success: b, c, p
\end{minipage} & \begin{minipage}[b]{\linewidth}\raggedright
Records not refused, control \ensuremath{\rightarrow} arm
\end{minipage} & \begin{minipage}[b]{\linewidth}\raggedright
Refusal wall: b, c, p
\end{minipage} \\
\midrule\noalign{}
\endhead
\bottomrule\noalign{}
\endlastfoot
adaptive\_l1 & Layer 1, not armed & 71 & 0, 1, 0.50 & 5 \ensuremath{\rightarrow} 18 of 213 & 1, 4, 0.19 \\
adaptive\_l2 & Layer 2, not armed & 32 & 0, 0, 1.00 & 3 \ensuremath{\rightarrow} 10 of 96 & 1, 3, 0.31 \\
adaptive\_l3 & Layer 3, armed & 71 & 0, 0, 1.00 & 5 \ensuremath{\rightarrow} 17 of 213 & 1, 2, 0.50 \\
\end{longtable}
}
}

\emph{Note.} One batch, one environment and the same generation layer for all arms. b counts items on which only the control had the outcome, c items on which only the arm had it; exact McNemar, one-sided in the direction of the arm (Equation 4.8). Records are three repeats per declared item.

\begin{center}
\includegraphics[width=0.93\linewidth,keepaspectratio,alt={Figure 5.6 Records not refused by the control and by each adaptive arm, with only the generation layer armed}]{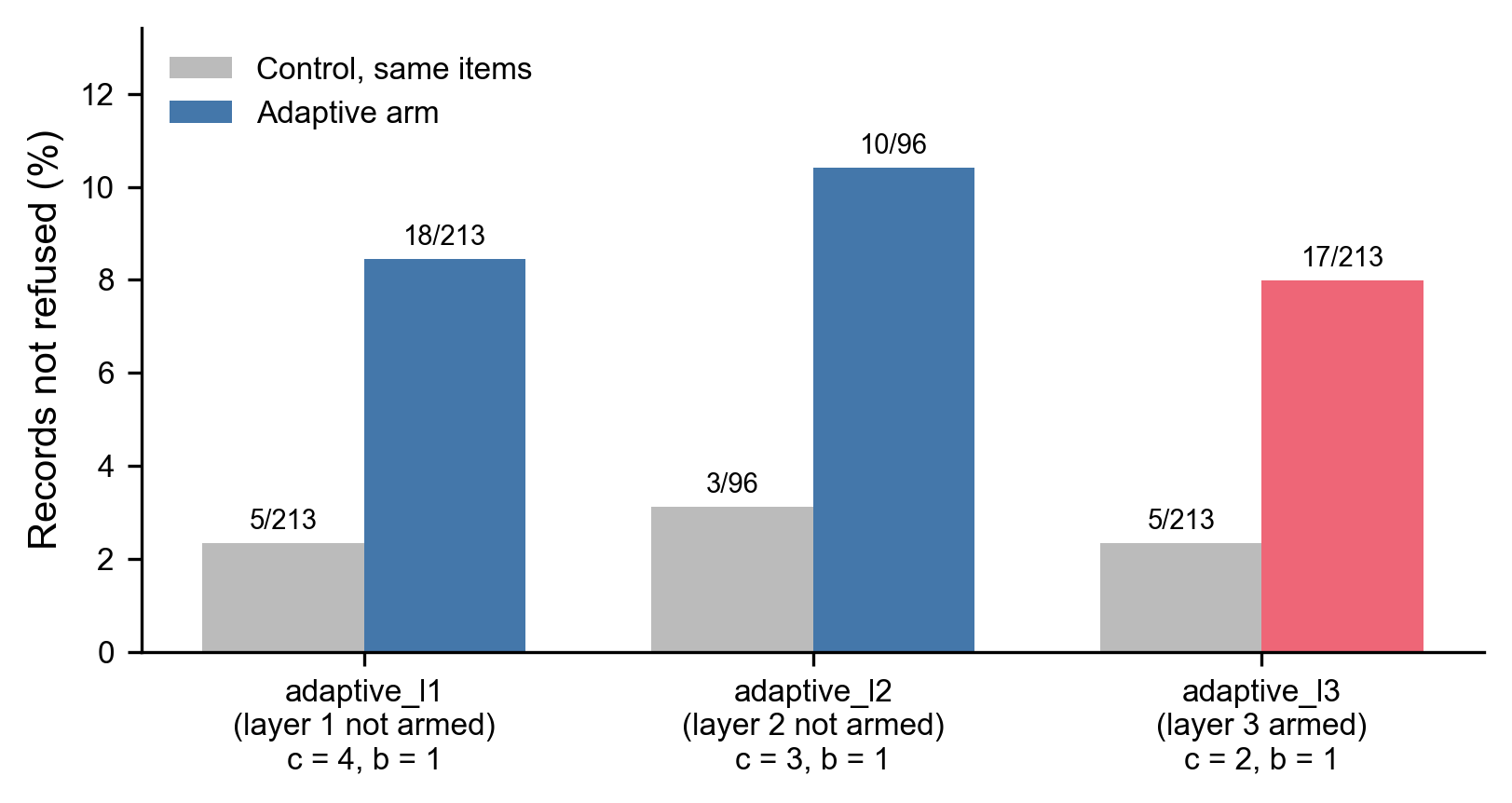}
\end{center}

{\small\textbf{Figure 5.6} Records not refused by the control and by each adaptive arm, with only the generation layer armed}

\emph{Note.} Grey bars are the control on the same items as the arm; c and b are the item-level discordant counts of the refusal wall (Table 5.9).

Neither outcome was rejected for any arm. With the observed b, rejection required c \ensuremath{\geq} 5 for H7 and c \ensuremath{\geq} 7 for the refusal wall, whereas adaptive\_l3 reached 0 and 2. The generation-layer arm released twelve more records than the control (5 to 17 of 213), but the negative control adaptive\_l1 released thirteen more on the same items (5 to 18), and the item-level counts c run 4, 3 and 2, smallest for the only arm whose target layer was armed. Under the rule fixed before the run, the movement is not attributable to the layer-3 construction.

One further observation bears on SQ2. All residual successes under the retrieval layer occurred with no injected chunk cited: the 5 strict successes of \emph{retrieval} and the 4 of \emph{retrieval+\allowbreak{}graph} in the registered batch, and 9 and 11 of 216 poisoned records in the correction batches, whereas strict success never occurs on the clean index (0 of 450). These counts are a lower bound on reach through the graph descriptions. For two such items from the 25 October batch, a capture of the full context over three draws each found the poisoned claim in the entity or relationship sections in all six draws (relationship six, entity four) and in the chunk section in none: the extracted descriptions carry poisoned content into the prompt by a route on which none of the three layers acts.

\subsection{5.7 Uneven Harms (SQ4)}\label{uneven-harms-sq4}

\subsubsection{5.7.1 Safety-Consequence Weighting (H5)}\label{safety-consequence-weighting-h5}

With weights of 3, 2 and 1 for life-safety, compliance and general items, the benchmark\textquotesingle s total weight is 386. Across the 97 usable runs the difference \(D\) (Equation 4.7) had a mean and median of +0.005; it was positive in 71 runs, zero in 9 and negative in 17. Errors were thus slightly concentrated on safety-critical items, but the tilt lies within its uncertainty: of the 95 runs with raw records, 87 have an item-clustered 95 percent interval containing zero, 8 lie above it and none below (Figure 5.7). The weights moved the error rate by about 4 percent where, by the arithmetic bound of Equation 4.7, they could have moved it by about 17 (Appendix D.3).

\begin{center}
\includegraphics[width=0.93\linewidth,keepaspectratio,alt={Figure 5.7 The severity tilt (D = textSW-ER - textER) for each run, with its item-clustered 95 percent bootstrap interval}]{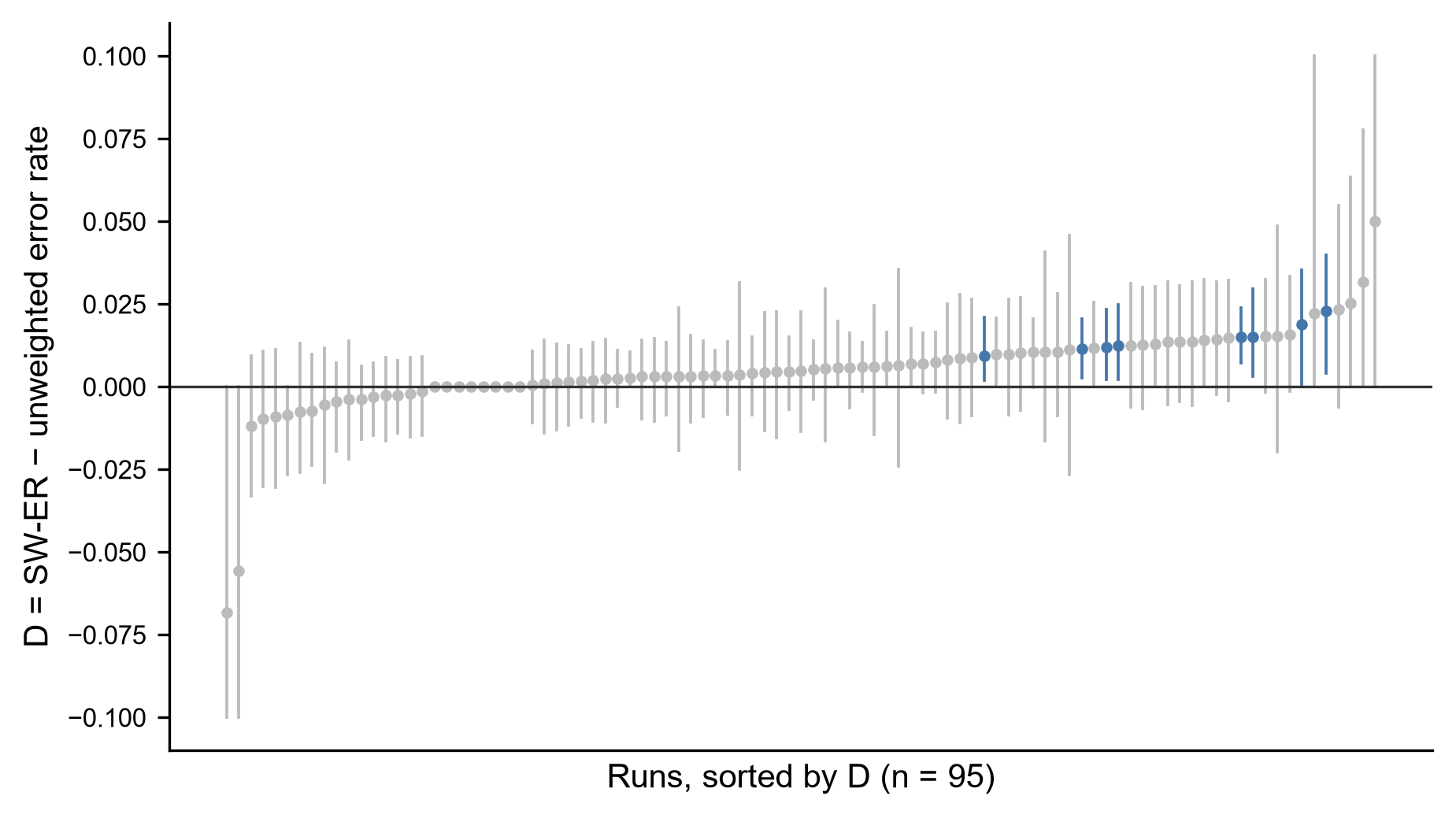}
\end{center}

{\small\textbf{Figure 5.7} The severity tilt \(D = \text{SW-ER} - \text{ER}\) for each run, with its item-clustered 95 percent bootstrap interval}

\emph{Note.} Blue intervals lie wholly above zero, grey contain zero; none lies wholly below. Each interval resamples whole items (2,000 draws). Two runs without raw records are omitted.

\subsubsection{5.7.2 Directional Bias (H6)}\label{directional-bias-h6}

The paired poison set has loosening and tightening versions for 130 targets; the generating model refused to write the second version for three, leaving 127 complete pairs. Both directions succeeded at the same rate, 0.181: in 13 pairs both versions succeeded, in 94 neither, and in 20 only one, split 10 and 10 (two-sided exact McNemar p = 1.000; 95 percent interval for the difference {[}\ensuremath{-}0.073, +0.074{]}). Direction changed individual outcomes but favored neither direction overall.

\subsubsection{5.7.3 Confidence and Correctness (H8)}\label{confidence-and-correctness-h8}

Confidence is read in two ways: the lexical certainty of the answer and, where the generation layer ran, the NLI model\textquotesingle s confidence in its verdict. Table 5.10 compares the clean and poisoned undefended configurations.

{\small\textbf{Table 5.10} Accuracy and expressed certainty, clean and poisoned corpus}

{\def\LTcaptype{none} 
{\small
\begin{longtable}[]{@{}
  >{\raggedright\arraybackslash}p{(\linewidth - 12\tabcolsep) * \real{0.2097}}
  >{\raggedright\arraybackslash}p{(\linewidth - 12\tabcolsep) * \real{0.1129}}
  >{\raggedright\arraybackslash}p{(\linewidth - 12\tabcolsep) * \real{0.1290}}
  >{\raggedright\arraybackslash}p{(\linewidth - 12\tabcolsep) * \real{0.1613}}
  >{\raggedright\arraybackslash}p{(\linewidth - 12\tabcolsep) * \real{0.0968}}
  >{\raggedright\arraybackslash}p{(\linewidth - 12\tabcolsep) * \real{0.1290}}
  >{\raggedright\arraybackslash}p{(\linewidth - 12\tabcolsep) * \real{0.1613}}@{}}
\toprule\noalign{}
\begin{minipage}[b]{\linewidth}\raggedright
Configuration
\end{minipage} & \begin{minipage}[b]{\linewidth}\raggedright
Answers scored
\end{minipage} & \begin{minipage}[b]{\linewidth}\raggedright
Strict accuracy
\end{minipage} & \begin{minipage}[b]{\linewidth}\raggedright
Share of strong assertions
\end{minipage} & \begin{minipage}[b]{\linewidth}\raggedright
Hedges per 100 words
\end{minipage} & \begin{minipage}[b]{\linewidth}\raggedright
P(strong \textbar{} correct)
\end{minipage} & \begin{minipage}[b]{\linewidth}\raggedright
P(strong \textbar{} incorrect)
\end{minipage} \\
\midrule\noalign{}
\endhead
\bottomrule\noalign{}
\endlastfoot
Clean, no defense & 444 & 0.896 & 0.628 & 0.095 & 0.626 (n = 398) & 0.667 (n = 6) \\
Poisoned, no defense & 437 & 0.680 & 0.586 & 0.089 & 0.559 (n = 297) & 0.750 (n = 16) \\
\end{longtable}
}
}

Poisoning lowered strict accuracy by 21.6 percentage points but the share of strongly asserted answers by only 4.3 points (item-level paired difference \ensuremath{-}0.043, 95 percent interval {[}\ensuremath{-}0.099, +0.014{]}), and the hedge rate did not change. Wrong answers were not stated less assertively than right ones; the point estimate runs the other way (0.750 against 0.559), although with 16 wrong answers the intervals overlap. The verifier\textquotesingle s own confidence was no better guide: it ranked correct above incorrect answers only weakly where the retrieval layer was armed (AUROC 0.58) and inversely where it was not (0.21), because without that layer the poison entails the wrong answers it supports (Appendix D.4). \textbf{H8 is not supported}: expressed certainty did not track correctness under attack, as J. Liu et al. (2026) report for contradictory evidence.

The registered runs carried no self-rating, so a second-batch pass on the clean index and on A1\textquotesingle s fold-0 index at M = 5 asked the model to rate its confidence in its own answer from 1 to 5. Ratings were mapped to probabilities and compared with strict correctness by the expected calibration error and the AUROC (Equation 4.11), with intervals over the 125 item units and one draw shared by both arms (Table 5.11 and Figure 5.8).

{\small\textbf{Table 5.11} Second batch: self-rated confidence against strict correctness}

{\def\LTcaptype{none} 
{\small
\begin{longtable}[]{@{}
  >{\raggedright\arraybackslash}p{(\linewidth - 14\tabcolsep) * \real{0.1356}}
  >{\raggedright\arraybackslash}p{(\linewidth - 14\tabcolsep) * \real{0.1186}}
  >{\raggedright\arraybackslash}p{(\linewidth - 14\tabcolsep) * \real{0.1186}}
  >{\raggedright\arraybackslash}p{(\linewidth - 14\tabcolsep) * \real{0.1695}}
  >{\raggedright\arraybackslash}p{(\linewidth - 14\tabcolsep) * \real{0.1017}}
  >{\raggedright\arraybackslash}p{(\linewidth - 14\tabcolsep) * \real{0.1186}}
  >{\raggedright\arraybackslash}p{(\linewidth - 14\tabcolsep) * \real{0.1186}}
  >{\raggedright\arraybackslash}p{(\linewidth - 14\tabcolsep) * \real{0.1186}}@{}}
\toprule\noalign{}
\begin{minipage}[b]{\linewidth}\raggedright
Arm
\end{minipage} & \begin{minipage}[b]{\linewidth}\raggedright
Records
\end{minipage} & \begin{minipage}[b]{\linewidth}\raggedright
Refused
\end{minipage} & \begin{minipage}[b]{\linewidth}\raggedright
Rating unreadable
\end{minipage} & \begin{minipage}[b]{\linewidth}\raggedright
Scored
\end{minipage} & \begin{minipage}[b]{\linewidth}\raggedright
Ratings 3 / 4 / 5
\end{minipage} & \begin{minipage}[b]{\linewidth}\raggedright
ECE {[}95\% CI{]}
\end{minipage} & \begin{minipage}[b]{\linewidth}\raggedright
AUROC {[}95\% CI{]}
\end{minipage} \\
\midrule\noalign{}
\endhead
\bottomrule\noalign{}
\endlastfoot
Clean & 450 & 12 & 11 & 427 & 4 / 18 / 405 & 0.104 {[}0.064, 0.147{]} & 0.650 {[}0.556, 0.754{]} \\
Poisoned & 450 & 14 & 7 & 429 & 8 / 98 / 323 & 0.309 {[}0.224, 0.400{]} & 0.489 {[}0.427, 0.555{]} \\
\end{longtable}
}
}

\emph{Note.} Refusals and unreadable ratings are held out of both measures. Poisoned minus clean, paired: \ensuremath{\Delta}ECE = +0.205 {[}0.129, 0.284{]}, \ensuremath{\Delta}AUROC = \ensuremath{-}0.161 {[}\ensuremath{-}0.285, \ensuremath{-}0.043{]}; 2,000 resamples of 125 units, none skipped; the intervals are descriptive. No rating of 1 or 2 occurred.

\begin{center}
\includegraphics[width=0.73\linewidth,keepaspectratio,alt={Figure 5.8 Fraction of strictly correct answers at each self-rated confidence level, clean and poisoned index}]{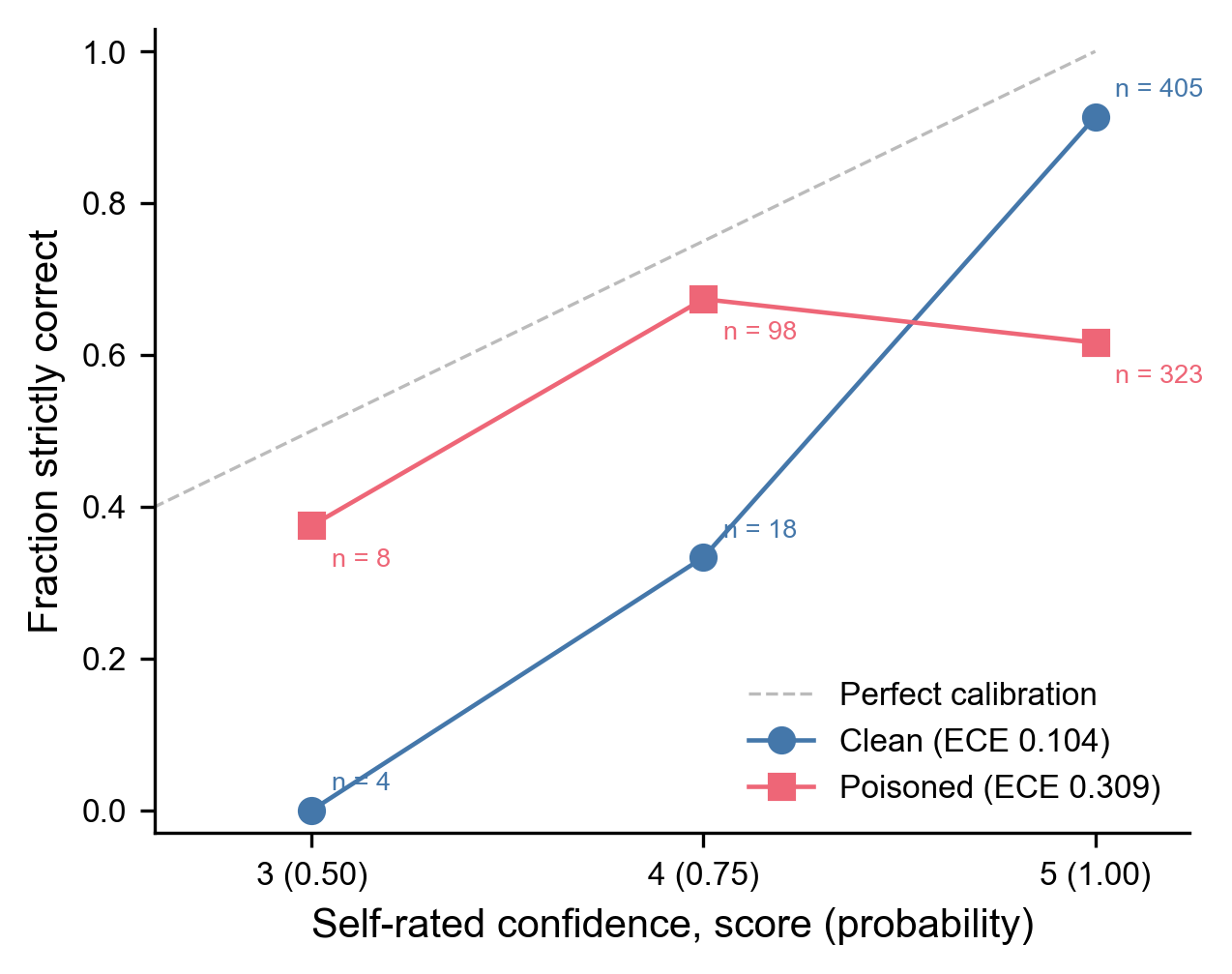}
\end{center}

{\small\textbf{Figure 5.8} Fraction of strictly correct answers at each self-rated confidence level, clean and poisoned index}

\emph{Note.} One point per rating level; n is the number of scored answers at that level. The dashed line is perfect calibration.

The model rated 95 percent of its clean and 75 percent of its poisoned answers at the top of the scale. Under poisoning, answers rated 5 were strictly correct in 61.6 percent of cases against 91.4 percent on the clean index, and this level carries most of the rise in calibration error; the rating no longer ranked correct above incorrect answers (AUROC 0.489).

\subsubsection{5.7.4 Natural Version Conflict}\label{natural-version-conflict}

Twelve further English questions probed version handling on the clean indexes, without any attacker. Five asked which instrument a value, class or date belongs to, and seven whether a provision already applied on the date of the run (12 September 2026), eighteen days before the commencement of 2026 amendments that are present in the two larger indexes but not in the experimental index (Table 5.12).

{\small\textbf{Table 5.12} Version handling on the clean corpora (counts)}

{\def\LTcaptype{none} 
{\small
\begin{longtable}[]{@{}
  >{\raggedright\arraybackslash}p{(\linewidth - 4\tabcolsep) * \real{0.3158}}
  >{\raggedright\arraybackslash}p{(\linewidth - 4\tabcolsep) * \real{0.4211}}
  >{\raggedright\arraybackslash}p{(\linewidth - 4\tabcolsep) * \real{0.2632}}@{}}
\toprule\noalign{}
\begin{minipage}[b]{\linewidth}\raggedright
Index
\end{minipage} & \begin{minipage}[b]{\linewidth}\raggedright
Presents a not-yet-in-force provision as current (error)
\end{minipage} & \begin{minipage}[b]{\linewidth}\raggedright
Attributes the provision to the correct instrument (accuracy)
\end{minipage} \\
\midrule\noalign{}
\endhead
\bottomrule\noalign{}
\endlastfoot
Experimental index (2 documents) & 0 / 5 & 0 / 2 \\
70-document index & 1 / 5 & 0 / 2 \\
192-document index & 1 / 5 & 0 / 2 \\
\end{longtable}
}
}

\emph{Note.} The two metrics have opposite polarity and are not pooled. The experimental index contains no amended edition, so its zero errors are structural.

On neither attribution item, in any index, did the system attribute the value to the governing instrument, and on the two larger indexes one in five temporal questions was answered by presenting a provision not yet in force as current. The failure is narrower than ``the system ignores versions'': answers named editions and dates, and the questions asking directly whether a provision was in force were answered correctly. The system retrieved the governing document but did not carry its commencement status into an answer that was not explicitly about status (item-level reading in Appendix A, Table A-9, and Appendix D.5). The confusion that A3 exploits deliberately is thus present without an attacker.

\subsection{5.8 Summary of Results}\label{summary-of-results}

Text-level poisoning was effective at a small budget; graph-level poisoning was less effective at the same budget. Fabricated amendment notices were repeated in every case and went undetected by value-based scoring, because the answers kept the correct value and the rubric checks only whether required values are present. The full defense met its registered criterion and six of the seven configurations met its threshold, but only the retrieval layer did so while preserving accuracy. The generation layer refused, and the graph layer had no measurable effect, for three reasons: its anchors cover about half the items (77 of 150, of which 67 are adjudicable for A2); the graph attack asserted an applicability rather than a contradicting number, which an anchor table cannot match; and the layer deletes only relation edges that conflict with an anchored value without filtering chunks, so poison recall stayed at 0.981 and the poisoned text still reached the generator. The layers interacted strongly. An unregistered legitimate amendment, like the poison, reached the prompt through the graph descriptions. The adaptive attacker did not gain. Wrong answers were not expressed with less certainty, self-rated confidence became less calibrated under poisoning, and on neither attribution item did the system name the governing instrument: it credited a value to an older edition containing the same number and dropped commencement dates, consistent with the absence of edition metadata in the chunks and descriptions it answers from.

\section{Chapter 6: Discussion}\label{chapter-6-discussion}

\subsection{6.1 What the Results Mean}\label{what-the-results-mean}

Three of the four sub-questions produced answers smaller, or less favorable to the proposal, than anticipated; because every reading was fixed in advance, these are results, not failed attempts at larger ones.

\subsection{6.2 The Graph Is a Second Surface, Not a Second Line of Defense}\label{the-graph-is-a-second-surface-not-a-second-line-of-defense}

The proposal, following the literature, treated graph-level poisoning as the more dangerous route: shared relations reach many queries, and GRAGPoison reported very high success on LightRAG (J. Liang et al., 2026). On this system, at a matched budget, the graph attack succeeded on fewer targets than the text attack (22 against 33 of 150). Under attack, then, poisoning aimed at the graph had the lower success rate, so the graph layer was not the softer of the two targets; that is narrower than saying the graph is a defense.

Whether structure blunts text injection needs a graph-free comparison, which the registered design did not contain. A second-batch run therefore used LightRAG\textquotesingle s naive mode, which retrieves chunks by vector similarity alone, on the clean index and on A1\textquotesingle s fold-0 index at M = 5, one pass per item, scored automatically. On the 75 targeted items strict accuracy fell from 0.9155 (65 of 71 answered) to 0.7067 (53 of 75), with 14 items lost and 2 gained (two-sided exact McNemar p = 0.0042); on the 75 untargeted items it fell from 0.9452 to 0.9437 (4 lost, 2 gained; p = 0.6875). The text attack therefore does not need the graph. What the graph adds or removes on top of that is not identified by this arm, which also differs from the hybrid runs in repetition (Section 9.3).

The graph also remains a channel. For two items whose full prompt context was captured, the poisoned claim appeared in the entity or relationship sections in all six draws and in the chunk section in none (Section 5.6). This was not intended: each layer was designed around the units the proposal named (candidate chunks, anchored relation edges, cited evidence), whereas the descriptions are summaries merged from several sources at indexing time that carry no per-document provenance; the gap was found only after the batches had run, and it leaves a route open for payloads written to survive extraction. This agrees with J. Liang et al.\textquotesingle s (2026) location of the damage at the extraction stage, with Wen et al.\textquotesingle s (2025) demonstration that rewriting a few words of the source text can change the graph built from it, and with Zhao et al.\textquotesingle s (2026) finding that adversarial graph content is retrieved more easily than it is adopted.

\subsection{6.3 One Criterion, Two Mechanisms}\label{one-criterion-two-mechanisms}

Applied to every configuration, the H4a threshold (Equation 3.1) is met by six of seven through two quite different mechanisms, and the criterion cannot tell them apart, because a refusal lowers the attack success rate exactly as a correct answer does. The generation layer meets it by refusing most queries, at a cost of 58 to 74 points of clean accuracy; the retrieval layer meets it by excluding the poison while answering nearly everything. The retrieval layer\textquotesingle s lack of a measurable cost has two parts. On the clean benchmark it holds by construction, because the trust term is the same constant for every manifest document and so cannot change their order (Section 5.5.2), although the non-inferiority margin was only 0.36 points, and a variant whose outlier term used a different graph-side path lost 32 points. On a legitimate amendment left outside the manifest, the second-batch false-positive test also found no measurable cost: of the 20 items that depend on the amendment, 17 were answered correctly when it was filtered and 16 when it was registered (C \ensuremath{-} B = \ensuremath{-}0.05; one discordant item, exact McNemar p = 1.0, a difference within noise), but for a reason that is no comfort (Section 6.5).

Under the usual definition, which this thesis adopted, a refusal is a failed attack (W. Zou et al., 2025; Equation 4.4), so any defense with an aggressive enough fallback reports a near-zero rate. None of the defenses reviewed in Chapter 2 reports its refusal share beside its success rate; where robust accuracy is reported, refusal can be inferred, and otherwise the published rate cannot separate detection from refusal. The remedy is cheap: report the answered share and the unconditional accuracy beside the rate. Here that separates the mechanisms at once: \emph{retrieval} reports ASR 0.023 with every query answered and 0.857 of poisoned records correct, \emph{all} reports ASR 0.000 with 0.20 answered and 0.167 correct. A compliance assistant that refuses three of four honest queries has not become resistant to poisoning; it has stopped being useful.

\subsection{6.4 Layer Interaction}\label{layer-interaction}

The factorial analysis showed that the layers do not compose additively. The retrieval layer contributes +0.435 alone and +0.148 with the generation layer armed, because it removes exactly the poisoned passages the generation layer would otherwise have had to reject, while the generation layer keeps withholding answers that would have been correct. A one-factor-at-a-time evaluation would have reported one of these numbers without revealing which question it answered. Layered defenses, including the proposal\textquotesingle s own, must therefore be evaluated as configurations. RAGuard, a two-layer defense for vector-based RAG, is one of the few that reports such an ablation: its adversarially trained retriever alone left residual attack success, whereas its counterfactual document filter alone, and both layers together, removed it (Kumar et al., 2026, pp. 2, 9). The factorial here extends that practice to three layers acting at different stages of a graph-based pipeline, where the effects proved strongly context-dependent.

\subsection{6.5 Provenance versus Content}\label{provenance-versus-content}

The most effective layer judged no content at all. It excluded documents absent from the hash manifest, which closed the chunk channel completely, in agreement with J. Liang et al.\textquotesingle s (2026) finding that a simple credibility annotation was the most effective defense they tested. It did not stop the attack completely: every residual success came through the graph descriptions (Section 5.6). A provenance filter also cannot distinguish a poisoned new document from a legitimate new amendment, and in a regulatory domain new amendments are precisely what users most need. The second-batch test shows how this plays out in LightRAG (Section 5.5.2): excluding an unregistered amendment cost nothing only because the filter never governed the amendment\textquotesingle s content, whose wording reached the prompt through the entity and relationship descriptions in 14 of 16 measurable items and through the chunk section in none. The filter is therefore both less costly and less protective than its design implies: it withholds a document\textquotesingle s chunks, while the descriptions extracted from that document at indexing time pass, whether the document is genuine or poisoned. Fei et al. (2026) identify a related blind spot in graph memories for LLM agents, where a defense that checks only the provenance of retrieved records cannot see an untrusted write to the graph structure that changes which records are selected. The problem thus moves from detection to admission, and admission must cover indexing as well as retrieval.

The anchor-based graph layer, which the proposal regarded as its most domain-specific contribution, had no measurable effect, for two reasons. Its coverage was limited: anchors are hand-extracted subject--relation--comparator--value--unit assertions, which explanatory, enumerative and rule-type answers do not fit, so the table covers 77 of 150 items and 67 after exemptions. And the attack did not need to violate an anchored value: A2 stated contradictory applicability rather than a contradictory number, and A4\textquotesingle s graph arm simply moved to an uncovered parameter.

\subsection{6.6 Harms beyond Attack Success Rate}\label{harms-beyond-attack-success-rate}

Three findings concern harms that a binary success rate does not register. The amendment-narrative cases showed that an answer can contain the correct value and still direct the reader to a fabricated one by declaring the correct value superseded, which value-based scoring counts as correct. The version-conflict probe showed that, even without an attacker, the system stated a not-yet-in-force rule as current when the question did not ask about status, so the ambiguity the narrative attack exploits is a standing property of the system. And poisoned answers were expressed with almost undiminished certainty, consistent with J. Liu et al. (2026). The second-batch self-rating puts a number on this (Section 5.7.3): calibration error rose from 0.104 to 0.309 under poisoning (paired difference +0.205, 95 percent interval {[}0.129, 0.284{]}), because the model kept rating three quarters of its poisoned answers at the top of the scale. Its stated confidence became a worse guide exactly when a guide was needed. Together these suggest that evaluation of regulatory RAG needs a version-aware correctness criterion that checks not only the value but the instrument and date it is attributed to. On severity and direction the data show only that errors leaned slightly toward life-safety provisions and that loosening and tightening split evenly.

These readings do not rest on one scorer. An independent judge agreed with the strict rubric on 50 of 58 explanatory answers, and where the author disagreed with the rubric, the rubric was always the stricter reader (Section 4.10). Accuracy under the strict rubric is therefore, if anything, a conservative reading of answer quality.

\subsection{6.7 A Proposed Improvement to the Defense Design}\label{a-proposed-improvement-to-the-defense-design}

These findings suggest a different allocation of effort. The retrieval layer should become an admission control: a signed manifest, with amendments admitted through an authenticated channel before indexing. The graph layer should move from contradiction detection to version arbitration, with anchor rows carrying instrument and commencement date. The generation layer should verify claims against the specific chunks cited, with a fallback calibrated on clean questions. And the entity and relationship descriptions need their own provenance, since they are the channel the poison used.

\section{Chapter 7: Conclusion}\label{chapter-7-conclusion}

This thesis asked how poisoned documents can make a LightRAG assistant for UK building regulations misstate a provision, and whether a three-layer defense can prevent it without harming honest answers. It answered with a pre-registered experiment on one instrumented system: a labeled 150-item benchmark, four attacks and a full factorial over three defensive layers.

The answer has four parts. First, text-level poisoning works at a small budget: five passages per target succeeded on 33 of 150 targets, and success rose with the budget although retrieval was saturated at one passage. Second, the graph was not the softer target: shared-relation poisoning succeeded on 22 of 150 targets at the same budget, the opposite of the registered prediction, although poisoned graph descriptions still reached the prompt through a channel no layer inspects. Third, the full defense met the registered criterion and six of the seven configurations met its threshold, but the generation layer did so by refusing most queries and losing most of its clean accuracy, while the retrieval layer did so with no measurable loss of clean accuracy, by a margin of 0.36 points; even a legitimate unregistered amendment was not lost, because its content, like the poison\textquotesingle s, reached the generator through the graph descriptions. The layers interacted so strongly that per-layer contributions cannot be stated. Fourth, the harms beyond the success rate were real but hard to measure: a fabricated amendment notice was repeated in every case while value-based scoring judged most answers correct, wrong answers were expressed with nearly undiminished certainty, the model\textquotesingle s self-rated confidence became markedly less calibrated under poisoning, and on neither attribution item did the system name the instrument that governs a provision.

Against its objectives, the study characterized the attack surface, implemented four attacks and three defensive layers, built a reusable benchmark, and evaluated outcomes beyond binary success; the bounds on these claims are in Chapter 9.

The study contributes three claims that extend the existing literature. On one system under one rubric, the graph is a second attack surface rather than a second line of defense. A layered defense\textquotesingle s components do not have separately reportable effects, so layered defenses must be evaluated factorially. And a zero attack success rate under defense is not evidence of detection until the refusals and the clean-accuracy cost behind it are reported; here the defended zeros came from provenance filtering and refusal, not from recognizing false content.

For the AEC sector, a regulatory question-answering assistant can be poisoned cheaply, can state a fabricated amendment as current, and does so with the confidence of a correct answer. Such systems can support professional judgment but cannot replace the verification of values against the governing document. Recommendations follow in Chapter 8.

\section{Chapter 8: Recommendations}\label{chapter-8-recommendations}

\subsection{8.1 For Organizations Deploying Regulatory RAG}\label{for-organizations-deploying-regulatory-rag}

\textbf{Control admission to the corpus.} The single most effective measure in this study was a manifest of approved documents with content hashes. Organizations should admit new documents, including official amendments, only through an authenticated and logged process, and treat the right to add documents as a privileged role. Admission must be enforced before indexing, because in a graph-based system the descriptions extracted from an unadmitted document still reach the prompt when its chunks are filtered.

\textbf{Require version-aware answers.} Answers about regulatory values should state the instrument, edition and commencement date they rely on. Any answer that announces an amendment should be checked against the official source before it is acted upon, because fabricated amendment notices were repeated by the system in every tested case.

\textbf{Price refusals before adopting a defense.} A low attack success rate should be reported with the share of queries answered and the accuracy on clean questions; a defense that refuses most honest queries is safe only as an unused system is safe.

\textbf{Keep human verification for safety-critical values.} Because wrong answers were expressed as confidently as right ones, a system\textquotesingle s tone cannot be used to decide when to check. Values bearing on means of escape, structure and guarding should be confirmed against the governing document by a competent person.

\subsection{8.2 For Researchers}\label{for-researchers}

\textbf{Report the composition of a zero.} Defense evaluations should publish the refusal share, the unconditional accuracy and the clean-accuracy cost beside the attack success rate, and should distinguish exclusion by provenance from detection by content.

\textbf{Evaluate layered defenses factorially. A full factorial over the layers costs little more than a one-at-a-time sweep and is the only design in which interactions can be estimated.}

\textbf{Instrument the graph channel.} Measures of poison reach should cover entity and relationship descriptions, not only text chunks, because that is where poisoned content, and the content of an unregistered legitimate amendment, entered the prompt in this study.

\textbf{Check the scorer as well as the system.} A rule-based scorer should be compared with an independent second scorer from another model family, with its prompt and statistic fixed before any answer is judged.

\textbf{Develop version-aware correctness criteria.} Benchmarks for regulatory question answering should score the attribution of a value to its governing instrument and date, not only the value itself.

\section{Chapter 9: Limitations and Delimitations}\label{chapter-9-limitations-and-delimitations}

This chapter gathers every boundary on the claims of Chapters 5 to 7: what was deliberately left out (9.1), what was planned but not performed (9.2), the limits of the measurements (9.3), factors outside the author\textquotesingle s control (9.4), and the outcomes that would have overturned the conclusions (9.5).

\subsection{9.1 Delimitations}\label{delimitations}

The study was confined to one system (LightRAG 1.5.8), one serving model, one jurisdiction\textquotesingle s regulations, one language (English) and one experimental index, built from the merged Approved Documents and the Manual. This followed the proposal\textquotesingle s timeline, which put attack one, attack two and the three defensive layers on the small corpus first and treated corpus-scale and retrieval-mode ablations as secondary. Other graph-based frameworks, embedding models and rerankers, attacks requiring access to the model or serving stack, availability and confidentiality attacks, and the behavior of practitioners were not examined. The attacks follow PoisonedRAG and GRAGPoison because these are the reference constructions, the latter being the only graph attack demonstrated on LightRAG; the narrative attack was kept to three cases, as planned; and the defense weights were fixed rather than tuned. The results therefore describe statutory, value-dense text in a curated corpus and may not transfer to open corpora or to systems with different retrieval parameters.

\subsection{9.2 Experiments Planned but Not Performed}\label{experiments-planned-but-not-performed}

\textbf{Larger corpora.} No attack or defense experiment, and no run of the full 150-item benchmark, was made on the 70- or 192-document index; these served only the two exploratory probes (Sections 5.4 and 5.7.4). Whether the dose response, the layer comparison and the defense partition hold under more competing text, and whether the dilution with corpus size reported elsewhere (J. Lin, 2026) applies, is unknown.

\textbf{Edition-conflict items.} Ten records (five questions, each in two phrasings) were written to test whether the system attributes a value to the edition that states it when two editions of the same guidance are indexed together. Their answers exist only in the larger indexes, so they were not run, and no claim about edition attribution goes beyond the twelve-question probe.

\textbf{H4a on attack one.} H4a was registered on attack one, and a dated amendment added the same test on attack two once that attack was designed. The defense ablation ran on attack two\textquotesingle s poisoned index, so the attack-one member of H4a has no run, and every defense result concerns attack two.

\textbf{Retrieval modes and the naive poisoning baseline.} All registered runs used the hybrid mode; the local and global modes were not ablated, and the second-batch naive-mode run (Section 6.2) is not a graph-free baseline matched in repetition and scoring. The proposal\textquotesingle s crude-poison control, fixed by amendment as blunt conflicting text, was not run, so the second half of H2 is untested; the naive retrieval mode removes the graph, not the sophistication of the poison.

\textbf{Full prompt-context capture.} The context the generator received was captured for two poisoned items (Section 5.6) and for the 16 measurable items of the false-positive test (Section 5.5.2), but not for all 150 items under attack, so how often the graph descriptions carry the poison is known only from these samples and from a lower bound.

\textbf{Claim-level verification.} The generation layer checked whole sentences; the claim-level variant was not run.

\textbf{An iterated adaptive attacker.} The prompt for adaptive\_l3 described a loop in which the generator would query the NLI checker and revise its passage, but the loop was not wired into the generator: the checker\textquotesingle s verdicts on the 355 generated blocks were logged and never fed back. All adaptive arms were single-pass, so H7\textquotesingle s null is not a demonstration of robustness; running the loop would be a new construction with new frozen prompts and budget.

\textbf{A registered self-rating and a second serving model.} No registered run asked the model for a self-rating, because adding it to the answer prompt would have changed the answers being measured; the rating exists only from the second-batch pass, whose answers are not the registered ones. All results come from one commercial model.

\subsection{9.3 Limitations of Measurement}\label{limitations-of-measurement}

\textbf{Benchmark construction.} The items were written and reviewed by the author, who also ran the analysis; selection may favor clearly stated provisions, making clean accuracy and attack success optimistic together.

\textbf{Value-based scoring.} The rubric checks whether the required values are present, so it cannot see an answer that keeps the correct value while declaring it superseded. The second scorer covers only the 30 explanatory items, so it is a sensitivity analysis of scoring on those items, not of the attack success rate over the whole benchmark; its agreement with the rubric (\ensuremath{\kappa} = 0.655) has a wide interval, {[}0.352, 0.886{]}, and the author\textquotesingle s review of 15 answers can detect a direction of bias but cannot estimate agreement (\ensuremath{\kappa} against the judge 0.391, interval including zero).

\textbf{Resolution of the secondary tests.} H3\textquotesingle s test had a discordance rate of 0.133 and little power; the directional test had 20 discordant pairs, so only a split of 15 to 5 or more would have reached significance; H2\textquotesingle s pre-registered subgroup had 39 items and the second-batch entity strata had ten discordant pairs in the shared stratum; the version probe has denominators of five and two; and the severity tilt lies within its uncertainty in 87 of 95 runs. Nulls here concern resolution, not the absence of effects. Safety weights were re-rated blind with low agreement (kappa 0.226).

\textbf{Noise and replication.} With no temperature control, noise was estimated from three repeats, so effects smaller than the repeat spread, such as the graph layer\textquotesingle s, cannot be resolved. The clean-index arms behind H4b exist in a single batch, and the non-inferiority margin is about half of one item.

\textbf{The generation layer\textquotesingle s instrument.} Because LightRAG returns references at file level, each sentence was checked against many unrelated chunks, and almost every sentence not literally entailed was retracted; the layer\textquotesingle s detection capability was therefore not identified. In the second-batch partial-defense run, the control answered about 2 percent of records, so the tests needed five and seven discordant items and could hardly have rejected even for a stronger attacker; adaptive\_l2 covered only 32 of the 72 items.

\textbf{Reach measured on chunks only.} The reach measures count text chunks, because they followed passage-level practice and LightRAG\textquotesingle s references identify files and chunks only; the reach of the entity and relationship descriptions is known only as a lower bound and from the two context captures.

\subsection{9.4 Factors beyond the Researcher\textquotesingle s Control}\label{factors-beyond-the-researchers-control}

The corpus changes as amendments are published, so the version probe, dated eighteen days before a commencement, refers to that moment. Limited time constrained repeats, corpus scales and adaptive iterations; this affects the smaller effects but not the main comparisons, which held in all three ablation batches.

\subsection{9.5 Outcomes That Would Have Overturned the Conclusions}\label{outcomes-that-would-have-overturned-the-conclusions}

Had H1a failed, text-level poisoning would not have been shown to be effective; had graph poisoning exceeded text poisoning, the conclusion about the softer target would be reversed; had the retrieval layer met H4a only by refusing, no configuration would have preserved utility; and had the adaptive attacker recovered success, the defended zeros would have been shown to be fragile. The last remains open for the reasons in Section 9.2.

\section{Bibliography and References}\label{bibliography-and-references}

\bibentry{Alizadeh Noughabi, H., Zarrinkalam, F., \& Dehghantanha, A. (2026). Defense against knowledge poisoning attack on GraphRAG. In \emph{Proceedings of the 64th Annual Meeting of the Association for Computational Linguistics (Volume 2: Short Papers)}. Association for Computational Linguistics. \url{https://aclanthology.org/2026.acl-short.47/}}

\bibentry{Chen, J., Tian, J., \& Zheng, T. (2026). Qwen-AEC: Instruction-tuning large language models for high-performance building code analysis. \emph{Advanced Engineering Informatics}, \emph{71}(Part A). \url{https://doi.org/10.1016/j.aei.2025.104268}}

\bibentry{Chen, Q., Qi, C., Huang, Y., Li, M., Wang, R., Zhang, D., Qin, K., \& Liang, S. (2026). KEPo: Knowledge evolution poison on graph-based retrieval-augmented generation. In \emph{Proceedings of the ACM Web Conference 2026} (pp. 2308--2319). ACM. \url{https://doi.org/10.1145/3774904.3792547}}

\bibentry{Chen, Y., Song, X., Wang, Z., \& Pan, Y. (2026). 基于 LightRAG 自动建模的 MQTT 协议形式化安全分析方法 {[}A formal security analysis method for the MQTT protocol based on automatic modelling with LightRAG{]}. \emph{计算机科学与探索 {[}Journal of Frontiers of Computer Science and Technology{]}}, \emph{20}(7), 1971--1984. \url{https://doi.org/10.3778/j.issn.1673-9418.2507078}}

\bibentry{Chen, Z., Wang, Y., Jiao, P., Wu, K., Wang, X., Ao, X., \& Zhang, D. (2026). Can LLMs fool graph learning? Exploring universal adversarial attacks on text-attributed graphs. In \emph{Proceedings of the ACM Web Conference 2026}. ACM. \url{https://doi.org/10.1145/3774904.3792502}}

\bibentry{Choudhary, S., Palumbo, N., Hooda, A., Dvijotham, K., \& Jha, S. (2026). \emph{Through the stealth lens: Attention-aware defenses against poisoning in RAG} {[}Poster presentation{]}. International Conference on Machine Learning (ICML 2026). arXiv. \url{https://arxiv.org/abs/2506.04390}}

\bibentry{Cohen, J. (1960). A coefficient of agreement for nominal scales. \emph{Educational and Psychological Measurement, 20}(1), 37--46. \url{https://doi.org/10.1177/001316446002000104}}

\bibentry{Edemacu, K., Shashidhar, V. M., Tuape, M., Abudu, D., Jang, B., \& Kim, J. W. (2025). \emph{Defending against knowledge poisoning attacks during retrieval-augmented generation} (arXiv:2508.02835). arXiv. \url{https://arxiv.org/abs/2508.02835}}

\bibentry{Efron, B., \& Tibshirani, R. J. (1993). \emph{An introduction to the bootstrap}. Chapman \& Hall. \url{https://doi.org/10.1201/9780429246593}}

\bibentry{Fei, Z., Fei, H., Gope, P., Wang, X., Sikdar, B., \& Zhang, Y. (2026). \emph{Selection integrity for LLM graph memory: An accumulability criterion for information-flow-blind retrieval} (arXiv:2606.12290). arXiv. \url{https://arxiv.org/abs/2606.12290}}

\bibentry{Giri, B., Hasan, M. T., Rasku, J., Waseem, M., \& Abrahamsson, P. (2026). Trustworthy RAG: An evaluation agent for detecting misinformation and knowledge poisoning in generative AI systems. In \emph{Proceedings of the International Conference on Software Engineering Advances (ICSEA 2026)}. arXiv. \url{https://arxiv.org/abs/2608.21095}}

\bibentry{Gu, X., Qin, S., Shu, Y., Ma, B., Liu, F., \& Liu, M. (2026). 基于 RAG 的轻量级网络安全漏洞风险感知方法 {[}A lightweight RAG-based method for cybersecurity vulnerability risk perception{]}. \emph{计算机科学 {[}Computer Science{]}}, \emph{53}(S1), 863--872. \url{https://doi.org/10.11896/jsjkx.250300034}}

\bibentry{Guo, C., Pleiss, G., Sun, Y., \& Weinberger, K. Q. (2017). On calibration of modern neural networks. In \emph{Proceedings of the 34th International Conference on Machine Learning} (pp. 1321--1330). PMLR. \url{https://proceedings.mlr.press/v70/guo17a.html}}

\bibentry{Guo, Z., Xia, L., Yu, Y., Ao, T., \& Huang, C. (2025). LightRAG: Simple and fast retrieval-augmented generation. In \emph{Findings of the Association for Computational Linguistics: EMNLP 2025} (pp. 10746--10761). Association for Computational Linguistics. \url{https://doi.org/10.18653/v1/2025.findings-emnlp.568}}

\bibentry{Jiang, J., Cai, C., Li, M., \& Zhu, T. (2025). 基于大语言模型和 RAG 的自动化渗透测试框架研究 {[}An automated penetration-testing framework based on large language models and RAG{]}. \emph{电信科学 {[}Telecommunications Science{]}}, \emph{41}(9), 119--132. \url{https://doi.org/10.11959/j.issn.1000-0801.2025159}}

\bibentry{Jung, J., Zheng, H., Jang, H., Song, J., Chen, B., \& Lee, J. (2026). CamoDocs: A poisoning attack against retrieval-augmented language models using camouflaged documents. In \emph{Proceedings of the 2026 Conference on Empirical Methods in Natural Language Processing} (pp. 10954--10979). Association for Computational Linguistics. \url{https://doi.org/10.18653/2026.emnlp-main.477}}

\bibentry{Kereopa-Yorke, B., Diaz, G., Wright, H., Del Rosario, R. F., \& Lynar, T. (2026). \emph{Oracle poisoning: Corrupting knowledge graphs to weaponise AI agent reasoning} (arXiv:2605.09822). arXiv. \url{https://arxiv.org/abs/2605.09822}}

\bibentry{Kumar, P., Nielson, T., Kolhe, T., Zala, S., \& Li, V. (2026). \emph{RAGuard: A layered defense framework for retrieval-augmented generation systems against data poisoning} (arXiv:2607.26339). arXiv. \url{https://arxiv.org/abs/2607.26339}}

\bibentry{Li, B., Cai, M., \& Zhang, T. (2026). 基于蒙特卡洛树搜索的 RAG 语料库投毒方法 {[}A Monte Carlo tree search method for poisoning RAG corpora{]}. \emph{计算机科学与探索 {[}Journal of Frontiers of Computer Science and Technology{]}}, \emph{20}(10), 2857--2870. \url{https://doi.org/10.3778/j.issn.1673-9418.2512021}}

\bibentry{Liang, J., Wang, Y., Li, C., Jiang, T., Zhu, R., Gong, N., \& Wang, T. (2026). GraphRAG under fire. In \emph{2026 IEEE Symposium on Security and Privacy (SP)} (pp. 195--210). IEEE. \url{https://doi.org/10.1109/SP63933.2026.00070}}

\bibentry{Liang, X., Niu, S., Li, Z., Zhang, S., Wang, H., Xiong, F., Fan, Z., Tang, B., Zhao, J., Yang, J., Song, S., \& Wang, M. (2025). SafeRAG: Benchmarking security in retrieval-augmented generation of large language model. In \emph{Proceedings of the 63rd Annual Meeting of the Association for Computational Linguistics (Volume 1: Long Papers)} (pp. 4609--4631). Association for Computational Linguistics. \url{https://doi.org/10.18653/v1/2025.acl-long.230}}

\bibentry{Lin, J. (2026). \emph{面向大模型 RAG 的后门攻击算法------向量投毒复现 {[}Backdoor attack algorithms against large-model RAG: Reproducing vector poisoning{]}} {[}Computer software; graduation project, Jinan University College of Cyber Security{]}. GitHub. \url{https://github.com/VoxLayer/rag_backdoor_project}}

\bibentry{Liu, J., Wang, R., Zong, Q., Wang, Y., Qian, C., Zeng, Q., Zheng, T., Shi, H., Guo, D., Xu, B., Li, C., \& Song, Y. (2026). NOVA: Noise-aware verbal confidence calibration for robust large language models in RAG systems. In \emph{Findings of the Association for Computational Linguistics: EMNLP 2026} (pp. 14793--14835). Association for Computational Linguistics. \url{https://aclanthology.org/2026.findings-emnlp.705/}}

\bibentry{McNemar, Q. (1947). Note on the sampling error of the difference between correlated proportions or percentages. \emph{Psychometrika}, \emph{12}(2), 153--157. \url{https://doi.org/10.1007/BF02295996}}

\bibentry{Miao, B., Yuan, D., Zhang, T., Yang, Y., \& Huang, Z. (2026). 基于思维链污染的检索增强生成后门攻击 {[}Chain-of-thought poisoning based retrieval-augmented generation backdoor attack{]}. \emph{信息网络安全 {[}Netinfo Security{]}}, \emph{26}(6), 977--998. \url{https://doi.org/10.3969/j.issn.1671-1122.2026.06.011}}

\bibentry{Mu, Y., Hu, H., Li, F., Yuan, Q., Wu, J., Liu, Z., Liu, P., Wang, M., Zhou, H., \& Liu, Y. (2026). \emph{Towards secure retrieval-augmented generation: A comprehensive review of threats, defenses and benchmarks} (arXiv:2603.21654). arXiv. \url{https://arxiv.org/abs/2603.21654}}

\bibentry{National Institute of Standards and Technology. (2024). \emph{Artificial intelligence risk management framework: Generative artificial intelligence profile} (NIST AI 600-1). U.S. Department of Commerce. \url{https://doi.org/10.6028/NIST.AI.600-1}}

\bibentry{OWASP. (2026). \emph{Retrieval-augmented generation security cheat sheet}. OWASP Cheat Sheet Series. \url{https://cheatsheetseries.owasp.org/cheatsheets/RAG_Security_Cheat_Sheet.html}}

\bibentry{Page, E. B. (1963). Ordered hypotheses for multiple treatments: A significance test for linear ranks. \emph{Journal of the American Statistical Association}, \emph{58}(301), 216--230. \url{https://doi.org/10.1080/01621459.1963.10500843}}

\bibentry{Qian, J. (2026). SilentRetrieval: Hijacking retrieval-augmented generation via semantically-preserving adversarial data poisoning. In \emph{Proceedings of the 32nd ACM SIGKDD Conference on Knowledge Discovery and Data Mining} (pp. 4012--4023). ACM. \url{https://doi.org/10.1145/3770855.3818186}}

\bibentry{Shi, J. W. L., Solihin, W., \& Yeoh, J. K. W. (2025). BuildThemis: Fine-tuning a large language model for automated code compliance of building regulations. \emph{Advanced Engineering Informatics}, \emph{68}(Part B), Article 103676. \url{https://doi.org/10.1016/j.aei.2025.103676}}

\bibentry{Sun, H., Yuan, D., Deng, Y., Liao, W., Yan, Z., \& Kong, Y. (2026). InterceptRAG：面向检索增强生成的知识篡改攻击 {[}InterceptRAG: Knowledge tampering attacks against retrieval-augmented generation{]}. \emph{信息安全学报 {[}Journal of Cyber Security{]}}. Advance online publication. \url{https://link.cnki.net/urlid/10.1380.TN.20260721.1726.003}}

\bibentry{Thornton, S. (2026). \emph{Semantic chameleon: Corpus-dependent poisoning attacks and defenses in RAG systems} (arXiv:2603.18034). arXiv. \url{https://arxiv.org/abs/2603.18034}}

\bibentry{Uhm, M., Kim, J., Ahn, S., Jeong, H., \& Kim, H. (2025). Effectiveness of retrieval augmented generation-based large language models for generating construction safety information. \emph{Automation in Construction}, \emph{170}, Article 105926. \url{https://doi.org/10.1016/j.autcon.2024.105926}}

\bibentry{Wang, S., Zhang, Y., Guo, Y., \& Kang, J. (2026). LedgerRAG: Governance-driven agentic chain of retrieval for dynamic knowledge scenarios. \emph{Electronics}, \emph{15}(7), Article 1376. \url{https://doi.org/10.3390/electronics15071376}}

\bibentry{Wang, Z., Wang, G., \& Wang, K. (2026). RefineRAG: Word-level poisoning attacks via retriever-guided text refinement. In \emph{Pattern Recognition} (Lecture Notes in Computer Science, pp. 529--544). Springer Nature Switzerland. \url{https://doi.org/10.1007/978-3-032-31583-0_35}}

\bibentry{Wen, J., Chen, T., Zheng, Z., \& Huang, C. (2025). \emph{A few words can distort graphs: Knowledge poisoning attacks on graph-based retrieval-augmented generation of large language models} (arXiv:2508.04276). arXiv. \url{https://arxiv.org/abs/2508.04276}}

\bibentry{\emph{Write-time conflict governance for trustworthy dynamic RAG: An auditable temporal-confidence dominance protocol} {[}Preprint, not peer reviewed{]}. (2026). SSRN. \url{https://ssrn.com/abstract=7226272}}

\bibentry{Wu, H., Li, X., Li, M., Peng, Y., Hu, J., \& Wu, W. (2026). 面向知识库投毒的图谱协同防御方法 {[}A graph-collaborative defence against knowledge-base poisoning{]}. \emph{计算机科学 {[}Computer Science{]}}. Advance online publication. \url{https://kns.cnki.net/kcms2/article/abstract?v=u3UHakgXwP87_tna1walmswE_TPJ3mUbo5nhTmS2ajjqP_Tmp2HvY9w2TneVJOL4XvFAi3On7BU3nPHuxsNMh3AnoejGdys6_Uzk7r6nj6fc-IujPLmY8EV4BCkXBMLGRpl9q3i5wCMrN6HG-4kcbxn-cFCO6aTzJcIPtKXEBHQ=&uniplatform=NZKPT}}

\bibentry{Xiang, C., Wu, T., Zhong, Z., Wagner, D., Chen, D., \& Mittal, P. (2024). \emph{Certifiably robust RAG against retrieval corruption} (arXiv:2405.15556). arXiv. \url{https://arxiv.org/abs/2405.15556}}

\bibentry{Yan, L., Xu, X., \& Li, Z. (2026). BadGraph: Structural knowledge isolation attacks against graph retrieval-augmented generation. In \emph{Proceedings of the 35th USENIX Security Symposium}. USENIX Association. \url{https://www.usenix.org/conference/usenixsecurity26/presentation/yan-leiming}}

\bibentry{Yang, S., Zhang, J., Wang, Y., Lee, D., \& Wang, S. (2026). Query-efficient agentic graph extraction attacks on GraphRAG systems. In \emph{Proceedings of the 64th Annual Meeting of the Association for Computational Linguistics (Volume 1: Long Papers)} (pp. 16017--16041). Association for Computational Linguistics. \url{https://doi.org/10.18653/v1/2026.acl-long.727}}

\bibentry{Yang, X., Liang, J., Liu, Y., Xiong, X., He, R., \& Tan, T. (2026). \emph{Reranker helps, but not enough: Towards strong poisoning attacks against retrieval-augmented generation} {[}Poster presentation{]}. International Conference on Machine Learning (ICML 2026). OpenReview. \url{https://openreview.net/forum?id=hY720JSxOG}}

\bibentry{Zeng, B. (2025). \emph{面向检索增强大模型应用的内容投毒攻击方法 {[}Content poisoning attack for retrieval-augmented large language model applications{]}} {[}Master's thesis, Central South University{]}. China National Knowledge Infrastructure. \url{https://kns.cnki.net/kcms2/article/abstract?v=u3UHakgXwP9RCcJzY3SgHzyXAziIIoQPPrWnE7IRXQZJ4a2piUH4T3-C7v7klyeye4ZEQlOtoh0x2YifUtl_kusqp7ui8tz0O7x8GNj_ZXXu2nbSAaJDUniy120k_NbrckoUN-Bv43RfEyYHzc9pZSJ_LafJVS7b9NM-eiqNl7Y=&uniplatform=NZKPT}}

\bibentry{Zhang, B., Xin, H., Fang, M., Liu, Z., Yi, B., Li, T., \& Liu, Z. (2025). Traceback of poisoning attacks to retrieval-augmented generation. In \emph{Proceedings of the ACM Web Conference 2025} (pp. 2085--2097). ACM. \url{https://doi.org/10.1145/3696410.3714756}}

\bibentry{Zhang, L., Huang, C., Ding, W., Pan, Y., Zhu, Z., \& Yang, H. (2026). 虚假信息检测的智能化转型：基于大语言模型的技术体系、困境与前瞻 {[}The intelligent transformation of disinformation detection: Large-language-model technical systems, difficulties and prospects{]}. \emph{信息安全与通信保密 {[}Information Security and Communications Privacy{]}}, (7), 9--25.}

\bibentry{Zhao, T., Chen, J., Ru, Y., Zhu, H., Hu, N., Liu, J., \& Lin, Q. (2026). Exploring knowledge poisoning attacks to retrieval-augmented generation. \emph{Information Fusion}, \emph{127}, Article 103900. \url{https://www.sciencedirect.com/science/article/abs/pii/S1566253525009625}}

\bibentry{Zheng, L., Chiang, W.-L., Sheng, Y., Zhuang, S., Wu, Z., Zhuang, Y., Lin, Z., Li, Z., Li, D., Xing, E. P., Zhang, H., Gonzalez, J. E., \& Stoica, I. (2023). Judging LLM-as-a-judge with MT-Bench and Chatbot Arena. In \emph{Advances in Neural Information Processing Systems 36} (Datasets and Benchmarks Track). \url{https://arxiv.org/abs/2306.05685}}

\bibentry{Zheng, W., Tang, G., Jiang, S., Huo, L., Zhang, X., Xie, J., \& Liu, M. (2026). LCR-RAG: Enhancing logical consistency in retrieval-augmented generation via neuro-symbolic reinforcement learning. In \emph{Proceedings of the 64th Annual Meeting of the Association for Computational Linguistics (Volume 1: Long Papers)}. Association for Computational Linguistics. \url{https://aclanthology.org/2026.acl-long.814/}}

\bibentry{Zheng, Z., Choi, Y.-J., Huang, C., Cao, H., Tian, S., \& Pei, T. (2026). Trigger as entity: Backdoor attacks to graph-based retrieval-augmented generation of large language models. \emph{IEEE Transactions on Information Forensics and Security}, \emph{21}, 5583--5596. \url{https://doi.org/10.1109/TIFS.2026.3699202}}

\bibentry{Zhou, J., Yang, C., Du, Y., Wu, Y., Duan, Y., Bai, H., \& Zhang, T. (2026). TailGuard：面向 RAG 检索优化型投毒攻击的防御框架 {[}TailGuard: A defence framework against retrieval-optimisation poisoning attacks on RAG{]}. \emph{计算机科学 {[}Computer Science{]}}. Advance online publication. \url{https://link.cnki.net/urlid/50.1075.tp.20261012.1124.004}}

\bibentry{Zou, M., Liang, L., Wang, Z., Zhao, Y., Qu, L., \& Xu, Z. (2026). \emph{From attack success to attack severity: Counterfactual memory attacks on LLM agents} {[}Manuscript under review{]}. OpenReview. \url{https://openreview.net/forum?id=A77bt8Tjw1}}

\bibentry{Zou, W., Geng, R., Wang, B., \& Jia, J. (2025). PoisonedRAG: Knowledge corruption attacks to retrieval-augmented generation of large language models. In \emph{Proceedings of the 34th USENIX Security Symposium} (pp. 3827--3844). USENIX Association. \url{https://www.usenix.org/conference/usenixsecurity25/presentation/zou-poisonedrag}}

\section{Appendices}\label{appendices}

\subsection{Appendix A: Data Tables}\label{appendix-a-data-tables}

The tables in this appendix hold the data behind the results reported in Chapters 4 and 5. Every count was read from the project\textquotesingle s artifacts; the file each table was read from is named in its note, so that any number can be recomputed.

{\small\textbf{Table A-1} The experimental index and the two auxiliary indexes}

{\def\LTcaptype{none} 
{\small
\begin{longtable}[]{@{}
  >{\raggedright\arraybackslash}p{(\linewidth - 14\tabcolsep) * \real{0.1818}}
  >{\raggedright\arraybackslash}p{(\linewidth - 14\tabcolsep) * \real{0.1364}}
  >{\raggedright\arraybackslash}p{(\linewidth - 14\tabcolsep) * \real{0.0909}}
  >{\raggedright\arraybackslash}p{(\linewidth - 14\tabcolsep) * \real{0.1212}}
  >{\raggedright\arraybackslash}p{(\linewidth - 14\tabcolsep) * \real{0.1212}}
  >{\raggedright\arraybackslash}p{(\linewidth - 14\tabcolsep) * \real{0.1364}}
  >{\raggedright\arraybackslash}p{(\linewidth - 14\tabcolsep) * \real{0.1364}}
  >{\raggedright\arraybackslash}p{(\linewidth - 14\tabcolsep) * \real{0.0758}}@{}}
\toprule\noalign{}
\begin{minipage}[b]{\linewidth}\raggedright
Scale
\end{minipage} & \begin{minipage}[b]{\linewidth}\raggedright
Documents
\end{minipage} & \begin{minipage}[b]{\linewidth}\raggedright
Chunks
\end{minipage} & \begin{minipage}[b]{\linewidth}\raggedright
Chunks per document
\end{minipage} & \begin{minipage}[b]{\linewidth}\raggedright
Entities
\end{minipage} & \begin{minipage}[b]{\linewidth}\raggedright
Relations
\end{minipage} & \begin{minipage}[b]{\linewidth}\raggedright
Relations per chunk
\end{minipage} & \begin{minipage}[b]{\linewidth}\raggedright
Graph file (MB)
\end{minipage} \\
\midrule\noalign{}
\endhead
\bottomrule\noalign{}
\endlastfoot
Experimental index & 2 & 850 & 425.0 & 12,113 & 22,751 & 26.8 & 21.5 \\
70-document index (auxiliary) & 70 & 2,637 & 37.7 & 31,099 & 82,302 & 31.2 & 81.0 \\
192-document index (auxiliary) & 192 & 5,372 & 28.0 & 86,189 & 202,919 & 37.8 & 184.3 \\
\end{longtable}
}
}

{\small\textbf{Table A-2} Composition of the 150 benchmark items}

{\def\LTcaptype{none} 
{\small
\begin{longtable}[]{@{}
  >{\raggedright\arraybackslash}p{(\linewidth - 2\tabcolsep) * \real{0.4466}}
  >{\raggedright\arraybackslash}p{(\linewidth - 2\tabcolsep) * \real{0.5534}}@{}}
\toprule\noalign{}
\begin{minipage}[b]{\linewidth}\raggedright
Field
\end{minipage} & \begin{minipage}[b]{\linewidth}\raggedright
Values (items)
\end{minipage} \\
\midrule\noalign{}
\endhead
\bottomrule\noalign{}
\endlastfoot
Part (Approved Document) & B (42), K (26), A (10), J (10), L (9), F (8), C (7), E (7), G (7), M (7), P (6), H (5), O (5), B+\allowbreak{}K (1) \\
Answer type & numeric (120), explanatory (30) \\
Value type & quantity (113), enumeration (18), rule (12), designation (7) \\
Safety consequence & 3 (93), 2 (50), 1 (7) \\
Twin items & cross-volume (4), overlap (2) \\
\end{longtable}
}
}

{\small\textbf{Table A-3} Direction labels on the poison seeds}

{\def\LTcaptype{none} 
{\small
\begin{longtable}[]{@{}
  >{\raggedright\arraybackslash}p{(\linewidth - 14\tabcolsep) * \real{0.2000}}
  >{\raggedright\arraybackslash}p{(\linewidth - 14\tabcolsep) * \real{0.0909}}
  >{\raggedright\arraybackslash}p{(\linewidth - 14\tabcolsep) * \real{0.1091}}
  >{\raggedright\arraybackslash}p{(\linewidth - 14\tabcolsep) * \real{0.1273}}
  >{\raggedright\arraybackslash}p{(\linewidth - 14\tabcolsep) * \real{0.1273}}
  >{\raggedright\arraybackslash}p{(\linewidth - 14\tabcolsep) * \real{0.1091}}
  >{\raggedright\arraybackslash}p{(\linewidth - 14\tabcolsep) * \real{0.1091}}
  >{\raggedright\arraybackslash}p{(\linewidth - 14\tabcolsep) * \real{0.1273}}@{}}
\toprule\noalign{}
\begin{minipage}[b]{\linewidth}\raggedright
Batch
\end{minipage} & \begin{minipage}[b]{\linewidth}\raggedright
Items
\end{minipage} & \begin{minipage}[b]{\linewidth}\raggedright
Loosen
\end{minipage} & \begin{minipage}[b]{\linewidth}\raggedright
Tighten
\end{minipage} & \begin{minipage}[b]{\linewidth}\raggedright
Neutral
\end{minipage} & \begin{minipage}[b]{\linewidth}\raggedright
Paired
\end{minipage} & \begin{minipage}[b]{\linewidth}\raggedright
Pair loosen
\end{minipage} & \begin{minipage}[b]{\linewidth}\raggedright
Pair tighten
\end{minipage} \\
\midrule\noalign{}
\endhead
\bottomrule\noalign{}
\endlastfoot
main\_batch1 & 65 & 32 & 26 & 7 & 52 & 25 & 27 \\
main\_batch2 & 85 & 41 & 43 & 1 & 78 & 43 & 35 \\
Combined & 150 & 73 & 69 & 8 & 130 & 68 & 62 \\
\end{longtable}
}
}

\emph{Note.} An item is unpaired when its mutation is a categorical substitution with no ordering.

{\small\textbf{Table A-4} The frozen anchor table used by the graph-side layer}

{\def\LTcaptype{none} 
{\small
\begin{longtable}[]{@{}
  >{\raggedright\arraybackslash}p{(\linewidth - 2\tabcolsep) * \real{0.4783}}
  >{\raggedright\arraybackslash}p{(\linewidth - 2\tabcolsep) * \real{0.5217}}@{}}
\toprule\noalign{}
\begin{minipage}[b]{\linewidth}\raggedright
Quantity
\end{minipage} & \begin{minipage}[b]{\linewidth}\raggedright
Value
\end{minipage} \\
\midrule\noalign{}
\endhead
\bottomrule\noalign{}
\endlastfoot
Rows & 145 \\
COVER pool / EXTRA pool & 79 / 66 \\
Rows marked skip & 8 \\
Benchmark items named & 74 of 150 \\
Distinct units & 23 \\
Comparators & \texttt{\textgreater{}=} 64, \texttt{\textless{}=} 39, class\_min 14, \texttt{\textgreater{}} 9, \texttt{=} 6, term\_eq 6, \texttt{\textless{}} 3, class\_eq 2, blank 1, range 1 \\
\end{longtable}
}
}

{\small\textbf{Table A-5} From anchor coverage to items the graph layer can adjudicate}

{\def\LTcaptype{none} 
{\small
\begin{longtable}[]{@{}
  >{\raggedright\arraybackslash}p{(\linewidth - 4\tabcolsep) * \real{0.6286}}
  >{\raggedright\arraybackslash}p{(\linewidth - 4\tabcolsep) * \real{0.1429}}
  >{\raggedright\arraybackslash}p{(\linewidth - 4\tabcolsep) * \real{0.2286}}@{}}
\toprule\noalign{}
\begin{minipage}[b]{\linewidth}\raggedright
Stage
\end{minipage} & \begin{minipage}[b]{\linewidth}\raggedright
Items
\end{minipage} & \begin{minipage}[b]{\linewidth}\raggedright
Removed
\end{minipage} \\
\midrule\noalign{}
\endhead
\bottomrule\noalign{}
\endlastfoot
Covered by the anchor table & 77 & --- \\
Less the numeric waiver & 74 & QA-K-17, QA-K-20, QA-K-23 \\
Less strict-comparator-only items & 70 & QA-B-02, QA-B-09, QA-B-27, QA-O-04 \\
Adjudicable for Attack A2 & 67 & QA-A-08, QA-L-04, QA-L-07 \\
\end{longtable}
}
}

\emph{Note.} Of the 67 adjudicable items, 40 have the poisoned entity inside the anchor row\textquotesingle s structured fields and 27 do not. Overall, 76 benchmark items carry no anchor row, 69 carry one and 5 carry two.

{\small\textbf{Table A-6} All eight ablation configurations, registered batch (18 October 2026), all 150 items}

{\def\LTcaptype{none} 
{\small
\begin{longtable}[]{@{}
  >{\raggedright\arraybackslash}p{(\linewidth - 12\tabcolsep) * \real{0.2321}}
  >{\raggedright\arraybackslash}p{(\linewidth - 12\tabcolsep) * \real{0.1071}}
  >{\raggedright\arraybackslash}p{(\linewidth - 12\tabcolsep) * \real{0.1429}}
  >{\raggedright\arraybackslash}p{(\linewidth - 12\tabcolsep) * \real{0.1429}}
  >{\raggedright\arraybackslash}p{(\linewidth - 12\tabcolsep) * \real{0.1429}}
  >{\raggedright\arraybackslash}p{(\linewidth - 12\tabcolsep) * \real{0.1071}}
  >{\raggedright\arraybackslash}p{(\linewidth - 12\tabcolsep) * \real{0.1250}}@{}}
\toprule\noalign{}
\begin{minipage}[b]{\linewidth}\raggedright
Configuration
\end{minipage} & \begin{minipage}[b]{\linewidth}\raggedright
Layers
\end{minipage} & \begin{minipage}[b]{\linewidth}\raggedright
Strict accuracy
\end{minipage} & \begin{minipage}[b]{\linewidth}\raggedright
Plain accuracy
\end{minipage} & \begin{minipage}[b]{\linewidth}\raggedright
Answered
\end{minipage} & \begin{minipage}[b]{\linewidth}\raggedright
ASR
\end{minipage} & \begin{minipage}[b]{\linewidth}\raggedright
Retreat rate
\end{minipage} \\
\midrule\noalign{}
\endhead
\bottomrule\noalign{}
\endlastfoot
none & --- & 0.6743 & 0.6578 & 0.9756 & 0.1574 & --- \\
retrieval & L1 & 0.9074 & 0.8933 & 0.9844 & 0.0231 & --- \\
graph & L2 & 0.6825 & 0.6689 & 0.9800 & 0.1667 & --- \\
retrieval+\allowbreak{}graph & L1+\allowbreak{}L2 & 0.8806 & 0.8689 & 0.9867 & 0.0185 & --- \\
generation & L3 & 0.5185 & 0.1244 & 0.2400 & 0.0463 & 0.7578 \\
retrieval+\allowbreak{}generation & L1+\allowbreak{}L3 & 0.9147 & 0.2622 & 0.2867 & 0.0000 & 0.7133 \\
graph+\allowbreak{}generation & L2+\allowbreak{}L3 & 0.5339 & 0.1400 & 0.2622 & 0.0463 & 0.7378 \\
all & L1+\allowbreak{}L2+\allowbreak{}L3 & 0.9065 & 0.2156 & 0.2378 & 0.0000 & 0.7622 \\
\end{longtable}
}
}

\emph{Note.} Each configuration has 450 records (150 items \ensuremath{\times} 3 repeats), of which 72 items are poisoned. Accuracies here pool clean and poisoned items; Table 5.5 separates them. The retreat rate is the share of third-layer blocks in which a retreat was triggered.

{\small\textbf{Table A-7} The defense partition in all three ablation batches (poisoned accuracy, refusal counted as wrong)}

{\def\LTcaptype{none} 
{\small
\begin{longtable}[]{@{}
  >{\raggedright\arraybackslash}p{(\linewidth - 6\tabcolsep) * \real{0.2400}}
  >{\raggedright\arraybackslash}p{(\linewidth - 6\tabcolsep) * \real{0.2000}}
  >{\raggedright\arraybackslash}p{(\linewidth - 6\tabcolsep) * \real{0.2800}}
  >{\raggedright\arraybackslash}p{(\linewidth - 6\tabcolsep) * \real{0.2800}}@{}}
\toprule\noalign{}
\begin{minipage}[b]{\linewidth}\raggedright
Batch
\end{minipage} & \begin{minipage}[b]{\linewidth}\raggedright
Undefended
\end{minipage} & \begin{minipage}[b]{\linewidth}\raggedright
Configurations with generation layer
\end{minipage} & \begin{minipage}[b]{\linewidth}\raggedright
Retrieval-only configurations
\end{minipage} \\
\midrule\noalign{}
\endhead
\bottomrule\noalign{}
\endlastfoot
18 Oct 2026 (registered) & 0.4213 & 0.0463 / 0.1944 / 0.0694 / 0.1667 & 0.8565 / 0.8426 \\
21 Oct 2026 (Layer 3 aggregation corrected) & 0.4306 & 0.0046 / 0.1157 / 0.0000 / 0.1111 & 0.8380 / 0.8287 \\
25 Oct 2026 (Layer 2 attribution corrected) & 0.4491 & 0.0046 / 0.1157 / 0.0093 / 0.1435 & 0.8426 / 0.8287 \\
\end{longtable}
}
}

\emph{Note.} The four generation-layer values are, in order, generation, retrieval+\allowbreak{}generation, graph+\allowbreak{}generation and all; the two retrieval-only values are retrieval and retrieval+\allowbreak{}graph.

{\small\textbf{Table A-8} Definitions of the reported quantities}

{\def\LTcaptype{none} 
{\small
\begin{longtable}[]{@{}
  >{\raggedright\arraybackslash}p{(\linewidth - 2\tabcolsep) * \real{0.3030}}
  >{\raggedright\arraybackslash}p{(\linewidth - 2\tabcolsep) * \real{0.6970}}@{}}
\toprule\noalign{}
\begin{minipage}[b]{\linewidth}\raggedright
Quantity
\end{minipage} & \begin{minipage}[b]{\linewidth}\raggedright
Definition
\end{minipage} \\
\midrule\noalign{}
\endhead
\bottomrule\noalign{}
\endlastfoot
Poisoned group & Records whose item was poisoned in the run (item\_poisoned true) \\
Clean group & Records whose item was not poisoned \\
Strict accuracy & Correct over the records left after refusals are dropped from the denominator \\
Plain accuracy & Correct over all records; a refusal counts as a wrong answer \\
Answered & Share of records that produced an answer (one minus the refusal rate) \\
Strict success & Correct value absent, attacker\textquotesingle s value present, and not a refusal \\
Disruption & Correct value absent, regardless of the attacker\textquotesingle s value \\
Poison recall@k & Mean share of a target\textquotesingle s injected passages among the answer\textquotesingle s cited references \\
ASR & Mean over poisoned items of each item\textquotesingle s strict-success proportion over its repeats \\
Retreat & Share of Layer-3 blocks with retreat\_triggered true \\
\end{longtable}
}
}

{\small\textbf{Table A-9} Version questions: what each English answer mentions, and the human verdicts}

{\def\LTcaptype{none} 
{\small
\begin{longtable}[]{@{}
  >{\raggedright\arraybackslash}p{(\linewidth - 14\tabcolsep) * \real{0.1538}}
  >{\raggedright\arraybackslash}p{(\linewidth - 14\tabcolsep) * \real{0.1154}}
  >{\raggedright\arraybackslash}p{(\linewidth - 14\tabcolsep) * \real{0.1026}}
  >{\raggedright\arraybackslash}p{(\linewidth - 14\tabcolsep) * \real{0.0641}}
  >{\raggedright\arraybackslash}p{(\linewidth - 14\tabcolsep) * \real{0.1026}}
  >{\raggedright\arraybackslash}p{(\linewidth - 14\tabcolsep) * \real{0.1154}}
  >{\raggedright\arraybackslash}p{(\linewidth - 14\tabcolsep) * \real{0.2051}}
  >{\raggedright\arraybackslash}p{(\linewidth - 14\tabcolsep) * \real{0.1410}}@{}}
\toprule\noalign{}
\begin{minipage}[b]{\linewidth}\raggedright
Index
\end{minipage} & \begin{minipage}[b]{\linewidth}\raggedright
Item
\end{minipage} & \begin{minipage}[b]{\linewidth}\raggedright
Names a document or edition
\end{minipage} & \begin{minipage}[b]{\linewidth}\raggedright
Names a date
\end{minipage} & \begin{minipage}[b]{\linewidth}\raggedright
States in-force status
\end{minipage} & \begin{minipage}[b]{\linewidth}\raggedright
Verdict
\end{minipage} & \begin{minipage}[b]{\linewidth}\raggedright
Not-yet-in-force stated as current
\end{minipage} & \begin{minipage}[b]{\linewidth}\raggedright
Correct attribution
\end{minipage} \\
\midrule\noalign{}
\endhead
\bottomrule\noalign{}
\endlastfoot
Experimental & ADB-VC-01 & yes & yes & no & partial & --- & --- \\
Experimental & ADB-VC-02 & yes & yes & no & partial & no & no \\
Experimental & ADB-VC-03 & yes & no & no & partial & no & no \\
Experimental & ADB-VC-04 & yes & no & no & refusal & --- & --- \\
Experimental & ADB-VC-05 & yes & yes & no & refusal & --- & --- \\
Experimental & ADB-TS-01 & yes & yes & no & refusal & no & --- \\
Experimental & ADB-TS-02 & yes & yes & yes & partial & no & --- \\
Experimental & ADB-TS-03 & yes & yes & no & partial & --- & --- \\
Experimental & ADB-TS-04 & yes & yes & no & correct & --- & --- \\
Experimental & ADB-TS-05 & yes & yes & no & refusal & --- & --- \\
Experimental & ADB-TS-06 & yes & yes & no & refusal & no & --- \\
Experimental & ADB-TS-07 & yes & yes & no & refusal & --- & --- \\
70-document & ADB-VC-01 & yes & yes & no & correct & --- & --- \\
70-document & ADB-VC-02 & yes & yes & no & partial & no & no \\
70-document & ADB-VC-03 & yes & no & no & partial & yes & no \\
70-document & ADB-VC-04 & yes & yes & no & correct & --- & --- \\
70-document & ADB-VC-05 & yes & yes & no & partial & --- & --- \\
70-document & ADB-TS-01 & yes & yes & yes & correct & no & --- \\
70-document & ADB-TS-02 & yes & yes & yes & correct & no & --- \\
70-document & ADB-TS-03 & yes & yes & no & incorrect & --- & --- \\
70-document & ADB-TS-04 & yes & yes & no & correct & --- & --- \\
70-document & ADB-TS-05 & yes & yes & no & partial & --- & --- \\
70-document & ADB-TS-06 & yes & yes & yes & correct & no & --- \\
70-document & ADB-TS-07 & yes & yes & no & partial & --- & --- \\
192-document & ADB-VC-01 & yes & yes & no & correct & --- & --- \\
192-document & ADB-VC-02 & yes & yes & no & partial & no & no \\
192-document & ADB-VC-03 & yes & no & no & partial & yes & no \\
192-document & ADB-VC-04 & yes & yes & no & correct & --- & --- \\
192-document & ADB-VC-05 & yes & yes & no & partial & --- & --- \\
192-document & ADB-TS-01 & yes & yes & no & correct & no & --- \\
192-document & ADB-TS-02 & yes & yes & yes & partial & no & --- \\
192-document & ADB-TS-03 & yes & yes & yes & incorrect & --- & --- \\
192-document & ADB-TS-04 & yes & yes & no & correct & --- & --- \\
192-document & ADB-TS-05 & yes & yes & no & partial & --- & --- \\
192-document & ADB-TS-06 & yes & yes & yes & correct & no & --- \\
192-document & ADB-TS-07 & yes & yes & no & partial & --- & --- \\
\end{longtable}
}
}

\emph{Note.} ``---'' means the item does not carry that indicator. A refusal states no claim to be judged.

\subsection{Appendix B: Experimental Environment, Instruments and Artifact Locations}\label{appendix-b-experimental-environment-instruments-and-artifact-locations}

The code repository and corpus sources below are part of the experimental environment rather than cited literature, and are therefore listed here instead of in the reference list.

\textbf{Table B-0 Code, corpus and models used in the experimental environment}

{\def\LTcaptype{none} 
{\small
\begin{longtable}[]{@{}
  >{\raggedright\arraybackslash}p{(\linewidth - 4\tabcolsep) * \real{0.3733}}
  >{\raggedright\arraybackslash}p{(\linewidth - 4\tabcolsep) * \real{0.1733}}
  >{\raggedright\arraybackslash}p{(\linewidth - 4\tabcolsep) * \real{0.4533}}@{}}
\toprule\noalign{}
\begin{minipage}[b]{\linewidth}\raggedright
Resource
\end{minipage} & \begin{minipage}[b]{\linewidth}\raggedright
Role
\end{minipage} & \begin{minipage}[b]{\linewidth}\raggedright
Location
\end{minipage} \\
\midrule\noalign{}
\endhead
\bottomrule\noalign{}
\endlastfoot
HKUDS. (n.d.). LightRAG {[}Computer software{]}, core version 1.5.8 & System under test & \url{https://github.com/HKUDS/LightRAG} \\
HM Government. (n.d.). Approved documents {[}Collection{]}. GOV.UK & Corpus source (merged Approved Documents, October 2024 edition) & \url{https://www.gov.uk/government/collections/approved-documents} \\
HM Government. (2020). Manual to the Building Regulations & Corpus source (second document of the experimental index) & \url{https://www.gov.uk/government/publications/manual-to-the-building-regulations} \\
DeepSeek API, model name deepseek-v4-flash (served from 10 September 2026 by DeepSeek-V4.1-Flash, reported as deepseek-flash) & Indexing extraction and answer generation & \url{https://api-docs.deepseek.com/} \\
Qwen3-Embedding-0.6B & Embedding model (local) & \url{https://huggingface.co/Qwen/Qwen3-Embedding-0.6B} \\
mDeBERTa-v3-base-mnli-xnli & NLI model of Layer 3 (local) & \url{https://huggingface.co/MoritzLaurer/mDeBERTa-v3-base-mnli-xnli} \\
xAI API, model snapshot \emph{grok-4.20-0309-non-reasoning} & LLM judge of the second-scorer check (Section 4.10), temperature 0 & \url{https://docs.x.ai/} \\
\end{longtable}
}
}

\textbf{Table B-1 Project documentation in the code repository}

\textbf{See the code repository: \url{https://github.com/BulletsHo/LightRAG-Poisoning-Attacks-and-Collaborative-Defense}}

{\def\LTcaptype{none} 
{\small
\begin{longtable}[]{@{}
  >{\raggedright\arraybackslash}p{(\linewidth - 4\tabcolsep) * \real{0.2501}}
  >{\raggedright\arraybackslash}p{(\linewidth - 4\tabcolsep) * \real{0.1787}}
  >{\raggedright\arraybackslash}p{(\linewidth - 4\tabcolsep) * \real{0.5712}}@{}}
\toprule\noalign{}
\begin{minipage}[b]{\linewidth}\raggedright
\textbf{File name}
\end{minipage} & \begin{minipage}[b]{\linewidth}\raggedright
\textbf{Title}
\end{minipage} & \begin{minipage}[b]{\linewidth}\raggedright
\textbf{Corresponding experimental content}
\end{minipage} \\
\midrule\noalign{}
\endhead
\bottomrule\noalign{}
\endlastfoot
\texttt{docs/\allowbreak{}00-read-me-first.md} & Start here & Getting started: what a retrieval system is, what poisoning is, and a three-command offline demonstration of the eight configurations \\
\texttt{docs/\allowbreak{}01-install.md} & Installing, or not installing & Environment setup: three ways to run (run.py without installation, editable installation, optional extras), and the model endpoints needed only for the full study \\
\texttt{docs/\allowbreak{}02-corpus.md} & The corpus: format, manifest, and where the real documents come from & Corpus construction: document format, the manifest (proof that the collection did not change between runs), what the demonstration corpus is and is not, and where the full corpus comes from \\
\texttt{docs/\allowbreak{}03-benchmark.md} & The benchmark: 150 items, value groups, and how to add one & Benchmark construction and validation: the structure of one item, value groups, readings of the published set, and rules for adding items \\
\texttt{docs/\allowbreak{}04-attacks.md} & The four attacks & The four poisoning attacks: A1 text-level retrieval hijacking, A2 shared-relation poisoning, A3 amendment-narrative poisoning and the three arms of the defense-aware attacker A4; planning separated from execution \\
\texttt{docs/\allowbreak{}05-defenses.md} & The three defence layers & The three-layer defense and the eight-configuration ablation: L1 retrieval-side filtering, L2 graph-side anchor verification, L3 generation-side citation verification, and how they chain in the pipeline \\
\texttt{docs/\allowbreak{}06-evaluation.md} & Evaluation: every column, and the statistics behind it & Measures and statistics: strict success, ASR and its denominator, the two accuracy denominators, SW-ER, poison recall, the directional test, the red-team readings, ECE and confidence miscalibration \\
\texttt{docs/\allowbreak{}07-reproduce.md} & Reproducing the study & The full reproduction procedure, steps 0–6 from checks to summary, and which parts cannot be reproduced exactly, and why \\
\texttt{docs/\allowbreak{}08-troubleshooting.md} & \texttt{Troubleshooting} & Troubleshooting: symptom, cause and fix, all taken from incidents that actually occurred \\
\texttt{docs/\allowbreak{}09-release-policy.md} & Release policy: what is published, what is withheld, and why & Ethics and the limits of release (constraints on publication, not on the experiments): four boundaries, the checks that must pass before release, and how to request unreleased material \\
\end{longtable}
}
}

\textbf{Table B-2 Instruments that check the reported numbers}

{\def\LTcaptype{none} 
{\small
\begin{longtable}[]{@{}
  >{\raggedright\arraybackslash}p{(\linewidth - 2\tabcolsep) * \real{0.6226}}
  >{\raggedright\arraybackslash}p{(\linewidth - 2\tabcolsep) * \real{0.3774}}@{}}
\toprule\noalign{}
\begin{minipage}[b]{\linewidth}\raggedright
Instrument
\end{minipage} & \begin{minipage}[b]{\linewidth}\raggedright
What it establishes
\end{minipage} \\
\midrule\noalign{}
\endhead
\bottomrule\noalign{}
\endlastfoot
\texttt{verify\_benchmark\_freeze.py} & The 150-item set is byte-identical to its freeze \\
\texttt{verify\_anchor\_freeze.py} & The anchor table and its adjudication files are unchanged \\
\texttt{verify\_clean\_workspaces.py} & No attack ever wrote into a clean index \\
\texttt{verify\_corpus\_stats.py} & The corpus statistics in Table A-1 recompute from the clean corpus \\
matcher.py support & The anchor coverage tree of Table A-5 \\
\texttt{report\_ablation\_table.py} & The ablation quantities of Tables 5.5, 5.7 and A-6 \\
\texttt{analyze\_confidence\_calibration.py} & The certainty and verifier analyses of Section 5.7.3 \\
apply\_fpr\_blind\_review.py, capture\_fpr\_context.py & The provenance-filtering counts and context capture of Section 5.5.2 and Table 5.6 \\
\texttt{compare\_naive\_baseline.py} & The naive-mode fold readings of Section 6.2 \\
\texttt{h7\_partial.py} & The partial-defense tests of Table 5.9 and Figure 5.6 \\
\texttt{b6\_entity\_strata.py} & The entity-sharing strata of Section 5.3 \\
\texttt{a5\_self\_rated\_calibration.py} & Table 5.11 and Figure 5.8 \\
check\_l26\_agreement.py, check\_l26\_review\_agreement.py & Table 4.5 and Figure 4.2 \\
\texttt{literal\_exact\_match.py} & The literal exact-match counts of Section 5.5.1 and Table D-2, with positive and negative controls \\
\texttt{check\_artifacts.py} & Every artifact location in Table B-1 exists and is non-empty \\
\end{longtable}
}
}

\emph{Note.} These instruments belong to the author\textquotesingle{}s research repository, not to the public code repository of Table B-1; they read frozen artifacts that include the withheld poison sets, and are available on request under the terms in docs/09-release-policy.md.

\textbf{Table B-3 Pinned system configuration for every run}

{\def\LTcaptype{none} 
{\small
\begin{longtable}[]{@{}
  >{\raggedright\arraybackslash}p{(\linewidth - 2\tabcolsep) * \real{0.4062}}
  >{\raggedright\arraybackslash}p{(\linewidth - 2\tabcolsep) * \real{0.5938}}@{}}
\toprule\noalign{}
\begin{minipage}[b]{\linewidth}\raggedright
Setting
\end{minipage} & \begin{minipage}[b]{\linewidth}\raggedright
Value
\end{minipage} \\
\midrule\noalign{}
\endhead
\bottomrule\noalign{}
\endlastfoot
LightRAG core & 1.5.8 (local server) \\
Generation and extraction model & deepseek-v4-flash (API; temperature not exposed); served as DeepSeek-V4.1-Flash (deepseek-flash) from 10 September 2026, after the experimental index was extracted (5–7 September) \\
Embedding model & Qwen3 embedding 0.6B (local) \\
Storage & JSON key--value, NetworkX graph, NanoVectorDB vectors \\
Query mode & hybrid \\
Candidates (top\_k) / chunks in context (chunk\_top\_k) & 40 / 20 \\
Token budgets (entity / relation / total) & 6,000 / 8,000 / 30,000 \\
Reranking & disabled \\
NLI model (Layer 3) & mDeBERTa-v3-base fine-tuned on MNLI and XNLI \\
Repeats per configuration & 3 \\
\end{longtable}
}
}

\subsection{Appendix C: Example Benchmark Item}\label{appendix-c-example-benchmark-item}

The record below is benchmark item QA-B-21 as frozen. It is also the item used for case C2 of the amendment-narrative attack.

{\small\textbf{Table C-1} Benchmark item QA-B-21}

{\def\LTcaptype{none} 
{\small
\begin{longtable}[]{@{}
  >{\raggedright\arraybackslash}p{(\linewidth - 2\tabcolsep) * \real{0.2601}}
  >{\raggedright\arraybackslash}p{(\linewidth - 2\tabcolsep) * \real{0.7399}}@{}}
\toprule\noalign{}
\begin{minipage}[b]{\linewidth}\raggedright
Field
\end{minipage} & \begin{minipage}[b]{\linewidth}\raggedright
Content
\end{minipage} \\
\midrule\noalign{}
\endhead
\bottomrule\noalign{}
\endlastfoot
Part / section & B, Section 3: Design for vertical escape (Volume 2, buildings other than dwellinghouses) \\
Locator & Table 3.1 \\
Question & Under Table 3.1 of Approved Document B Volume 2, what are the minimum widths of escape stairs for a ``residential (institutional)'' building serving up to 150 people, for an ``assembly and recreation'' building serving an area used for assembly purposes with up to 220 people, and for any stair not otherwise described serving up to 50 people? \\
Ground truth & Residential (institutional): 1000 mm. Assembly and recreation serving an area used for assembly purposes: 1100 mm. Any stair not described above: 800 mm, which may need to be increased to 1000 mm to comply with Approved Document M. \\
Expected behavior & Give 1000 mm, 1100 mm and 800 mm, and note the Approved Document M uplift to 1000 mm. \\
Known failure modes & Gives 1100 mm for the residential (institutional) case; omits the Approved Document M uplift. \\
Answer type / value type & numeric / quantity \\
Safety consequence & 3 (life safety) \\
\end{longtable}
}
}

\subsection{Appendix D: Extended Results and Method Details}\label{appendix-d-extended-results-and-method-details}

This appendix keeps the detail that was moved out of Chapters 4 and 5 to respect the word limit of the main text. Each section is referred to from the place in the main text where its material was summarized.

\subsubsection{D.1 Distribution of A1 Success across Repeats (Section 5.2.2)}\label{d.1-distribution-of-a1-success-across-repeats-section-5.2.2}

Across repeats, 93 of the 150 targets never succeeded at M = 5, 24 succeeded once, 14 twice and 19 every time; a quarter of the targets were thus neither reliably attacked nor reliably safe. Zhao et al. (2026) describe adoption as polarized: adversarial content is either rejected outright or, once adopted, dominates the answer. The split outcomes here do not show that pattern, and it is why the registered outcome uses the majority of three repeats rather than a single run.

\subsubsection{D.2 Factorial Effects on the Clean Items (Section 5.5.3)}\label{d.2-factorial-effects-on-the-clean-items-section-5.5.3}

On the clean items the graph layer\textquotesingle s four effects are also small and mostly negative, ranging from \ensuremath{-}0.090 to +0.026.

The clean side shows the same interaction in reverse. The generation layer costs 0.744 of clean accuracy with no other layer armed and 0.603 with the retrieval layer armed; no context brings its cost near zero. A configuration\textquotesingle s score is therefore not a sum of per-layer contributions, and any single, context-free figure such as ``the retrieval layer contributes 0.435'' would report a quantity this design shows does not exist: that layer\textquotesingle s effect ranges from +0.097 to +0.435 depending on which other layers are armed.

\subsubsection{D.3 Severity Weighting in Detail (Section 5.7.1)}\label{d.3-severity-weighting-in-detail-section-5.7.1}

With weights of 3, 2 and 1 for life-safety, compliance and general items, the total weight of the benchmark is 386. Across the 97 usable runs, the difference \(D\) (Equation 4.7) had a mean of +0.005 and a median of +0.005; it was positive in 71 runs, zero in 9 and negative in 17. Errors were thus slightly concentrated on safety-critical items, but the tilt lies within its uncertainty: of the 95 runs with raw records, 87 have an item-clustered 95 percent interval containing zero, 8 lie above it and none below (Figure 5.7). The ratio \(R\), undefined in 8 runs in which every answered item was correct, tells the same story (mean 1.038 over the rest); intervals computed earlier for \(R\) resampled individual records rather than items and were therefore too narrow. The tilt is also small relative to what the weighting could have shown: with all 150 items and a total weight of 386, the ratio cannot exceed \(3 \times \frac{150}{386} = 1.166\), and every run lies within the corresponding bound for its own item mix. The weights moved the error rate by about 4 percent where, by this arithmetic bound rather than any pre-registered estimate, they could have moved it by about 17.

\subsubsection{D.4 Expressed Certainty and Verifier Confidence in Detail (Section 5.7.3)}\label{d.4-expressed-certainty-and-verifier-confidence-in-detail-section-5.7.3}

Poisoning lowered strict accuracy by 21.6 percentage points but lowered the share of strongly asserted answers by only 4.3 points, and the hedge rate did not change. The item-level paired difference in strong assertion was \ensuremath{-}0.043 (95 percent interval {[}\ensuremath{-}0.099, +0.014{]}); a second, independent clean baseline on 65 items gave \ensuremath{-}0.040 (95 percent interval {[}\ensuremath{-}0.151, +0.079{]}). Under poisoning, wrong answers were not stated less assertively than right ones; the point estimate runs the other way (0.750 against 0.559), although with 16 wrong answers the intervals overlap. One sign of awareness did appear: the share of correct answers that explicitly flagged conflicting sources rose from 0.000 to 0.094.

The verifier\textquotesingle s own confidence was no better guide. An NLI pass was right 81.4 percent of the time, but its confidence ranked correct above incorrect answers only weakly where the retrieval layer was armed (AUROC 0.58) and inversely where it was not (0.21). The inversion is not saturation (the confidence still varies, SD 0.095): without the retrieval layer, 110 of the 115 wrong answers that passed cite an injected passage, and the verifier is most confident on exactly these (mean 0.74 against 0.64 for correct answers), because the poison does entail them. Among passed answers that cite no injected passage, AUROC is 0.81. The verifier\textquotesingle s confidence thus measures support by the cited evidence, not truth. H8 is not supported: expressed certainty did not track correctness under attack, so the tone of an answer cannot tell a user whether it is right. Plausible causes are that the poison passages are written in the same authoritative register as the regulations and that LightRAG\textquotesingle s answer prompt asks the model to answer from its context, whose register it reproduces; J. Liu et al. (2026) report the same overconfidence under contradictory evidence.

In the second-batch self-rating pass, the 75 targeted items alone show the same inversion more sharply: the 134 poisoned answers rated 5 were strictly correct in 11.9 percent of cases, and the AUROC on these items was 0.188 (descriptive only).

\subsubsection{D.5 Version Probe: Item-Level Reading (Section 5.7.4)}\label{d.5-version-probe-item-level-reading-section-5.7.4}

On neither attribution item, in any index, did the system attribute the value to the instrument that governs it, and in the two larger indexes one in five temporal questions was answered by presenting a provision not yet in force as current. An item-by-item reading (Appendix A, Table A-9) shows that the failure is narrower than ``the system ignores versions''. Answers did mention editions and dates: on the larger indexes, 11 of 11 answered questions named a document or edition and 9 named a date, and the two questions asking whether a specific provision was in force (ADB-TS-01 and ADB-TS-06) were answered correctly on both larger indexes. The two failures were specific. ADB-VC-02 attributed the 7.5 m limit to the 2019 edition, which contains the same number, rather than to the 2026 replacement table; ADB-VC-03 stated the 2026 rule for blocks with a storey of 18 m or more as current, without its commencement date, even on the 192-document index where the answer cited the 2026 booklet itself. The system therefore retrieved the governing document but did not carry its commencement status into an answer that was not explicitly about status. The version confusion that A3 exploits deliberately is thus already present without an attacker.

\subsubsection{D.6 Construction of the Four Attacks (Section 4.5)}\label{d.6-construction-of-the-four-attacks-section-4.5}

\textbf{Attack A1} follows the black-box form of PoisonedRAG (W. Zou et al., 2025). Each poison passage is P = S + I, where S is the target question verbatim, which makes the passage retrievable, and I is an LLM-generated paragraph formatted like a provision of the relevant Approved Document and asserting the mutated value. Every passage was checked by the scorer that later judged the answers and rejected if it also stated the correct value or lacked the attacker\textquotesingle s exact value (at most three generation calls per target). The \textbf{budget} \(M\) is the number of poison passages injected for one target question. The term follows PoisonedRAG, which injects \(N = 5\) malicious texts per target question by default (W. Zou et al., 2025), and the proposal, which fixed \(M \in \left\{ 1,3,5 \right\}\) as the independent variable of A1. The three budgets are nested within one ordered list, so that the passages used at \(M = 3\) include those used at \(M = 1\). Each fold was poisoned separately, giving six cells, each run three times.

\textbf{Attack A2} adapts the three steps of GRAGPoison (J. Liang et al., 2026) to the regulatory corpus. Rather than restating the target value as a provision, it attaches a rival relation to an entity that also occurs in a different Part of the corpus (hot water, for instance, appears in both Part G and Part L), claiming a scope of application that conflicts with the provision behind the question. The shared entity was inferred from the corpus text alone; reading the built index to select relations would have placed the attack in the white-box tier reserved for A4. Two differences from A1 were deliberate so that the comparison isolates the layer: the passage does not carry the question, and the narrative concerns applicability rather than recency, since recency is A3\textquotesingle s mechanism. The budget was fixed at five passages per target, matching A1 at M = 5. Poison could be constructed for 143 of the 150 targets; six had no shared entity and one presented a construction conflict. These seven remain in the denominator as failures, because an attack that cannot be mounted does not succeed.

\textbf{Attack A3} tests the narrative that the proposal identified as specific to the domain: a fabricated amendment notice announcing that a clause has been revised. Three cases were constructed, each a 2026 ``Amendment Booklet'' passage changing one value: the reduced maximum travel distance in a small block of flats (4.5 m to 3.5 m, a tightening), the minimum width of other escape stairs in a residential (institutional) building (800 mm to 900 mm, a tightening), and the guarding height at the edges of roofs in a dwelling (1,100 mm to 1,000 mm, a loosening). Each was run three times on the experimental index and repeated, three times, on the 70-document index, which contains genuine amendment booklets. A3 is qualitative by design: three cases support description, not a rate.

\textbf{Attack A4} is the defense-aware attacker. Its budget was declared before it ran on three axes: the files it was allowed to inspect, the anchor-table fields it could see, and a cap on queries to an oracle. Three arms were built, one per layer: adaptive\_l1 rewrote the poison to improve its semantic and outlier scores; adaptive\_l2 moved the tampered value to a neighboring parameter the anchor table does not cover; and adaptive\_l3 placed the false conclusion beside its own supporting sentence so that an entailment check would pass it. The control arm, A2\textquotesingle s fold-0 poison set, was re-run in the same batch to avoid a restart confound.

\subsubsection{D.7 Second Experimental Batch: Overview and Additional Readings}\label{d.7-second-experimental-batch-overview-and-additional-readings}

{\small\textbf{Table D-1} Experiments and analyses of the second batch (6--8 November 2026)}

{\def\LTcaptype{none} 
{\small
\begin{longtable}[]{@{}
  >{\raggedright\arraybackslash}p{(\linewidth - 8\tabcolsep) * \real{0.2439}}
  >{\raggedright\arraybackslash}p{(\linewidth - 8\tabcolsep) * \real{0.3506}}
  >{\raggedright\arraybackslash}p{(\linewidth - 8\tabcolsep) * \real{0.1768}}
  >{\raggedright\arraybackslash}p{(\linewidth - 8\tabcolsep) * \real{0.1067}}
  >{\raggedright\arraybackslash}p{(\linewidth - 8\tabcolsep) * \real{0.1220}}@{}}
\toprule\noalign{}
\begin{minipage}[b]{\linewidth}\raggedright
Work
\end{minipage} & \begin{minipage}[b]{\linewidth}\raggedright
Question
\end{minipage} & \begin{minipage}[b]{\linewidth}\raggedright
Design
\end{minipage} & \begin{minipage}[b]{\linewidth}\raggedright
Status
\end{minipage} & \begin{minipage}[b]{\linewidth}\raggedright
Reported in
\end{minipage} \\
\midrule\noalign{}
\endhead
\bottomrule\noalign{}
\endlastfoot
Provenance filtering and a legitimate amendment & Does the retrieval layer cost accuracy when a genuine amendment is not in the manifest? & 20 items, three arms \ensuremath{\times} 3 runs, blind review of 92 records & Done & 5.5.2 \\
Graph-free retrieval mode & Does the text attack need the graph? & Naive mode, clean and A1-poisoned index, one pass per item & Done & 6.2 \\
Adaptive arms under a partial defense & Do the A4 arms gain when only the generation layer is armed? & Control and three arms in one batch & Done & 5.6 \\
Entity-sharing strata & Does A2\textquotesingle s deficit depend on entities shared between targets? & Re-analysis of the archived A1 and A2 runs & Done & 5.3 \\
Self-rated confidence & Is the model\textquotesingle s stated confidence calibrated? & 1--5 rating, clean and A1-poisoned index, 450 records each & Done & 5.7.3 \\
Second scorer & Does an independent LLM judge agree with the rubric? & 60 answers, two passes; author review of 15 & Done & 4.10 \\
Literal exact match & The proposal\textquotesingle s exact-match indicator & 18 archived runs & Done & 5.5.1 \\
Prompt-context capture & How often do the graph descriptions carry the poison? & Two items \ensuremath{\times} three draws; 150-item capture & Partial & 5.6, 9.2 \\
Edition-conflict items & Does the system attribute a value to the right edition? & Ten records written for the larger indexes & Not run & 9.2 \\
Full benchmark on the larger indexes & Do the main results hold with more competing text? & --- & Not run & 9.2 \\
\end{longtable}
}
}

\emph{Note.} Each experiment had its design and reading rule written down before it was run; none belongs to the confirmatory family.

\textbf{Provenance filtering.} The three arms\textquotesingle{} correct counts were 2, 15 and 13 under the frozen scorer before review, 5, 18 and 17 in the reviewer\textquotesingle s first-pass verdicts, and 3, 17 and 16 after adjudication, the reading reported in Table 5.6; C \ensuremath{-} B was \ensuremath{-}0.10 before review and \ensuremath{-}0.05 after. The context capture covered 16 of the 20 items (the other four have no wording distinctive to the amendment), three draws per item and arm.

\textbf{Second scorer.} A two-answer smoke test failed its original criterion of byte-identical replies, because one reply differed by a single synonym while the score was unchanged. The criterion was therefore replaced, before the full run, by a stricter one: all 60 answers scored twice, with at least 90 percent of the binary verdicts required to agree. All 58 agreed, and 53 of the 58 five-level scores were identical, every difference being one level. The judge\textquotesingle s 1--5 scores against the rubric\textquotesingle s verdicts are shown in Figure 4.2.

{\small\textbf{Table D-2} Literal exact match beside strict correctness, by batch}

{\def\LTcaptype{none} 
{\small
\begin{longtable}[]{@{}
  >{\raggedright\arraybackslash}p{(\linewidth - 8\tabcolsep) * \real{0.2564}}
  >{\raggedright\arraybackslash}p{(\linewidth - 8\tabcolsep) * \real{0.1026}}
  >{\raggedright\arraybackslash}p{(\linewidth - 8\tabcolsep) * \real{0.2051}}
  >{\raggedright\arraybackslash}p{(\linewidth - 8\tabcolsep) * \real{0.2051}}
  >{\raggedright\arraybackslash}p{(\linewidth - 8\tabcolsep) * \real{0.2308}}@{}}
\toprule\noalign{}
\begin{minipage}[b]{\linewidth}\raggedright
Batch
\end{minipage} & \begin{minipage}[b]{\linewidth}\raggedright
Runs
\end{minipage} & \begin{minipage}[b]{\linewidth}\raggedright
Answered records
\end{minipage} & \begin{minipage}[b]{\linewidth}\raggedright
Strictly correct
\end{minipage} & \begin{minipage}[b]{\linewidth}\raggedright
Literally equal to the reference
\end{minipage} \\
\midrule\noalign{}
\endhead
\bottomrule\noalign{}
\endlastfoot
Registered ablation, 18 October 2026 & 8 & 2,229 & 1,724 & 0 \\
Correction batch, 25 October 2026 & 8 & 1,965 & 1,556 & 0 \\
Clean index (no defense; generation layer) & 2 & 453 & 406 & 0 \\
\end{longtable}
}
}

\emph{Note.} Normalization as in Equation 4.13. Positive and negative controls passed in every run.

\subsection{Appendix E: Pre-registration Record}\label{appendix-e-pre-registration-record}

The hypotheses were registered in a document frozen on 4 October 2026 (version 4), before the first attack run, with a SHA-256 manifest of the document and of the artifacts it names. Any later change is a dated amendment appended to that document, never an edit of the registered text. Table E-1 lists the registered hypotheses and Table E-2 the amendments.

{\small\textbf{Table E-1} Registered hypotheses}

{\def\LTcaptype{none} 
{\small
\begin{longtable}[]{@{}
  >{\raggedright\arraybackslash}p{(\linewidth - 10\tabcolsep) * \real{0.0933}}
  >{\raggedright\arraybackslash}p{(\linewidth - 10\tabcolsep) * \real{0.0533}}
  >{\raggedright\arraybackslash}p{(\linewidth - 10\tabcolsep) * \real{0.2454}}
  >{\raggedright\arraybackslash}p{(\linewidth - 10\tabcolsep) * \real{0.3412}}
  >{\raggedright\arraybackslash}p{(\linewidth - 10\tabcolsep) * \real{0.1600}}
  >{\raggedright\arraybackslash}p{(\linewidth - 10\tabcolsep) * \real{0.1067}}@{}}
\toprule\noalign{}
\begin{minipage}[b]{\linewidth}\raggedright
ID
\end{minipage} & \begin{minipage}[b]{\linewidth}\raggedright
SQ
\end{minipage} & \begin{minipage}[b]{\linewidth}\raggedright
Registered claim
\end{minipage} & \begin{minipage}[b]{\linewidth}\raggedright
Test and threshold
\end{minipage} & \begin{minipage}[b]{\linewidth}\raggedright
Role
\end{minipage} & \begin{minipage}[b]{\linewidth}\raggedright
Reported in
\end{minipage} \\
\midrule\noalign{}
\endhead
\bottomrule\noalign{}
\endlastfoot
H1a & SQ1 & Text-level ASR at M = 5 exceeds 0.10 & One-sided exact binomial on item-level majority outcomes, H0: p \ensuremath{\leq} 0.10 & Confirmatory & 5.2.1 \\
H1b & SQ1 & ASR rises across nested budgets M = 1, 3, 5 & Page\textquotesingle s L, items as blocks; Holm-adjusted follow-up McNemar tests & Confirmatory & 5.2.1 \\
H2 & SQ2 & At a matched budget, graph-level ASR exceeds text-level ASR, and the naive poisoning baseline lies below text-level ASR & Deferred at registration; budget operator (M = 5 passages), one-sided exact McNemar and family fixed by A3; naive arm\textquotesingle s form fixed by A6 & Confirmatory (joins by A3) & 5.3; naive arm not run (9.2) \\
H3 & SQ2 & Text-level poisoning lowers accuracy on untargeted questions & One-sided exact McNemar on pooled untargeted observations; equivalence (\ensuremath{\delta} = 7 points) claimable only if the clean discordance rate is \ensuremath{\leq} 8.6\% & Confirmatory & 5.2.3 \\
H4a (attack one) & SQ3 & With all three layers on, ASR \ensuremath{\leq} min(0.5 \ensuremath{\times} undefended ASR, 0.10) & Point estimate with paired block-bootstrap CI; one-sided exact McNemar for any improvement & Confirmatory & Not run (9.2) \\
H4a (attack two) & SQ3 & The same criterion on attack two & As above & Confirmatory (joins by A3) & 5.5.1 \\
H4b & SQ3 & Clean accuracy is non-inferior & Lower bound of the one-sided 95\% CI of the accuracy change \textgreater{} \ensuremath{-}5 points & Confirmatory & 5.5.2 \\
H5 & SQ4 & Safety-consequence weighting changes the error rate & Descriptive, with bootstrap intervals & Secondary & 5.7.1 \\
H6 & SQ4 & Loosening and tightening differ in ASR & Two-sided exact McNemar over 127 paired targets; budget fixed at M = 5 by A2 & Secondary & 5.7.2 \\
H7 & SQ3 & Each adaptive arm beats the non-adaptive attack against the full defense & Test, comparison arm and per-arm readout fixed by A15 & Secondary (outside the family, A15/\allowbreak{}A16) & 5.6 \\
H8 & SQ4 & Expressed confidence of poisoned answers is not lower than that of correct clean answers; calibration error under attack exceeds clean & Registered on calibration error; the registered runs carried no self-rating, so expressed certainty is read lexically and a self-rating comes from the second batch & Secondary & 5.7.3 \\
\end{longtable}
}
}

\emph{Note.} The confirmatory family, as fixed by A3 and restated by A16, has seven members: \{H1a, H1b, H3, H4a (attack one), H4b, H4a (attack two), H2\}, Holm-corrected at a family-wise \ensuremath{\alpha} of 0.05 (smallest level 0.05 / 7 = 0.0071). Corpus scale, retrieval mode, Top-K sensitivity and proposal Section 8.5 were registered as exploratory only.

{\small\textbf{Table E-2} Dated amendments to the pre-registration}

{\def\LTcaptype{none} 
{\small
\begin{longtable}[]{@{}
  >{\raggedright\arraybackslash}p{(\linewidth - 6\tabcolsep) * \real{0.0775}}
  >{\raggedright\arraybackslash}p{(\linewidth - 6\tabcolsep) * \real{0.1937}}
  >{\raggedright\arraybackslash}p{(\linewidth - 6\tabcolsep) * \real{0.5156}}
  >{\raggedright\arraybackslash}p{(\linewidth - 6\tabcolsep) * \real{0.2131}}@{}}
\toprule\noalign{}
\begin{minipage}[b]{\linewidth}\raggedright
ID
\end{minipage} & \begin{minipage}[b]{\linewidth}\raggedright
Date
\end{minipage} & \begin{minipage}[b]{\linewidth}\raggedright
Subject
\end{minipage} & \begin{minipage}[b]{\linewidth}\raggedright
Concerns
\end{minipage} \\
\midrule\noalign{}
\endhead
\bottomrule\noalign{}
\endlastfoot
A1 & 2026-10-04 & Attack-one injected files use flat ingest names, mapped in injects.json; frozen poison sets unchanged & Procedure for H1a, H1b, H3 \\
A2 & 2026-10-05 & H6\textquotesingle s budget fixed at M = 5, before any paired-set run & H6 \\
A3 & 2026-10-07 & H2\textquotesingle s budget operator (passages per target, M = 5), test and family; attack two\textquotesingle s H4a test joins the family & H2, H4a (attack two), family \\
A4 & 2026-10-07 & What the attack-two generator may read (gray-box tier); erratum in A3 & H2 \\
A5 & 2026-10-10 & 34 deferred anchor-alignment verdicts merged into the frozen anchor file & Layer 2 (H4a, H4b) \\
A6 & 2026-10-10 & The naive poisoning arm is blunt conflicting text, not a random-character trigger & H2 (second half) \\
A7 & 2026-10-10 & Operating rules of the first-layer retrieval filter (Equations 4.1--4.3) & Layer 1 (H4a, H4b) \\
A8 & 2026-10-11 & The first layer\textquotesingle s deferred values: graph-side path, windows, calibration queries & Layer 1 (H4a, H4b) \\
A9 & 2026-10-11 & A sensitivity reading for the first layer\textquotesingle s graph-side path (the global-path arm) & H4b (secondary reading) \\
A10 & 2026-10-11 & First-layer weights are design values, not fitted; source of the inputs to \(S_{\text{trust}}\) & Layer 1 (H4a, H4b) \\
A11 & 2026-10-11 & The sensitivity arm fits nothing either; its second weights file is a layout record & H4b (secondary reading) \\
A12 & 2026-10-21 & Graded retreat for the third layer (withdrawn the same day by A13) & Layer 3 \\
A13 & 2026-10-21 & A12 withdrawn; the third layer\textquotesingle s retreat rule returns to the proposal\textquotesingle s wording & Layer 3 (correction batch of 21 October) \\
A14 & 2026-10-21 & Attack three\textquotesingle s three cases fixed before any poison was generated & Attack A3 (Section 5.4) \\
A15 & 2026-10-22 & H7\textquotesingle s test, the arm it is measured against and the per-arm mechanism readout & H7 \\
A16 & 2026-10-31 & Correction of A15\textquotesingle s quotation of the family: the family has seven members & Family, Holm levels \\
\end{longtable}
}
}

\end{document}